\documentclass[12pt,preprint]{aastex}
\def\sgn{\mathop{\rm sgn}}					% defining the sgn(x) function

\begin{document}

  \shorttitle{SDSS Stripe 82 Variable Stars}
  \shortauthors{Bhatti, Richmond, Ford \& Petro}

  \title{Variable Point Sources in Sloan Digital Sky Survey Stripe 82.\\
    I. Project Description and Initial Catalog (0 h $\le \alpha \le$ 4 h)}

  \author{Waqas A. Bhatti\altaffilmark{1}, Michael W. Richmond\altaffilmark{2}, Holland
  C. Ford\altaffilmark{1}, Larry D. Petro\altaffilmark{3}}
  \altaffiltext{1}{Department of Physics \& Astronomy, Johns Hopkins University, 3400 N. Charles St.,
    Baltimore, MD 21218, email: waqas@pha.jhu.edu}
  \altaffiltext{2}{Department of Physics, Rochester Institute of Technology, 84 Lomb Memorial Dr., Rochester,
    NY 14623}
  \altaffiltext{3}{Space Telescope Science Institute, 3700 San Martin Dr., Baltimore, MD 21218}

%%%%%%%%%%%%%%
%% ABSTRACT %%
%%%%%%%%%%%%%%

  \begin{abstract}

    We report the first results of a study of variable point sources identified using multi-color time-series
    photometry from Sloan Digital Sky Survey (SDSS) Stripe 82, including data from the SDSS-II Supernova
    Survey, over a span of nearly ten years (1998--2007). We construct a light-curve catalog of 221,842 point
    sources in the RA 0 to 4 h half of Stripe 82, limited to $r = 22.0$ mag, that have at least 10 detections
    in the $ugriz$ bands and color errors $<$ 0.2 mag. These sources are then classified by color and by
    cross-matching them to existing SDSS catalogs of interesting objects. Inhomogenous ensemble differential
    photometry techniques are used to greatly improve our sensitivity to variability and reduce contamination
    by sources that appear variable due to large photometric noise or systematic effects caused by non-uniform
    photometric conditions throughout the survey. We use robust variable identification methods to extract
    6,520 variable candidates from this dataset, resulting in an overall variable fraction of $\sim 2.9\%$ at
    the level of $\sim 0.05$ mag variability. Despite the sparse and uneven time-sampling of the light-curve
    data, we discover 143 periodic variables in total. Due to period ambiguity caused by relatively poor phase
    coverage, we identify a smaller final set of 101 periodic variables with well-determined periods and
    light-curves. Among these are 55 RR Lyrae, 30 eclipsing binary candidates, and 16 high amplitude Delta
    Scuti variables. In addition to these objects, we also identify a sample of 2,704 variable quasars matched
    to the SDSS Quasar Catalog \citep{2007AJ....134..102S}, which make up a large fraction of our variable
    candidates. An additional 2,403 quasar candidates are tentatively identified and selected by their
    non-stellar colors and variability. A sample of 11,328 point sources that appear to be nonvariable given
    the limits of our variability sensitivity is also briefly discussed. Finally, we describe several
    interesting objects discovered among our eclipsing binary candidates, and illustrate the use of our
    publicly available light-curve catalog\footnote{Available at
    \url{http://shrike.pha.jhu.edu/stripe82-variables}} by tracing Galaxy halo substructure with our small
    sample of RR Lyrae variables.
  
  \end{abstract}


%%%%%%%%%%%%%%%%%%
%% INTRODUCTION %%
%%%%%%%%%%%%%%%%%%

  \section{Introduction}\label{sec_intro}

  The last fifteen years have yielded a wealth of new knowledge of many types of astronomical objects due to
  the prevalence of large scale surveys such as the Sloan Digital Sky Survey (SDSS;
  \citealp{2000AJ....120.1579Y}) and the Two-Micron All Sky Survey (2MASS;
  \citealp{2006AJ....131.1163S}). These surveys cover large areas of the sky, are deep, are sensitive to many
  different types of objects, and most importantly, have uniform data reduction and characterization. Powerful
  data access tools made available to the community allow efficient dissemination and data-mining, and allow
  large populations of objects to be studied all at once.

  The next generation of astronomical surveys will include a powerful new tool: the exploration of the
  time-domain. Many classes of astronomical objects vary over differing timescales, so a uniform approach that
  covers all of these timescales while covering large portions of the sky will produce a diverse catalog of
  variable objects that may be studied in much the same manner as the static sky has been studied using SDSS
  and 2MASS. Two such projects in advanced stages of planning and development are the Pan-STARRS project
  (\citealp{2002SPIE.4836..154K}; currently undergoing commissioning) and the Large Synoptic Survey Telescope
  (LSST; \citealp{2002SPIE.4836...10T,2008arXiv0805.2366I}; expected to start observations in 2015).

  On a smaller scale, the variable sky has been explored by many successful projects, including OGLE (towards
  the Galactic Bulge; \citealp{2002AcA....52....1U}), MACHO (Galactic Bulge and Magellanic Clouds;
  \citealp{2001ApJS..136..439A}), and ASAS (all-sky to V = 15.0; \citealp{2002AcA....52..397P}). Photometric
  surveys for transiting planets in recent years, such as HAT \citep{2002PASP..114..974B}, TrES
  \citep{2004ApJ...613L.153A}, SuperWASP \citep{2006PASP..118.1407P}, and XO \citep{2006ApJ...648.1228M} have
  also resulted in studies of several classes of variable stars. In addition to these surveys, specialized
  searches for supernovae, including the CFHT Supernova Legacy Survey \citep{2005ASPC..342..466S} and the
  SDSS-II Supernova Survey \citep{2008AJ....135..338F}, can provide large and uniform datasets suitable for
  the identification and characterization of many different types of variable sources.

  Although the SDSS is largely a single-epoch survey, it covers some regions of the sky multiple times. The
  most prominent of these is a region on the celestial equator known as Stripe 82. This region has been
  surveyed repeatedly by the SDSS over many years, most recently by the SDSS-II Supernova Survey, and has been
  studied for variability by several authors. \citet{2007AJ....134.2236S} presented a catalog of 13,051
  variable star candidates discovered using observations of the Stripe carried out before the advent of the
  SDSS-II Supernova Survey. \citet{2008MNRAS.386..887B} published a light-curve catalog of Stripe 82,
  incorporating data from the first two years of the Supernova Survey. \citet{2008ApJ...684..635B} discovered
  a low-mass eclipsing binary in this region, using light-curves also generated from the first two years of
  the Supernova Survey. \citet{2008MNRAS.386..416B} reported another low mass eclipsing binary discovered in
  the footprint of Stripe 82, but using calibration data from 2MASS. More recently,
  \citet{2009MNRAS.398.1757W} and \citet{2009AJ....138..633K} have presented catalogs of RR Lyrae and M-dwarf
  flare stars present in Stripe 82 respectively.

  Here, we construct a light-curve catalog for a magnitude limited sample of point sources in SDSS Stripe 82,
  using data ranging from the first observations of this area of sky in 1998 to the high cadence observations
  carried out by the SDSS-II Supernova Survey (2005--2007). We use inhomogeneous ensemble differential
  photometry to remove systematic artifacts from these light-curves caused by variable photometric
  conditions. Point sources are then classified by color and by cross-matching to other catalogs of
  interesting objects in Stripe 82. The large number of observation epochs available in our dataset then
  allows robust identification of variable sources, especially periodic variables. 

  We first construct a light-curve catalog of variable point sources detected in Stripe 82. We then
  concentrate on three classes of periodic variables: eclipsing binaries, RR Lyrae, and Delta Scuti, and
  present periods and phase-folded light-curves for all such objects identified. In addition, we use the color
  and variability properties of quasi-stellar objects (QSOs) matched to the SDSS Quasar Catalog
  \citep{2007AJ....134..102S} to identify a new sample of variable candidate QSOs in our light-curve
  catalog. Finally, we extend the work of \citet{2007AJ....134..973I} by using a larger number of observation
  epochs and our robust variable extraction methods to identify a sample of objects that appear
  \emph{nonvariable} at the limits of our sensitivity.

  In this paper (the first of two), we first describe how we extract objects from the detection catalogs from
  the SDSS pipeline, and subsequently organize and generate an initial light-curve catalog for the RA 0 to 4 h
  half of Stripe 82 (Section \ref{sec_sdss}). We describe the inhomogenous ensemble differential photometry
  algorithms in detail (Section \ref{sec_ensemble}), and then discuss how point source classification and
  variable extraction are implemented (Sections \ref{sec_class_point} and \ref{sec_variables}). Two
  independent period search algorithms are described and the difficulties we face in searching for periodic
  variability in our sparse and unevenly sampled dataset are also outlined (Section \ref{sec_periods}). We
  then apply these variable extraction and period finding algorithms to objects in the RA 0 to 4 h half of
  Stripe 82, and describe the general properties of variables discovered after processing this initial
  light-curve catalog (Section \ref{sec_var_gen_prop}).

  We characterize the completeness and efficiency of our variable extraction pipeline by carrying out
  end-to-end simulations of the entire process (Section \ref{sec_efficiency}). Our period finding algorithms
  are analyzed in a similar manner and provide estimates for the efficiency and completeness of our periodic
  variable sample. Instructions on how to access our publicly available data are then provided (Section
  \ref{sec_access}). We then present and discuss the properties of eclipsing binaries, RR Lyrae, and Delta
  Scuti variables identified in this initial light-curve catalog (Section
  \ref{sec_periodic_variables}). Finally, we discuss samples of candidate QSOs and nonvariable objects also
  identified by our pipeline (Sections \ref{sec_sdss_qsos} and \ref{sec_nonvariables}). The second paper in
  this series will complete our light-curve catalog by adding objects and variables identified in the RA 20 to
  0 h half of Stripe 82, and extend the discussion of their properties in the context of a completely
  processed sample.

%%%%%%%%%%%%%%%%%%%%%%%%%%%%%%%%%%
%% THE SDSS-II SUPERNOVA SURVEY %%
%%%%%%%%%%%%%%%%%%%%%%%%%%%%%%%%%%

  \section{Data from SDSS Stripe 82}\label{sec_sdss}

  \subsection{Overview}\label{sec_sdss_overview}

  The Sloan Digital Sky Survey (SDSS) uses a dedicated 2.5-m telescope located at the Apache Point Observatory
  in New Mexico. During imaging, the telescope points at a fixed declination on the meridian and scans the sky
  using the time-delay and integrate mode that clocks the charge across the charge-coupled device arrays
  (CCDs) at the sidereal rate. The record of one particular scan of a $\sim 1.3\arcdeg$-wide \emph{strip} is
  called a \emph{run}. Two overlapping \emph{strips} make up one \emph{stripe}; as a result, each
  \emph{stripe} is $\sim 2.5\arcdeg$ wide. Each run exposes five CCDs in five filters, \emph{u}, \emph{g},
  \emph{r}, \emph{i}, and \emph{z} for approximately 54 seconds per pixel. One run is further broken up into
  \emph{fields}, which are $13\arcmin \times 10\arcmin$ each. Detections are fed through the SDSS photometric
  pipeline, are classified by morphological type, and are assigned magnitudes and associated
  uncertainties. Point source objects are well described by fitting a point spread function to the detection,
  resulting in so-called PSF magnitudes, while extended source objects are characterized by model
  magnitudes. Details of the SDSS photometry and classification algorithms may be found in
  \citet{2006AJ....131.2332G}, \citet{2002SPIE.4836..350L}, \citet{2002AJ....123..485S},
  \citet{2001AJ....122.2129H}, and references therein. Observations take place on nights that have seeing
  better than 1.7$\arcsec$ (FWHM), are moonless, and show little transparency variation due to cloud cover
  \citep{2000AJ....120.1579Y,2001AJ....122.2129H}.
  
\begin{figure}
    \plottwo{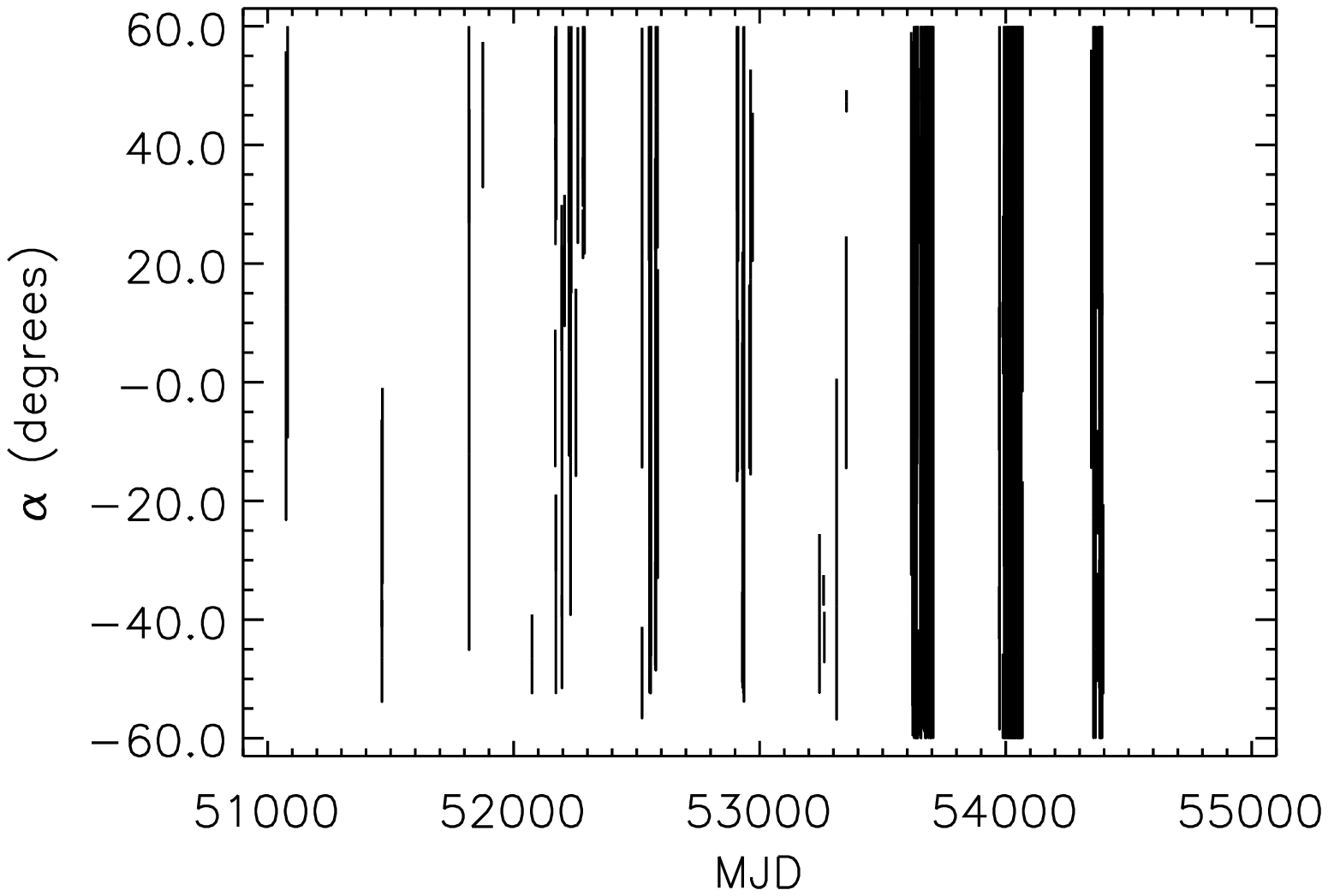}{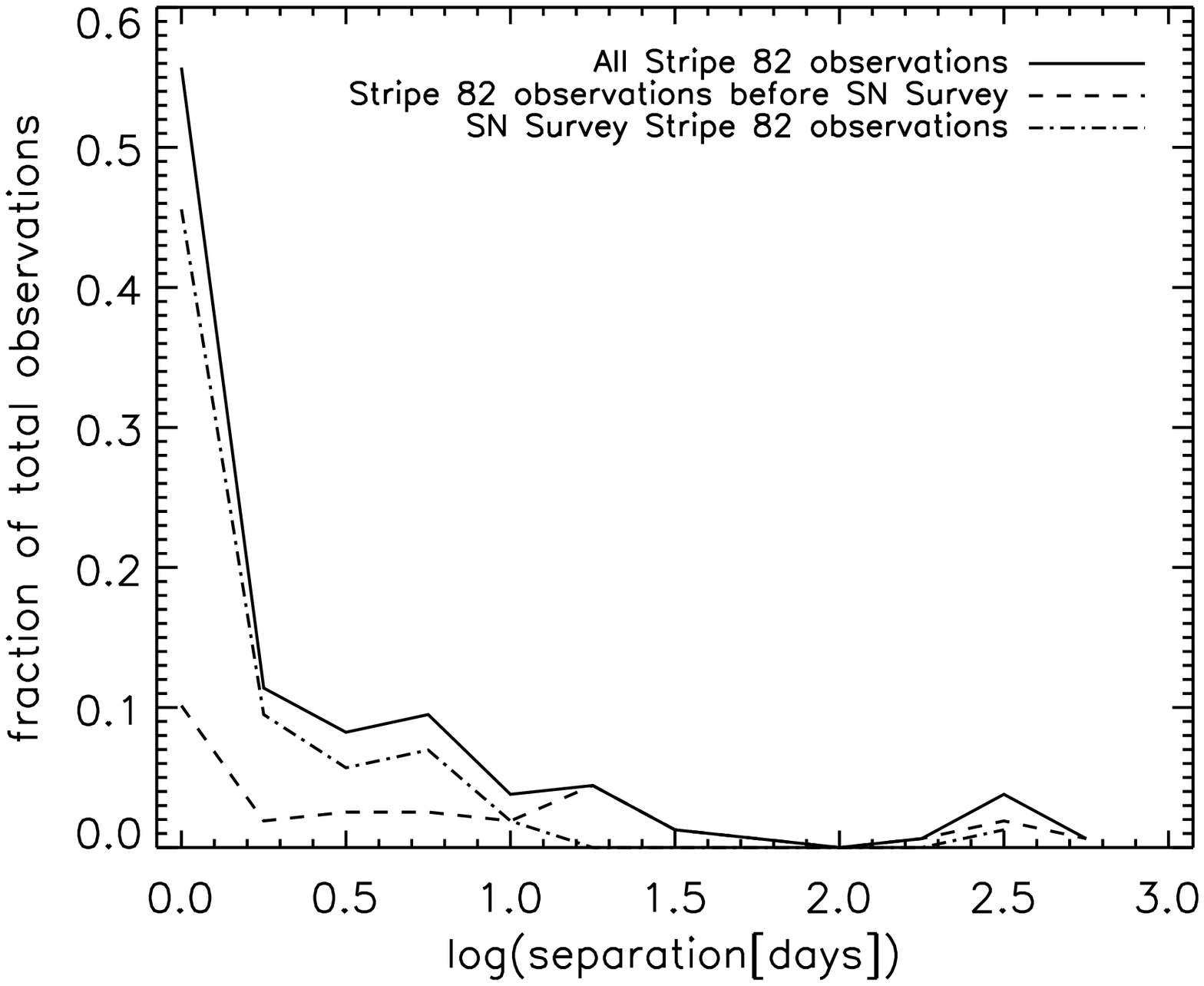}
    \caption{Left: Stripe 82 run temporal coverage vs. right ascension. The dense clusters at the right are
    the observation runs associated with the Supernova Survey. Note the large gaps between successive years of
    coverage of the Stripe. Right: Histogram of the separation in days between consecutive observations of
    Stripe 82. The bins are 0.25 dex wide. The solid line represents all observations of the Stripe; the
    dashed line represents observations of the Stripe before the SN Survey; the dot-dashed line represents
    observations of the Stripe carried out during the SN Survey. The high relative cadence of the SN Survey
    observations is apparent.}
    \label{fig_ras_runs}
  \end{figure}

  Stripe 82 is on the celestial equator, ranging from 20 h to 4 h in right ascension and -$1.27\arcdeg$ to
  $+1.27\arcdeg$ in declination, for a total area of about 300 sq. deg. The region is at high Galactic
  latitude, ranging from $-30\arcdeg$ to $-70\arcdeg$ in Galactic latitude and between $40\arcdeg$ and
  $200\arcdeg$ in Galactic longitude. The Stripe has been observed many times over the 10 years of operation
  of SDSS and SDSS-II, mostly during the fall (September -- December). Figure \ref{fig_ras_runs} (left panel)
  shows the temporal coverage of the Stripe as a function of time over all ten years of observation.

  Starting with commissioning runs in 2004, and full operations in 2005, Stripe 82 was observed at higher
  cadence by the SDSS-II Supernova Survey (hereafter, the SN Survey; \citealt{2008AJ....135..338F}). The
  nominal time between consecutive observations of a field was roughly 2 days (see Figure \ref{fig_ras_runs},
  right panel). The SN Survey was designed to obtain a uniform sample of medium-redshift ($z \sim 0.05-0.35$)
  Type Ia supernovae for the purposes of precision cosmology. The requirement for improved temporal coverage
  necessitated a compromise on the photometric quality. As a result, the SN Survey observed on nights that
  would not have been designated as photometric for the legacy SDSS. This relaxed photometric quality
  requirement poses a challenge for identification of `true' variables from the dataset, as we will note in
  Section \ref{sec_ensemble}.

\begin{figure}
    \plotone{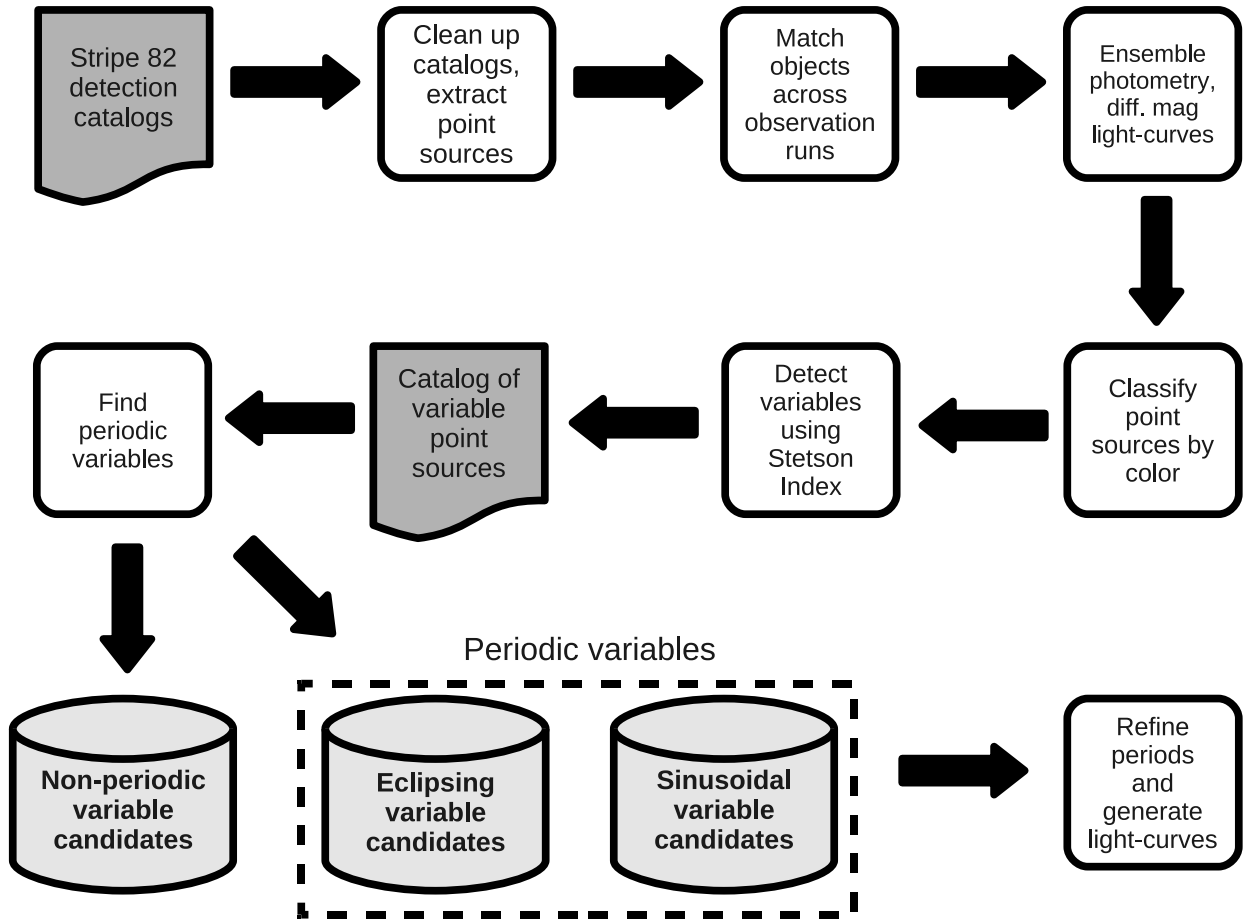}
    \caption{A schematic view of our processing pipeline. The initial extraction, reduction of object catalogs
    from the SN Survey, and ensemble photometry is described in Section \ref{sec_sdss}. Variables are detected
    using the methods presented in Section \ref{sec_variables}. We find periodic variables by running the
    period search algorithms outlined in Section \ref{sec_periods}. Once these are found, we refine their
    periods, and generate their phased light-curves and ephemerides. Non-periodic variables include long term
    variables (variable on scales of years), quasars, and flare stars.}
    \label{fig_pipeline}
  \end{figure}

  In our dataset, we have 242 runs from Stripe 82: 69 from observations of the Stripe before the SN Survey, 65
  for the first season of the SN Survey in 2005, 86 runs during the second season in 2006, and 22 runs during
  the final season in 2007\footnote{A full list of runs used is available at our catalog's website.}. We
  obtained the pre-SN Survey data for Stripe 82 using the Catalog Archive Server (CAS) SQL
  interface\footnote{\url{http://casjobs.sdss.org/CasJobs/}} to the SDSS database, downloading FITS files with
  lists of object detections per run along with their photometric properties. The data for the SN Survey was
  reduced by the SDSS photometric pipeline and made available on the SDSS Data Archive Server\footnote{See
  \url{http://das.sdss.org/www/html/imaging/dr-byRun-74.html}.} as calibrated object catalogs (tsObj FITS
  binary table files, one per field) and calibrated images (fpC FITS image files, one per field). We
  downloaded the calibrated object catalogs in form of tsObj FITS files for runs associated with the SN
  Survey, put them together with the object catalogs for the pre-SN Survey runs obtained earlier, and then
  processed all 242 runs through a multi-stage pipeline designed to extract suitable point sources and search
  for possible variables. An overview of the pipeline is given in Figure \ref{fig_pipeline}, while details are
  discussed below.

%%%%%%%%%%%%%%%%%%
%% PRE-ENSEMBLE %%
%%%%%%%%%%%%%%%%%%

  \subsection{Object Extraction and Pre-Ensemble Processing}\label{sec_pre_ensemble}

  The data fields extracted from the SN Survey data are listed in Table \ref{param_table}. We imposed a
  uniform set of quality conditions using status and quality flags available in the tsObj and CAS output FITS
  files\footnote{See \url{http://www.sdss.org/dr7/products/catalogs/flags.html} for details.}. We first
  required that a detected object: (1) was not marked as saturated (BRIGHT and SATURATED flags set to 0), (2)
  was not a deblended child of a nearby object (PARENT set to -1, or PARENTID set to 0), (3) had no deblended
  children itself (NCHILD set to 0), (4) had the EDGE flag set to 0. Second, we required a detected object's
  STATUS flags included SET, GOOD, OK\_RUN, OK\_SCANLINE, OK\_STRIPE all set to 1, in addition to either
  PRIMARY or SECONDARY set to 1. We explicitly discarded all objects that had a status flag of DUPLICATE set
  to 1, indicating a duplicate detection within the same observation run (in overlapping adjacent fields for
  example). Third, we ensured that an object had valid magnitudes recorded in all five bands by requiring that
  these measured values all be greater than -9999.0. Finally, we imposed a faint PSF magnitude limit of $z =
  21.0$. Although we extracted both PSF and fiber magnitudes for each object from the catalogs, we only used
  the PSF magnitudes for all of our subsequent processing and variability analysis.

  We then extracted only those objects classified by the SDSS photometric pipeline as point sources, by
  requiring the object TYPE flag be set to 6 (STAR). This left us with 57,406,616 point source detections over
  all runs. We then constructed a canonical match template for the entire Stripe using object detections in
  four high photometric quality SN Survey runs: North runs 5610 (MJD 53628) and 6430 (MJD 54011), and South
  runs 5776 (MJD 53669) and 6425 (MJD 54010). These runs were chosen because of acceptable seeing on these
  nights (median of $1.7\arcsec$) and full coverage of the Stripe from 20 h to 4 h in right
  ascension. Detections were matched between these runs using a 5.0$\arcsec$ radius, and multiple detections
  of the same object discarded. The final match template then contained all point sources detected at least
  once in any of these runs, for a total of 2,000,242 objects.

  We then matched all detections over all runs to the match template using a 5.0$\arcsec$ match radius. All
  detections grouped inside this match radius were considered detections of a single match template object,
  and constituted a \emph{match bundle}. The typical separation between neighboring point sources in our match
  template is $\sim$ 35$\arcsec$. This is comfortably larger than our chosen match radius, so there is little
  chance of a detection being mistakenly associated with a nearby match template object. We also note that the
  match radius sets a rough upper limit of $\sim 0.5\arcsec$ yr$^{-1}$ for the proper motion of any one match
  template object, assuming it is detected over all 10 years of temporal coverage.

\begin{figure}
    \plottwo{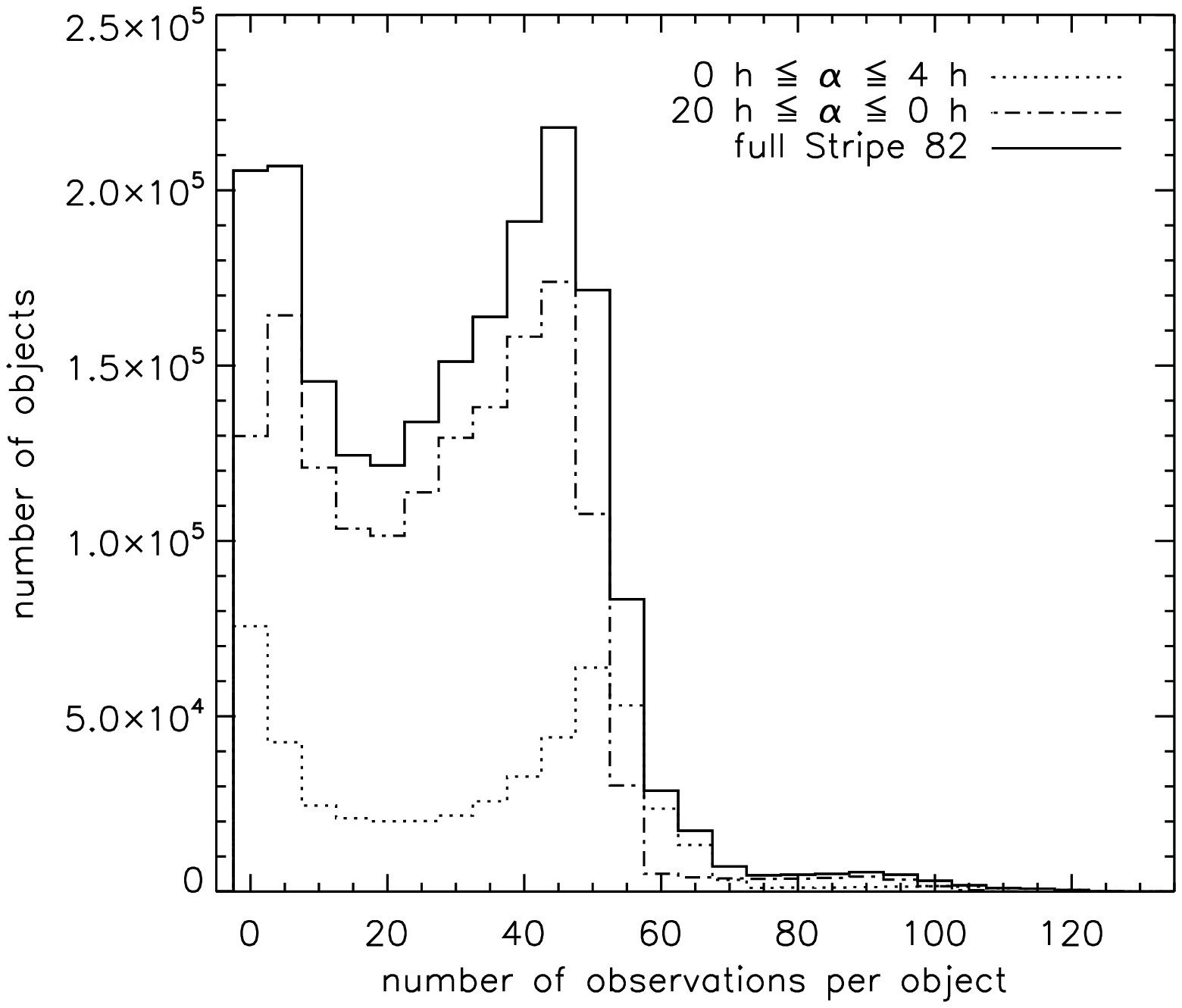}{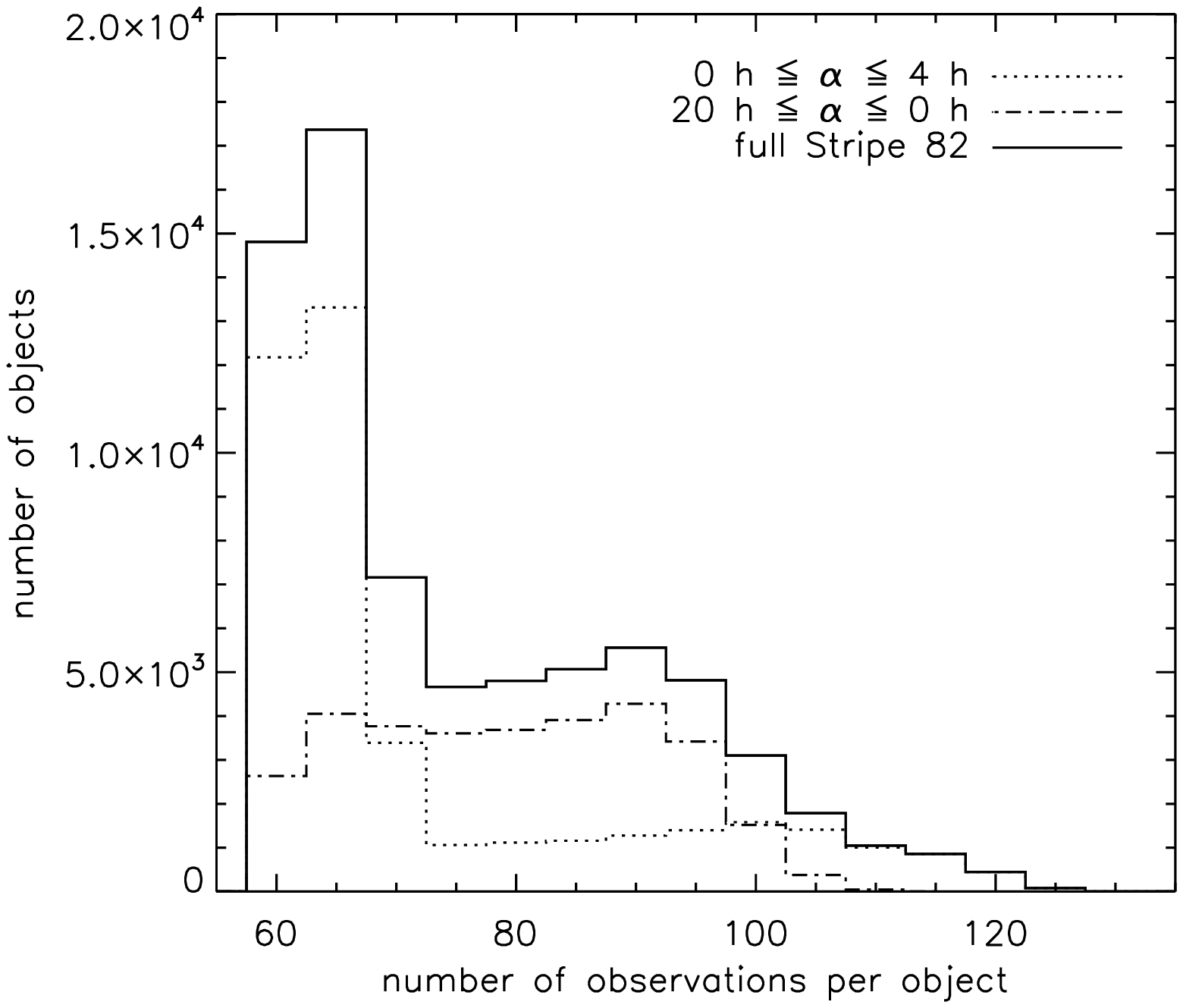}
    \caption{Left: Histogram of observations per object in the entire dataset spanning the years 1998 to
      2007. The median number of observations is 30. Right: Histogram of observations per object for objects
      with at least 60 observations.}
    \label{fig_obs_hist}
  \end{figure}

  Figure \ref{fig_obs_hist} presents the distribution of the number of detections per match bundle. There are
  473,759 objects (23.7$\%$ of the total) that have fewer than 10 detections in all 242 epochs over all 10
  years of available photometric data. Of these, $\sim$28.6\% have only a single detection, $\sim$35.4\% have
  extreme colors\footnote{Defined as $|u-g| > 3.0$ or $|g-r| > 3.0$ or $|r-i| > 3.0$ or $|i-z| > 3.0$.} most
  likely due to unreliable photometry, and $\sim$55.8\% are faint\footnote{Defined as $g$ and $r > 22.2$, $i >
  21.3$. These are the 95\% point source detection repeatability limits for these bands. See
  \url{http://www.sdss.org/dr7} for details.}. In contrast, only $\sim$0.2\% of objects with at least 10
  detections have extreme colors, while only $\sim$2.3\% of the objects with at least 10 detections are
  classified as faint. We, therefore, err on the side of caution and remove all objects with fewer than ten
  detections from any further consideration to keep contamination by unreliable photometry to a minimum. As a
  result, we have 1,526,483 objects with 55,610,252 total detections in our final point source light-curve
  catalog for Stripe 82. We carry this catalog forward to the ensemble photometry stage of our processing
  pipeline, described below.

%%%%%%%%%%%%%%%%%%%%%%%%%
%% ENSEMBLE PHOTOMETRY %%
%%%%%%%%%%%%%%%%%%%%%%%%%

  \subsection{Ensemble Differential Photometry}\label{sec_ensemble}

  As discussed earlier, the SN survey sacrifices some photometric quality to gain temporal
  coverage. Unfortunately, this means that observations are taken under all kinds of photometric conditions,
  including sub-optimal seeing, variable transparency, and at all lunar phases except for full moon. We note
  that the Stripe 82 runs before the SN Survey have more stringent constraints on these variables, and are
  generally of higher photometric quality. Even these observations, however, suffer from variable photometric
  conditions on time-scales of minutes to hours (see \citealt{2007AJ....134..973I} for details). We,
  therefore, face the challenge of reconciling the low cadence but relatively high quality measurements taken
  before the SN survey with measurements of lower quality at much higher cadence taken during the Survey.

\begin{figure}
    \plotone{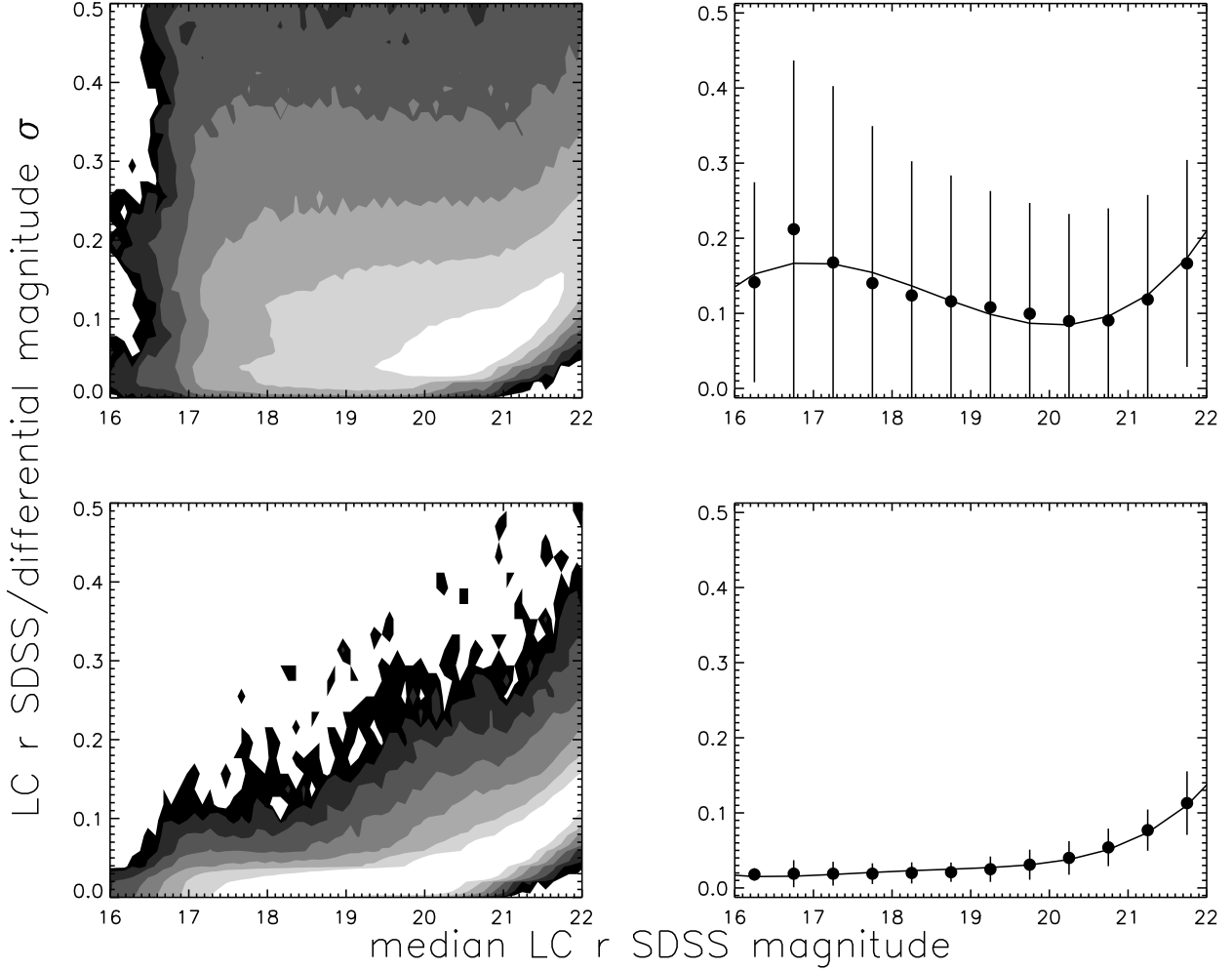}
    \caption{Top-left: SDSS $r$ median light-curve magnitude vs. SDSS $r$ light-curve standard
    deviation. Top-right: Median SDSS $r$ light-curve standard deviation plotted against median SDSS $r$ LC
    magnitude, binned in 0.5 magnitude bins, for $16.0 < r < 22.0$. Bottom-left: SDSS $r$ median light-curve
    mag vs. differential $r$ light-curve standard deviation. The scatter in the magnitude-$\sigma$ relation is
    much reduced. Bottom-right: Median differential $r$ light-curve standard deviation vs. median SDSS $r$ LC
    magnitude, binned in 0.5 magnitude bins, for $16.0 < r < 22.0$. `Error-bars' represent the rms scatter for
    each bin of the respective light-curve standard deviation distributions. The lines are fourth-degree
    polynomial fits. All density contours are in half-dex increments ranging from 1.0 to 3.0.}
    \label{fig_errmag_ensemble}
  \end{figure}

\begin{figure}
    \plottwo{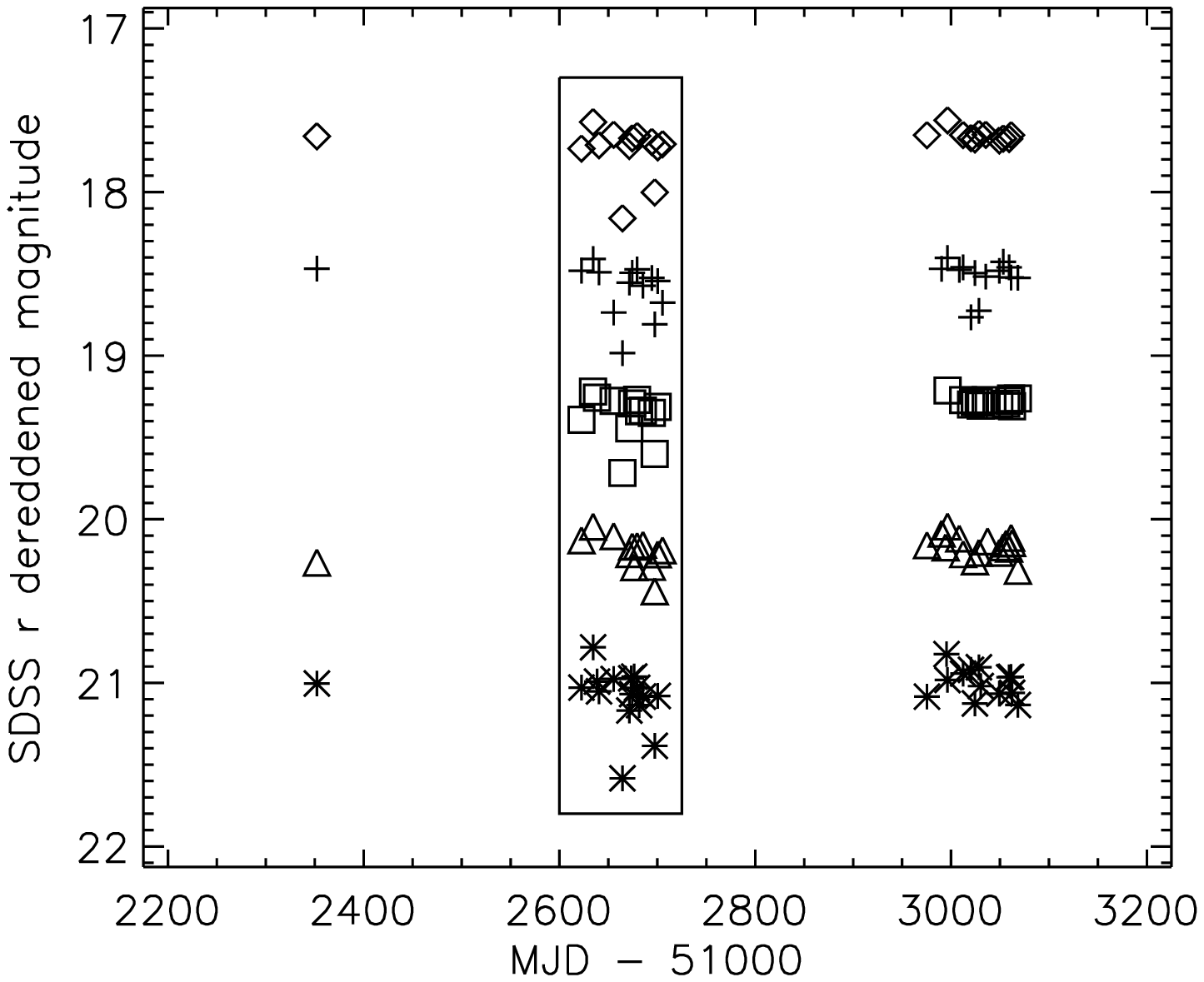}{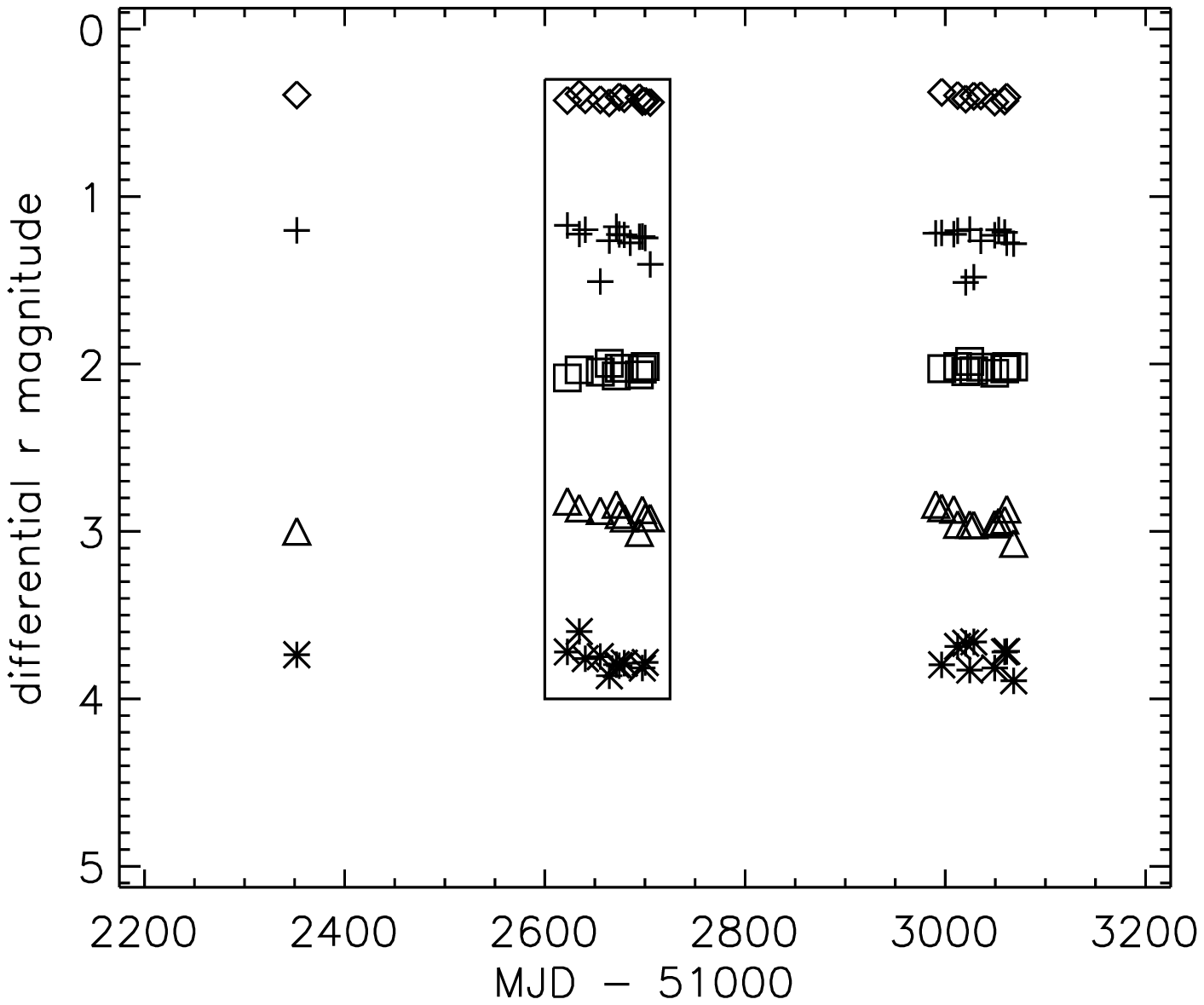}
    \caption{Left: SDSS \emph{r} mag light-curve for an M3 dwarf eclipsing binary candidate (plus symbols)
    compared to its four closest neighbors (diamond, square, triangle, and asterisk symbols). Note that all
    five objects appear to have a light-curve dip on the same observation date (outlined by the box); this is
    an example of false variability caused by systematic effects from varying photometric conditions. Right:
    differential \emph{r} light-curve after ensemble photometry carried out on the this field. Note that only
    the target object (plus symbols) appears to be variable: a possible eclipsing binary. The other objects no
    longer suffer from systematic effects that would have resulted in them being erroneously tagged as
    possible variables.}
    \label{fig_lcs_ensemble}
  \end{figure}

  In particular, identification of variable objects becomes problematic when large systematic effects caused
  by night-to-night sky and seeing variations are present. This is illustrated by an example relation between
  median SDSS light-curve magnitude and light-curve standard deviation (Fig \ref{fig_errmag_ensemble},
  top-left and top-right panels). This relation is often used to identify obvious variable objects; these lie
  above the general trend. It is difficult to identify true variables, however, when the relation itself has a
  scatter caused by magnitude measurements affected by differing photometric conditions. In addition, as seen
  in the left panel of Figure \ref{fig_lcs_ensemble}, these systematic effects usually appear as dips in
  light-curves on specific observation dates. These dips can be very large and affect measurements in all
  bands for all objects in a field, thus artificially increasing the light-curve standard deviation of even
  completely nonvariable objects.

  Such artifacts are usually removed in variability studies by the use of differential photometry. A target
  star is compared to an ensemble of comparison stars, and the measured magnitudes normalized to a weighted
  ensemble magnitude. The resulting differential magnitude for the target object removes any systematic
  variation in photometric conditions that affects all stars being observed on the same night. This technique
  is powerful and has been widely deployed, owing to its computational simplicity, as well as the very high
  photometric precisions that are possible. We, however, cannot employ this simple differential photometry
  method, because we are not guaranteed to have the same comparison stars in the ensemble for any one target
  star over all observations. Furthermore, we must ensure that all comparison stars are not intrinsic
  variables themselves; this is not possible \emph{a priori} when studying light-curves of $\sim 1.5$ million
  objects. We turn, therefore, to a differential photometry technique known as inhomogeneous ensemble
  photometry \citep{1992PASP..104..435H}. The general method is described immediately below; a discussion of
  its application to our dataset follows thereafter. 

  We first assume that most of the stars in a given field containing a target star are nonvariable, and that
  the measured magnitude of any one star $i$ at a time index value of $j$ is given by
  \begin{equation}\label{eq_ens_one}
    m(i,j) = m_0(i) + e(j),
  \end{equation}
  where $m(i,j)$ is the measured magnitude of a star, $m_0(i)$ is the `true' mean magnitude that would have
  been measured in the absence of photometric condition variations, and $e(j)$ represents the change in
  photometric zero-point caused at time index $j$ by these variations and is applied to all stars in the field
  at that time index. We must minimize the quantity $\beta$, given by
  \begin{equation}\label{eq_ens_two}
    \beta = \sum_{i=1}^{N_{stars}}\sum_{j=1}^{N_{obs}}\left[m(i,j) - m_0(i) - e(j)\right]^{2} w(i,j),
    \end{equation}
  where $w(i,j)$ is the weight associated with a measured magnitude $m(i,j)$ and is given by
  \begin{equation}\label{eq_ens_three}
    w(i,j) = w_1(j)w_2(i)w_3(i,j)w_4(i,j).
    \end{equation}
  The weights $w_1$, $w_2$, and $w_3$ are either zero or one, while $w_4$ is the statistical weight based on
  the uncertainty of measurement $\sigma[m(i,j)]$ for $m(i,j)$ and is calculated as $1/\sigma^2[m(i,j)]$. The
  weight $w_1 = 0$ if the observation at time index $j$ is to be excluded, $w_2 = 0$ if the star $i$ is to be
  excluded, and $w_3 = 0$ if a specific observation of star $i$ at time index $j$ is to be excluded from the
  photometric solution. Once this solution is computed, and we have obtained the `true' mean magnitudes
  $m_0(i)$ of all stars in the ensemble, along with the error terms $e(j)$, we can solve for the corrected
  magnitude $M(i,j)$ for any star $i$ at time index $j$:
  \begin{equation}\label{eq_ens_four}
    M(i,j) = m(i,j) - e(j).
  \end{equation}
  Finally, we can calculate the variance of the `true' mean magnitudes $\sigma^2[m_0(i)]$ using:
  \begin{equation}\label{eq_ens_five}
    \sigma^2[m_0(i)] = \frac{N_{obs}\sum_{j=1}^{N_{obs}}\left[M(i,j) - m_0(i)\right]^2 w(i,j)}
	  {(N_{obs}-1) \sum_{j=1}^{N_{obs}} w(i,j)}.
  \end{equation}
  The relation between $m_0(i)$ and $\sigma[m_0(i)]$ is equivalent to the usual magnitude-$\sigma$
  relation. Objects with large values of $\sigma[m_0(i)]$ relative to the general trend at that magnitude may
  be considered as possible variables.

  In practice, our inhomogeneous ensemble photometry implementation\footnote{See M. W. Richmond's website:
  \url{http://spiff.rit.edu/ensemble}.} normalizes the `true' mean magnitude $m_0(i)$ for a star $i$ to the
  `true' mean magnitude of the brightest star in the ensemble, resulting in effective differential magnitudes
  $M(i,j)$ for all of the stars. We use the weights $w_1$, $w_2$, and $w_3$ to take into account the presence
  or absence of comparison stars in the ensemble from night to night. To deal with the problem of comparison
  stars being possible variables, we run the photometric solution for a target three times, each time removing
  all comparison stars from the ensemble that appear to be variable. The ensemble itself is chosen such that
  it has at least 10 comparison stars observed within 10 seconds of the time of observation of the target star
  (so within 150 arcseconds of the position of the target star, assuming tracking at sidereal rate). The
  median number of comparison stars is $\sim 80$ per target star. This set of conditions ensures that we
  obtain the best possible differential magnitude for the target on each night that it is observed. At this
  stage, we also remove from any further consideration all target stars that do not have comparison star
  ensembles satisfying these conditions, since reliable differential photometry is not possible for these
  objects. In the processing of our RA 0 to 4 h light-curve catalog, this removes only $2,653$ objects ($\sim
  0.72\%$ of the total) from consideration.

  We show the impact of inhomogeneous ensemble photometry in Figures \ref{fig_errmag_ensemble} and
  \ref{fig_lcs_ensemble}. The relation between the median light-curve magnitude and the differential
  light-curve standard deviation (Fig \ref{fig_errmag_ensemble}, bottom-left and bottom-right panel) is seen
  to be much improved, with much of the extrinsic scatter removed. Furthermore, the light-curves of the target
  object and its four closest neighbors (Fig \ref{fig_lcs_ensemble}, right panel) no longer suffer from the
  systematic effects seen in the left panel.

  The fourth order polynomial fits to the empirical relation between the mean SDSS light-curve magnitude (in
  each band $ugriz$) and the differential magnitude light-curve standard deviation $\sigma$ calculated using
  differential magnitude light-curve data for 365,086 objects with at least 10 observations from our initial
  light-curve catalog (RA 0 to 4 h) are given below:
  \begin{eqnarray}\label{magerr_relations}
    \sigma(u) = 105.94 - 24.197u + 2.0621u^2 -0.07772u^3 + 0.00109u^4,\\
    \sigma(g) = 51.441 - 11.337g + 0.93529g^2 - 0.03423g^3 + 0.00047g^4,\\
    \sigma(r) = 48.684 - 10.831r + 0.90210r^2 - 0.03334r^3 + 0.00046r^4,\\
    \sigma(i) = 54.844 - 12.291i + 1.0315i^2 - 0.03842i^3 + 0.00054i^4,\\
    \sigma(z) = -69.446 + 15.591z - 1.0329z^2 + 0.04799z^3 - 0.00066z^4.
  \end{eqnarray}
  
  We generate differential magnitude light-curves in all five bands for all objects in our light-curve catalog
  and search them for variability. The large number of stars in our dataset, coupled with the numerous lookups
  required to build each target star's comparison ensemble, make this stage in our pipeline the most
  computationally intensive, and thus, the slowest.

%%%%%%%%%%%%%%%%%%%%%%%%%%%%%%%%%%%%%
%% CLASSIFICATION OF POINT SOURCES %%
%%%%%%%%%%%%%%%%%%%%%%%%%%%%%%%%%%%%%

  \subsection{Classification of Point Sources}\label{sec_class_point}

  We classify the point sources in our final light-curve catalog in two ways. First, we cross-match objects
  between our catalog and existing catalogs of interesting and possibly variable objects in the SDSS,
  especially those present in the Stripe 82 footprint. Second, we use several selection algorithms to sort
  objects into rough classification bins by their SDSS colors. In this way, we can select interesting
  populations of objects to study for variability without having to resort to spectra.

  Several authors have already compiled catalogs of interesting classes of point source objects discovered
  during SDSS operations. \citet{2007AJ....134..102S} reported the discovery of 77,429 quasars ranging in
  redshift from 0.08 to 5.41, covering 5,740 sq. deg of sky. Most quasars are suspected to be variable in some
  degree; our catalog of multi-color light-curves covering a base-line of nearly ten years may be used to test
  that assumption. Other catalogs that include possible variable objects include a catalog of 9,316
  spectroscopically confirmed white-dwarfs and 948 hot sub-dwarfs by \citet{2006ApJS..167...40E}, and the
  already mentioned SDSS Stripe 82 catalog of variable stars by \citet{2007AJ....134.2236S}, containing 13,051
  sources (hereafter referred to as the SDSS-I Variable Object Catalog). We also cross-match our catalog with
  the Stripe 82 Standard Star Catalog (containing $\sim$ 1 million objects) generated by
  \citet{2007AJ....134..973I}, to confirm the nonvariability of these objects and their suitability for use as
  faint standard stars. We use a 5.0$\arcsec$ radius to match between all of the preceding catalogs and our
  catalog of match template objects.

  In addition to classifying objects in this way, we also make use of the excellent color selection made
  possible by the five photometric bands used in the SDSS. Specifically, we use the median SDSS light-curve
  colors $u-g$, $g-r$, $r-i$, and $i-z$ for each object. We use color selection algorithms from three sources:
  (1) spectroscopic target selection color cuts from the SDSS-II Sloan Extension for Galactic Understanding
  and Exploration (SEGUE; \citealp{2009AJ....137.4377Y}) to select different types of stellar objects, (2)
  colors from \citet{2007AJ....134.2236S} to identify potential RR Lyrae, and (3) color cuts for M-dwarfs in
  \citet{2008AJ....135..785W}. Table \ref{color_table} presents the color classification schemes we use for
  our catalog. The SDSS photometric pipeline reports magnitudes measured for each detection, and the
  extinction in magnitudes at the position of the detection. We use these measurements and calculate the
  dereddened magnitudes for all objects. These are then used for all color selection cuts, except where noted
  in Table \ref{color_table}. We do not use dereddened colors for brown dwarfs and main-sequence/white-dwarf
  pair selection because these faint objects are likely to be nearby. We note that any one point source may be
  assigned multiple categories based on color; these are all noted in \emph{classification tags} associated
  with that point source in our catalog. Objects that are not matched to any of the catalogs listed above and
  cannot be classified using color are also noted and assigned a classification tag of \emph{unknown}.

  Results from the catalog cross-matching and color classification schemes above for the first release of our
  catalog are reported in Section \ref{sec_variable_properties}. We now turn to the identification of variable
  objects in our catalog.

%%%%%%%%%%%%%%%%%
%% VARIABILITY %%
%%%%%%%%%%%%%%%%%

  \section{Variable Objects in Stripe 82}

%%%%%%%%%%%%%%%%%%%%%%%%%%%%%%%%%%%%%%%%%%
%% EXTRACTION OF VARIABLE POINT SOURCES %%
%%%%%%%%%%%%%%%%%%%%%%%%%%%%%%%%%%%%%%%%%%

  \subsection{Extraction of Variable Point Sources}\label{sec_variables}

  We identify possible variable sources in two steps. The first involves using the results from the ensemble
  photometry stage of our pipeline, specifically the relation between the differential magnitude light-curve
  standard deviation $\sigma[m_0(i)]$ (Equation \ref{eq_ens_five}) and the `true' mean differential
  light-curve magnitude $m_0(i)$. As the ensemble differential photometry stage of our pipeline iterates over
  each target object, it generates a plot of this relation for all objects in its associated ensemble, and
  fits a second-degree polynomial to the empirical trend. Outliers more than 2-$\sigma$ away from the general
  trend are iteratively discarded to make the fit more robust.

\begin{figure}
    \plottwo{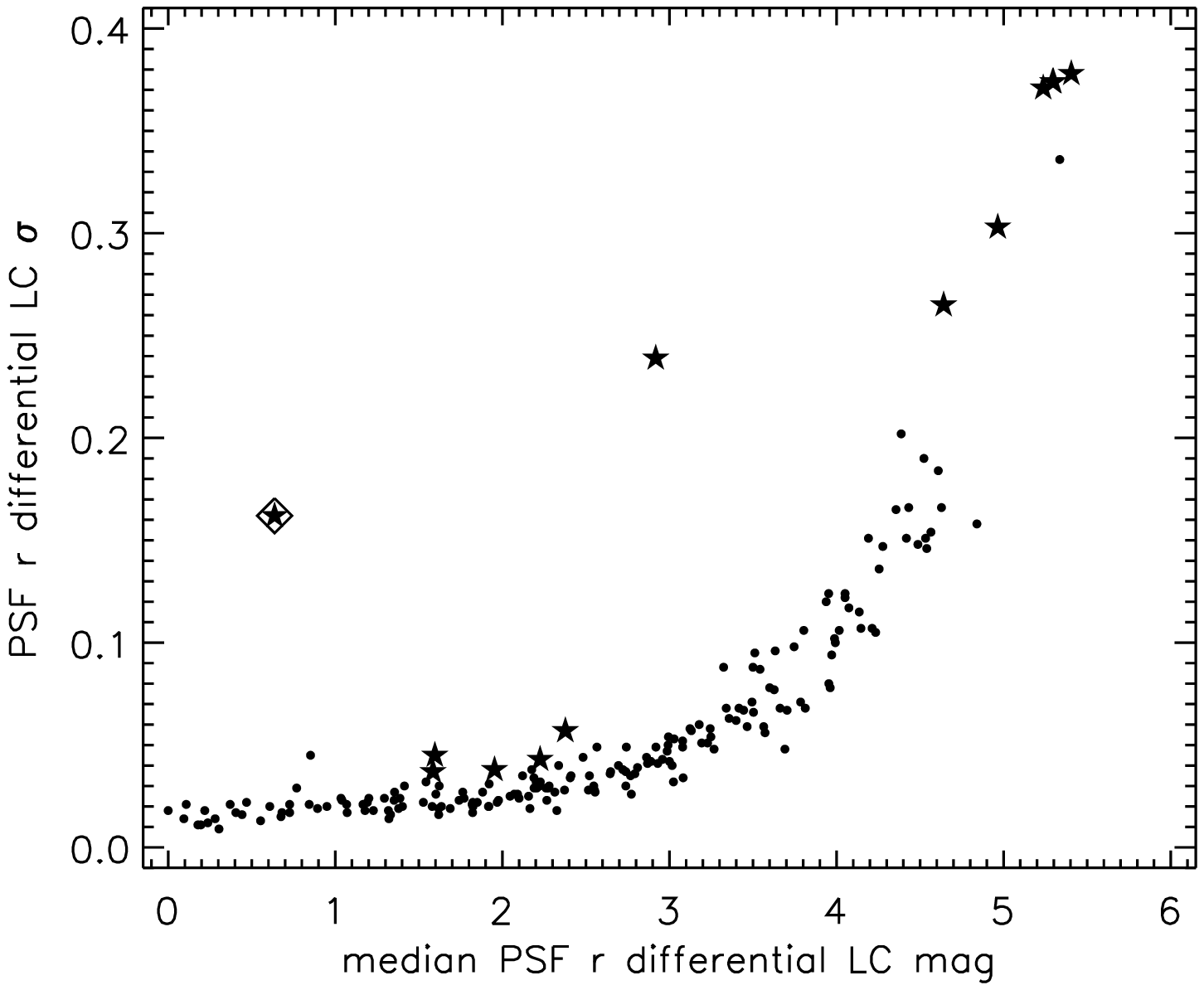}{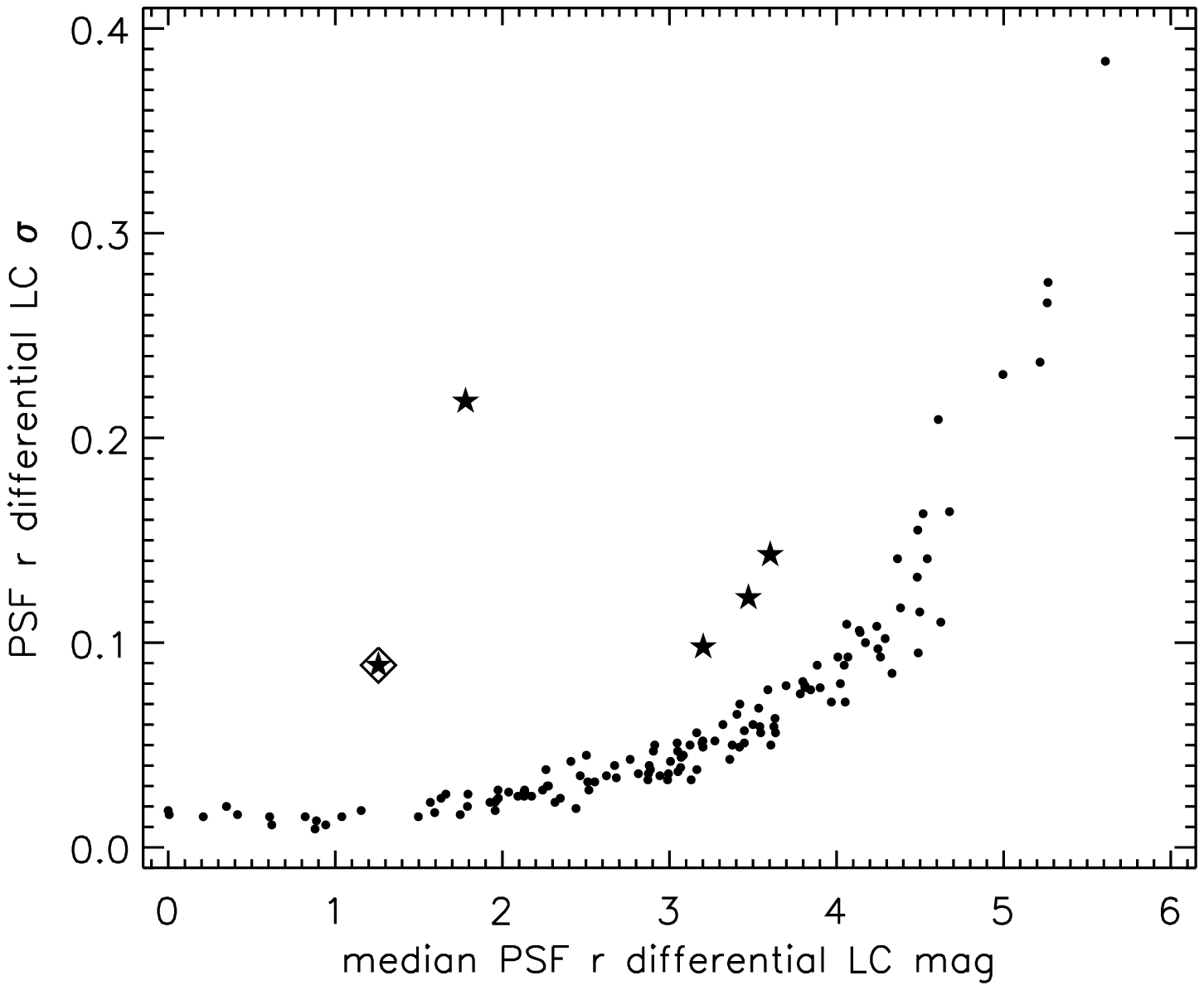}
    \caption{Examples of PSF $r$ differential magnitude-$\sigma$ trend plots generated by our ensemble
      photometry pipeline. These particular plots are for the comparison star ensemble fields of two objects
      eventually identified as periodic variables: MB875 (left panel) and MB32047 (right panel). Objects
      tagged by the ensemble photometry pipeline as \emph{tentative variables} are shown as star symbols and
      the two periodic variables in the field are shown as stars inside the diamond symbols.}
    \label{fig_ens_mag_err}
  \end{figure}

  This procedure is then repeated to obtain such plots and trends for the $u$, $g$, $r$, $i$, and $z$ bands
  separately. Examples of these plots are shown in Figure \ref{fig_ens_mag_err}. These are then used to tag
  the target object as a \emph{tentative variable} if: it lies at least 2-$\sigma$ above the general trend of
  the magnitude-standard deviation relation in (1) $u$ and $g$ bands simultaneously, or (2) $g$ and $r$ bands
  simultaneously, or (3) $r$ and $i$ bands simultaneously, or (4) $i$ and $z$ bands simultaneously. We also
  tag objects as tentative variables if they lie at least 1-$\sigma$ above the general trend in all three of
  the $r$, $i$, and $z$ bands simultaneously; this has the potential of picking up faint red variables in
  populations of K and M-dwarfs. This system, however, is not robust against false variability caused by large
  photometric noise for faint objects. These objects are likely to lie above the variability threshold due to
  their small numbers and large uncertainties, even in their differential magnitudes, and thus end up being
  erroneously tagged as tentative variables.

  The second step involved in identifying variable sources corrects for this type of false variability. We
  employ the Stetson variability index \citep{1996PASP..108..851S}. This is a measure of the correlation of
  simultaneous variability across a pair of bands. The Stetson variability index $J_{\alpha\beta}$ for two
  bands $\alpha$ and $\beta$ observed at the same time is
  \begin{equation}\label{eq_stet_one}
    J_{\alpha\beta} = \frac{\sum_{k=1}^{n} w_k \sgn(P_k^{\alpha\beta}) 
      \sqrt{|P_k^{\alpha\beta}|}}{\sum_{k=1}^{n} w_k},
  \end{equation}
  where $w_k$ is the weight assigned to $k$th pair of observations. The product of the normalized magnitude
  residuals $P_k$ for the $k$th pair of observations is given by
  \begin{equation}\label{eq_stet_two}
    P_k^{\alpha\beta} = \sqrt{\frac{n_\alpha}{n_\alpha-1}}\sqrt{\frac{n_\beta}{n_\beta-1}}
    \left(\frac{\alpha_k - \bar\alpha}{\sigma_{\alpha,k}}\right)
    \left(\frac{\beta_k - \bar\beta}{\sigma_{\beta,k}}\right).
  \end{equation}
  where $\alpha_k$ and $\beta_k$ are the magnitudes measured at time index $k$, $\sigma_{\alpha,k}$ and
  $\sigma_{\beta,k}$ are the errors associated with these measurements, $\bar\alpha$ and $\bar\beta$ are the
  weighted mean magnitudes, iterated as in \citet{1996PASP..108..851S} to avoid outliers, and $n$ is the
  number of observations. The Stetson variability index is large when the normalized magnitude residuals
  $P_k^{\alpha\beta}$ are correlated, as in the case of real variable sources. A nonvariable source, even one
  with a large light-curve standard deviation, will have uncorrelated magnitude measurements across pairs of
  bands. This will drive the Stetson index for such objects toward zero.

\begin{figure}
    \plotone{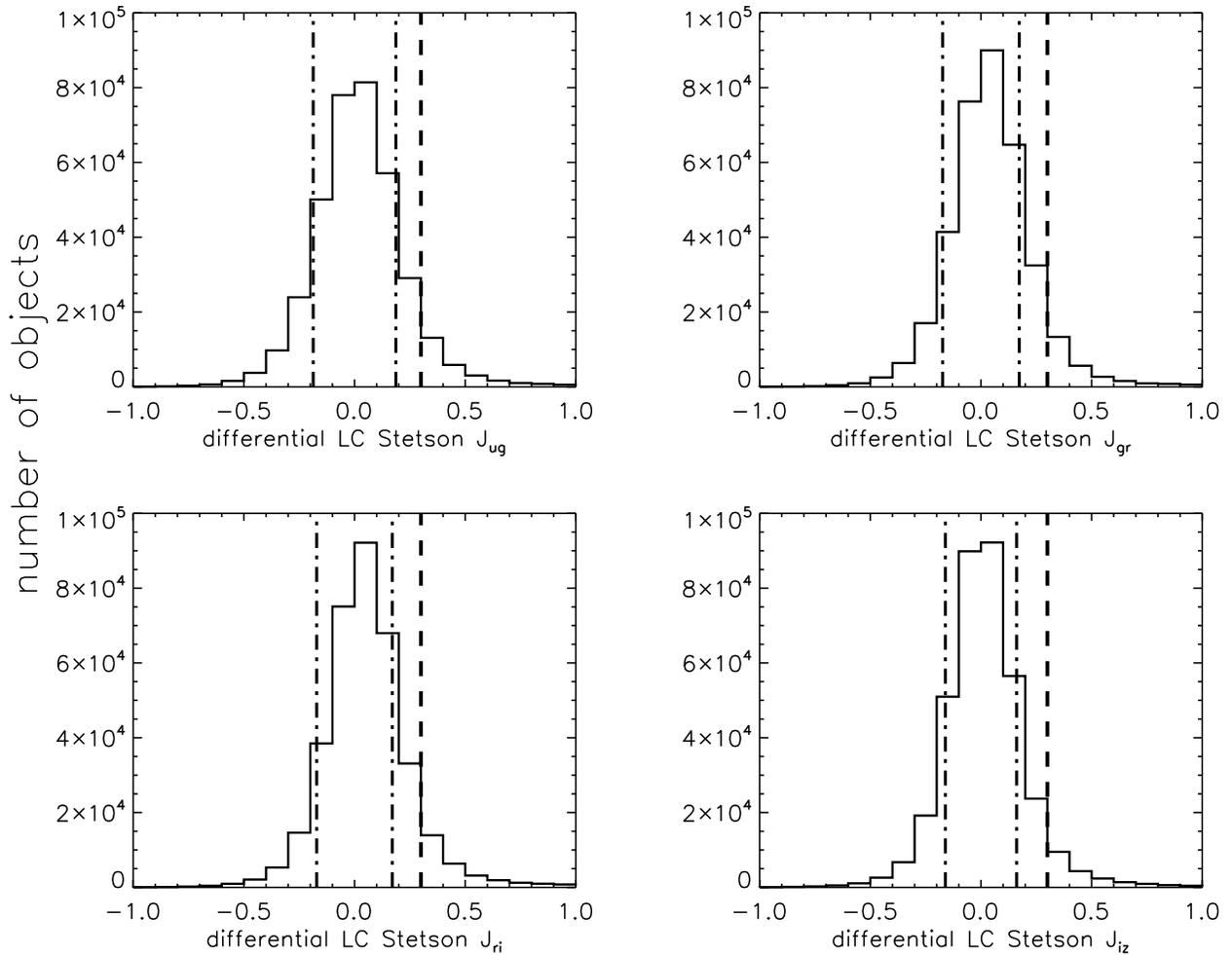}
    \caption{Stetson variability index histograms for all 365,086 objects with more than 10 detections in our
    RA 0 to 4 h light-curve catalog. The dot-dashed lines indicate the standard deviations of the
    distributions, while the dashed line indicates our threshold Stetson index value of 0.3.}
    \label{fig_stetson}
  \end{figure}

  We calculate the Stetson variability indices $J_{ug}$, $J_{gr}$, $J_{ri}$, and $J_{iz}$ using the
  differential $ugriz$ magnitude light-curves obtained from the ensemble photometry procedure described in
  Section \ref{sec_ensemble}. Taking into account the variability index values for obvious variables in the
  dataset identified by inspection of a large number of differential magnitude light-curves (and phased
  differential magnitude light-curves in the case of periodic variables), we set a threshold index value of
  0.3 to separate variable and nonvariable objects. Sources are therefore tagged as \emph{probable variables}
  if: (1) any one of $J_{ug}$, $J_{gr}$, $J_{ri}$, and $J_{iz}$ is greater than this threshold value, and (2)
  $J_{gr}$ and $J_{ri}$ are both greater than 0.0. Figure \ref{fig_stetson} shows the distributions of the
  calculated Stetson indices with their standard deviations (dot-dashed lines) and threshold Stetson index
  value (dashed line) marked. Variable objects are selected relatively efficiently using our threshold Stetson
  index value.  

\begin{figure}
    \plotone{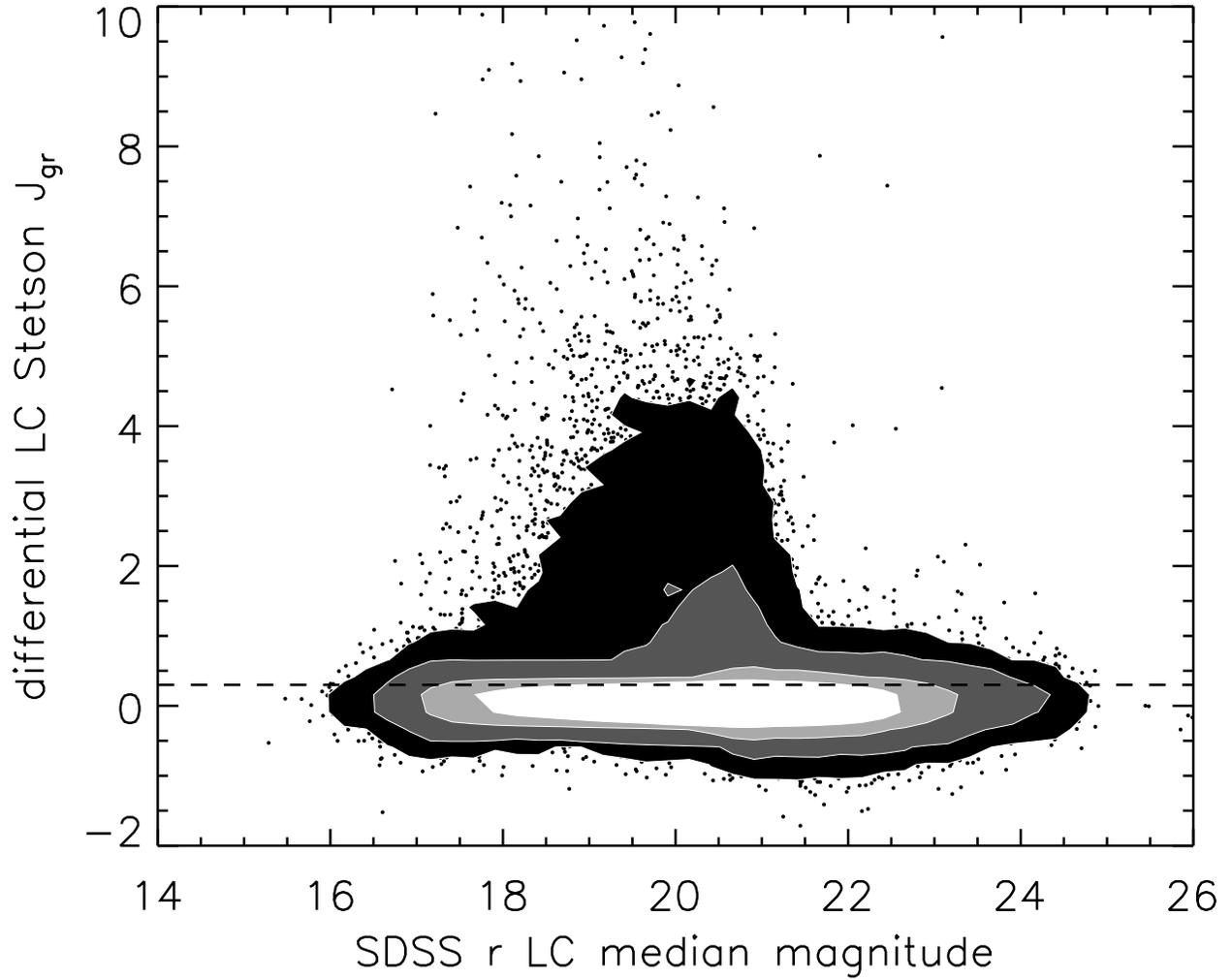}
    \caption{Stetson variability index $J_{gr}$ vs. SDSS light-curve median $r$ magnitude for all 365,086
    objects with more than 10 detections in our RA 0 to 4 h light-curve catalog. The density contours are in
    dex increments ranging from 1.0 to 3.5, and the dashed line shows the minimum threshold value for
    variability detection, $J_{gr} = 0.3$.}
    \label{fig_stetson_2}
  \end{figure}

  Figure \ref{fig_stetson_2} shows the distribution of Stetson $J_{gr}$ with median light-curve $r$ magnitude
  for all 365,086 objects with at least 10 observations in the RA 0 to 4 h light-curve catalog. The underlying
  distribution shows no discernable trend with magnitude, indicating that the ensemble photometry routines
  combined with the Stetson index as an indicator of variability are quite robust in variable
  selection. Furthermore, as we go to fainter magnitudes (beyond $r \sim 22.0$), the distribution of the
  Stetson index no longer shows an outlying population of variables, indicating that these fainter objects
  have no correlated variability detected and thus appropriately have smaller Stetson index values. This
  points out the effective faint magnitude limit of our search for variability (discussed further in Section
  \ref{sec_var_gen_prop} below). We quantify the efficiency and robustness of our variable detection routines
  in Section \ref{sec_efficiency} using synthetic light-curve catalogs and associated simulations of
  variability analysis.

  \emph{Probable variables} selected by this two part procedure can then be searched for periodic
  variability. If such variability exists, we generate estimates of the period, as described in Section
  \ref{sec_periods} below.

%%%%%%%%%%%%%%%%%%%%%
%% FINDING PERIODS %%
%%%%%%%%%%%%%%%%%%%%%

  \subsection{Finding Periods}\label{sec_periods}

  We search for periodicity among all objects marked as probable variables using two independent
  period-finding methods: the string length method of \citet{1983MNRAS.203..917D}, and a variation on the
  classic Lafler-Kinman phase-dispersion minimization algorithm \citep{1965ApJS...11..216L} used by
  \citet{1996PASP..108..851S}. Both of these methods attempt to minimize the sum of the dispersions of
  measurements ordered by phase for a test period (the `string length'), in order to produce the `smoothest'
  possible light-curve. Perhaps their most important feature is that they do not involve the calculation of
  sinusoidal components for a light-curve, as in the Lomb-Scargle periodogram \citep{1982ApJ...263..835S}, and
  thus do not pre-suppose such a shape when searching for periods. This makes them invaluable for studying a
  broad range of periodic variable types, as expected in our dataset. These two algorithms also do not involve
  binning consecutive light-curve measurements, as seen in more sophisticated period finding methods such as
  AoV \citep{1989MNRAS.241..153S}, and BLS \citep{2002A&A...391..369K}. BLS and AoV do not work well on
  datasets such as ours, which have light-curves with a very small number of unevenly distributed and sparse
  time-series measurements over a long baseline.

  The Dworetsky string length $L_{dw}$ calculated for a test period and a series of phase-ordered photometric
  measurements $m_{1}$ to $m_{N}$ is given by the relation
  \begin{equation}
    L_{dw} =  \sum_{i=1}^{N-1} \sqrt{(m^{\prime}_{i+1} - m^{\prime}_{i})^{2} + 
      (\phi_{i+1} - \phi_{i})^{2}} + 
    \sqrt{(m^{\prime}_{1} - m^{\prime}_{N})^{2} + 
      (\phi_{1} - \phi_{N} + 1)^{2}},
  \end{equation}
  where $\phi_i$ is the phase of observation $i$, and $m^\prime$ are modified magnitudes used to assign
  similar weights to the time and magnitude measurements, and are given by the relation
  \begin{equation}
    m_{i}^{\prime} = \left(m_i - m_{min}\right)/\left(2.0(m_{max} - m_{min})\right) - 0.25,
  \end{equation}
  where $m_{min}$ and $m_{max}$ are the minimum and maximum magnitude measurements in the timeseries,
  respectively. 

  The Stetson algorithm incorporates the weighting of individual light-curve measurements by their respective
  errors, thus, providing a more robust means of calculation of the string length. The Stetson string length
  $L_{st}$ calculated for a test period and a series of phase-ordered photometric measurements $m_{1}$ to
  $m_{N}$ is given by the relation
  \begin{equation}
    L_{st} = \frac{\sum_{i=1}^{N} w(i,i+1) |m_i - m_{i+1}|}{\sum_{i=1}^{N} w(i,i+1)},
  \end{equation}
  where $w$ are the weights assigned to magnitudes $m$, which in turn have measurement errors $\sigma$. The
  expression $w(i,i+1)$ is given by the relation
  \begin{equation}
    w(i,i+1) = \frac{1}{\left[\sigma_i^2 + \sigma_{i+1}^2\right]\left[(\sigma_{i+1} - \sigma_i) + 1/N\right]},
  \end{equation}
  where $\sigma_i$ is the measurement error associated with magnitude measurement $m_i$.
  
\begin{figure}
    \plottwo{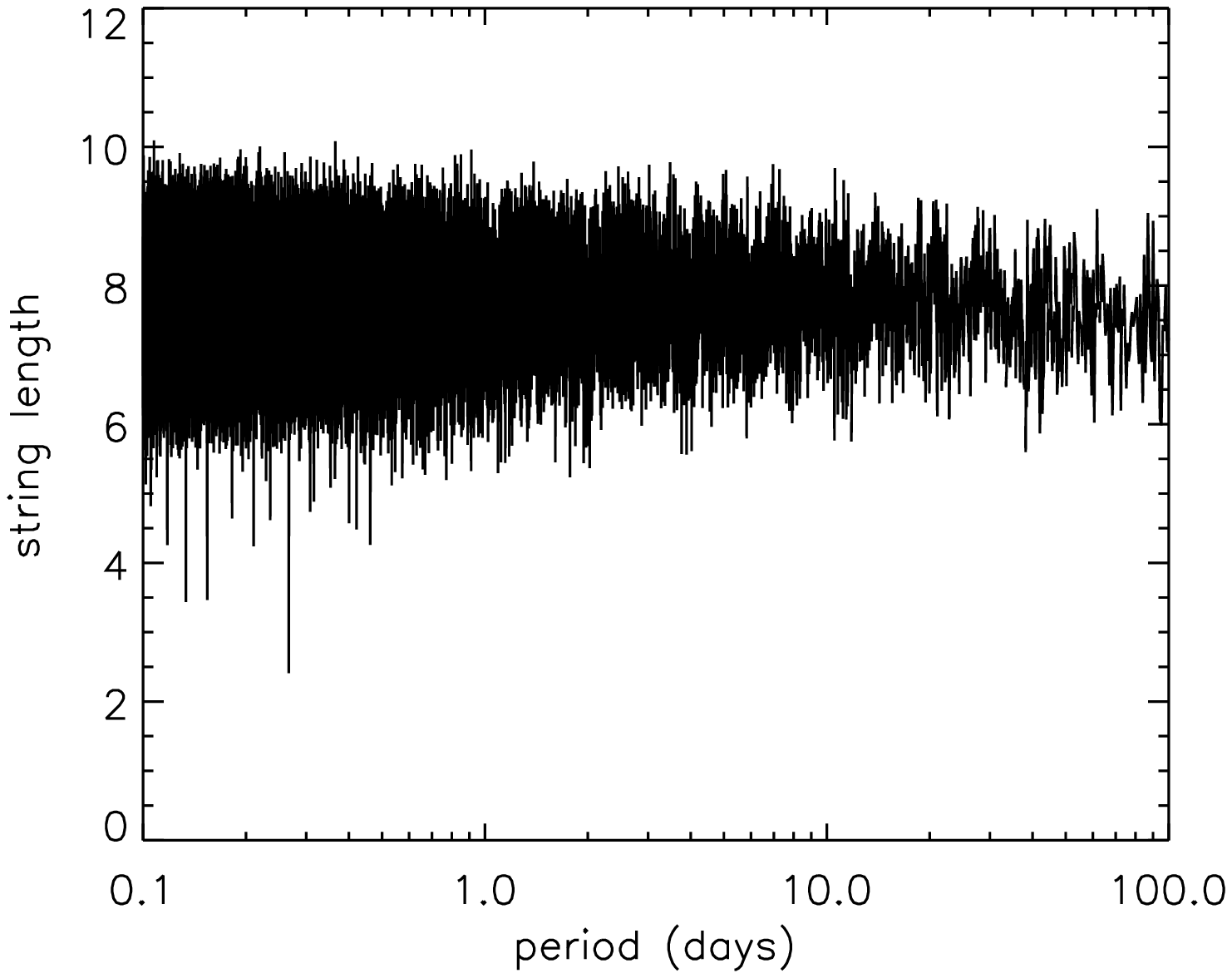}{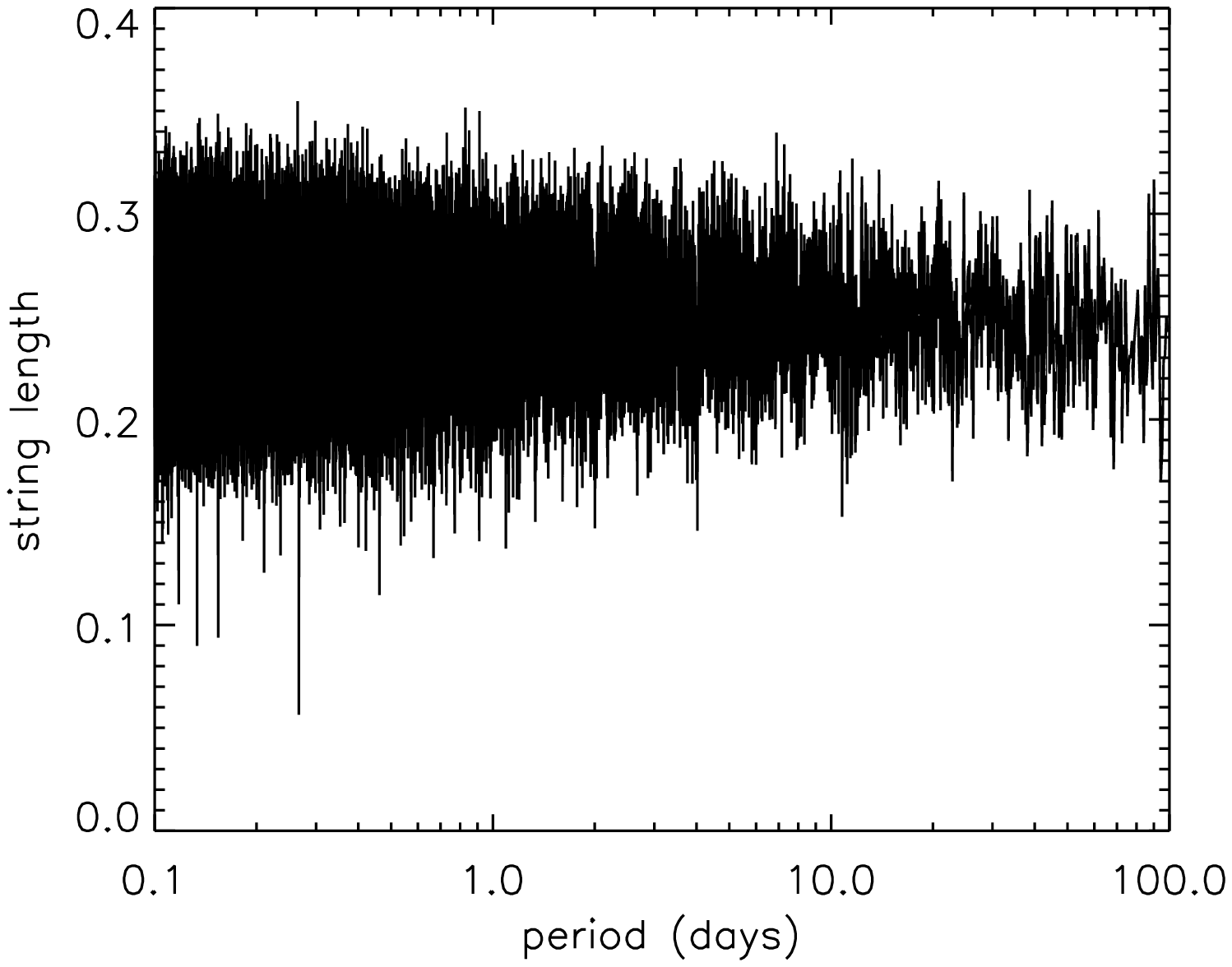} 
    \caption{Left: differential $g$-band Dworetsky string length diagram for an eclipsing binary candidate
    (MB6467 in Figure \ref{fig_nonm_ebs}). Right: differential $g$-band Stetson string length diagram for the
    same object. The smallest value of the string length in both cases indicates the most likely period, in
    this case; $\sim$0.2668 days.}
    \label{fig_strlens}
  \end{figure}

  We calculate $L_{dw}$ and $L_{st}$ using $griz$ differential magnitude light-curves for all objects tagged
  as \emph{probable variables} using a test period interval of 0.1 to 100.0 days (see Figure \ref{fig_strlens}
  for an example) and a frequency step-size of 0.0001 days$^{-1}$. The $u$-band differential magnitude
  light-curves have poor signal-to-noise for many of the objects in our dataset, and are therefore not used
  for string length calculations. Objects with light-curve variations on longer time-scales (i.e. years) do
  exist in our dataset, but we do not attempt to fit periods to these, mainly due to extremely poor phase
  coverage (typically 40--60 unevenly sampled measurements over 10 years). If, upon inspection of its phased
  light-curve, an object appears to be variable on a time-scale shorter than 0.1 day, we rerun the period
  finding routines using a test period interval of 0.01 to 10.0 days.

  For each object, we retain the test periods that result in the 20 shortest string lengths for both $L_{dw}$
  and $L_{st}$ independently, and then phase-fold the light-curves using these periods. The most likely period
  in each case is the one with the shortest string length. We require that the two period finding methods
  agree on the most likely period before accepting an object as a likely periodic variable. Further, we make
  use of the available multi-band photometric data by requiring that the most likely period reported by the
  string length algorithms be the same for each of the $gri$ light-curves in the case of non-M dwarf objects,
  and for each of the $riz$ light-curves for redder objects such as M-dwarfs. We then attempt to refine the
  period by rerunning the string length algorithms in a small period interval (typically 0.1 day) centered
  around the already determined most likely period. Finally, we visually inspect each phased light-curve
  obtained using this refined most likely period to assess its credibility, and place the object into one of
  three classification bins based on its light-curve shape: (1) eclipsing or ellipsoidal binary, (2)
  sinusoidal variable with an asymmetric light-curve, (3) and sinusoidal variable with a symmetric
  light-curve.

\begin{figure}
    \plotone{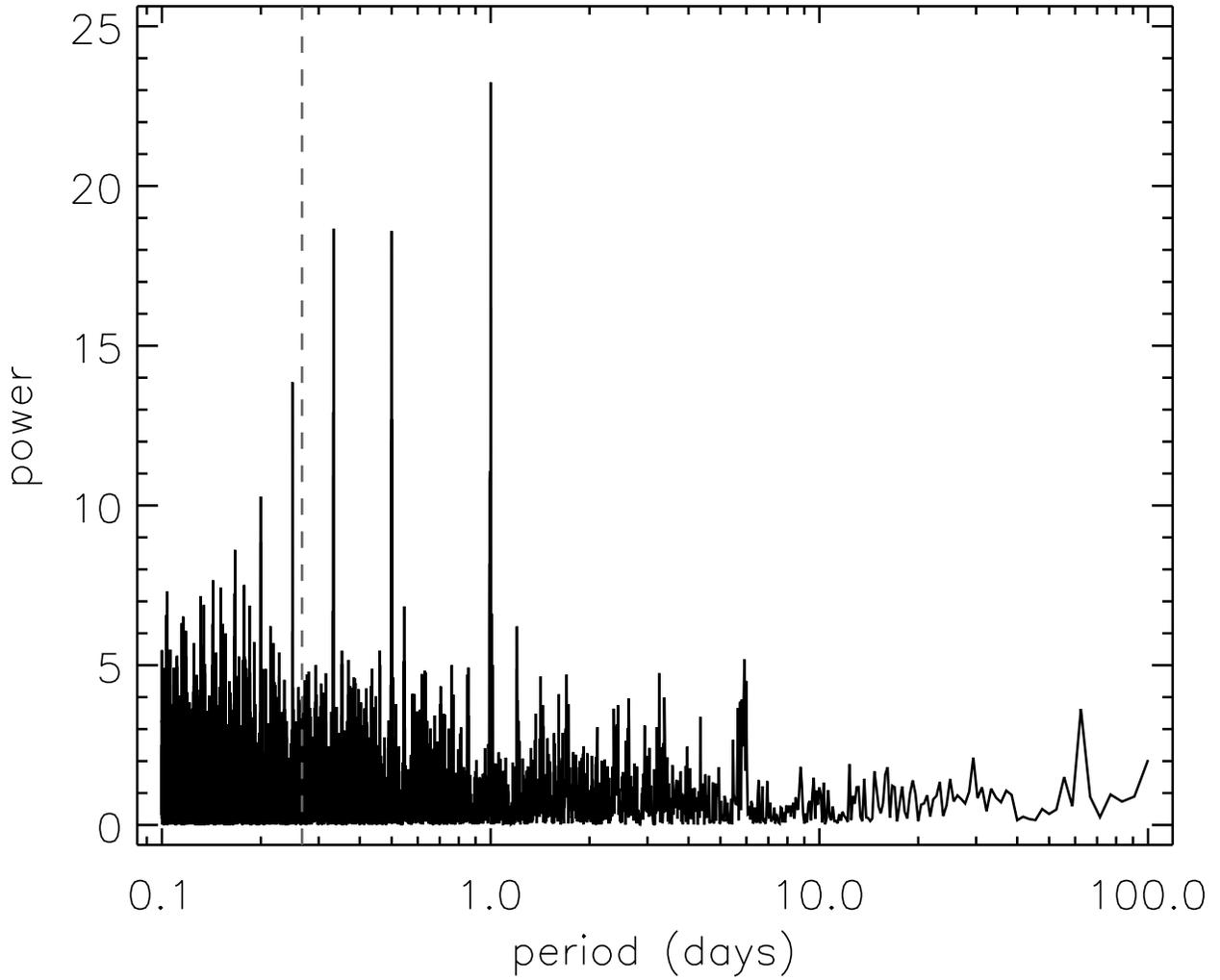}
    \caption{Lomb-Scargle periodogram for a typical spectral window function in the range 0.1 to 100.0
    days. Note the strong peaks near 1, 7, and 30 days and their aliases. This window function was calculated
    using the $g$-band differential magnitude light-curve from the same object as in Figure
    \ref{fig_strlens}. The period of this object is marked with the grey dashed line.}
    \label{fig_window_fn}
  \end{figure}

  As mentioned earlier, our dataset contains light-curves with a small number of measurements scattered
  unevenly over large baselines. A periodogram of the typical spectral window function for our data is shown
  in Figure \ref{fig_window_fn}. There is significant power in the peaks around 1.0, $\sim$ 7.0, and $\sim$
  30.0 days and their aliases, which are related to the three main duty cycles present in the SN Survey
  dataset. Fortunately, string length period-finding algorithms are relatively insensitive to these cycles (as
  evidenced by Figure \ref{fig_strlens}), but objects with periods close to the periods of these cycles suffer
  from severe aliasing, making it difficult to distinguish between the many likely periods. We also face the
  challenge of insufficient sampling near points of maximum variation in light-curves; for example, eclipsing
  binary candidates that do not have light-curve points that sample primary and secondary eclipses will suffer
  from ambiguity in period determination. Finally, objects that have large photometric noise even in their
  differential magnitude light-curves, such as faint red stars, will return many different string lengths that
  do not correspond to any `smooth' phased light-curve, thus making it impossible to determine their periods,
  even if evidence of periodicity is apparent in these objects' unphased light-curves.

%%%%%%%%%%%%%%%%%%%%%%%%%%%%%%%%%%%%%%%%%%%%%
%% DESCRIPTION OF THE CATALOG OF VARIABLES %%
%%%%%%%%%%%%%%%%%%%%%%%%%%%%%%%%%%%%%%%%%%%%%

  \section{The RA 0 h to 4 h Light-curve Catalog}\label{sec_variable_properties}

  \subsection{General Properties and Classification}\label{sec_var_gen_prop}

  The first release of our catalog covers the right ascension range 0 to 4 h. In this region, there are
  495,797 objects extracted from the SDSS dataset as described in Section \ref{sec_sdss} above. We restrict
  our attention to only those objects with at least 10 observations, and process these through our ensemble
  photometry pipeline discussed in Section \ref{sec_ensemble}. The resulting light-curve catalog contains
  365,086 point sources. We then remove objects that have large errors in color: $\sigma(u-g) > 0.2$,
  $\sigma(g-r) > 0.2$, $\sigma(r-i) > 0.2$, and $\sigma(i-z) > 0.2$. This pruning leaves 228,056 objects. The
  SDSS imaging pipeline (see \citealt{2002SPIE.4836..350L} and \citealt{2001ASPC..238..269L}) efficiently
  separates stars and galaxies up to $r = 21.5$ mag. To minimize the contamination fraction of galaxies in our
  sample, but at the same time remain sensitive to faint red variables, we impose a faint magnitude limit of
  $r = 22.0$ mag on the objects in our light-curve catalog. This leaves us with a final catalog of 221,842
  objects to consider for variability analysis.

  We then classify the objects in this light-curve catalog by color-selection and cross-matching against other
  catalogs as discussed in Section \ref{sec_class_point}. The results of this process are shown in Table
  \ref{type_table}. M-dwarfs make up the largest fraction of point sources by number; partly due to the
  initial single magnitude cutoff imposed at $z = 21.0$ mag (see Section \ref{sec_sdss}), and partly because
  of the intrinsic frequency of these low mass stars in the Galaxy. These objects are, therefore,
  overrepresented in our catalog to a significant degree. We also cross-match 4,196 objects classified as QSOs
  by \citet{2007AJ....134..102S}, 419 spectroscopically confirmed white dwarfs, and 15 hot subdwarfs from the
  catalog of \citet{2006ApJS..167...40E}. 165,591 point sources in our catalog are successfully cross-matched
  to objects classified as standard star candidates in \citet{2007AJ....134..973I}. Finally, we recover 3,972
  objects classified as Stripe 82 variable candidates by \citet{2007AJ....134.2236S}.

\begin{figure}
    \plottwo{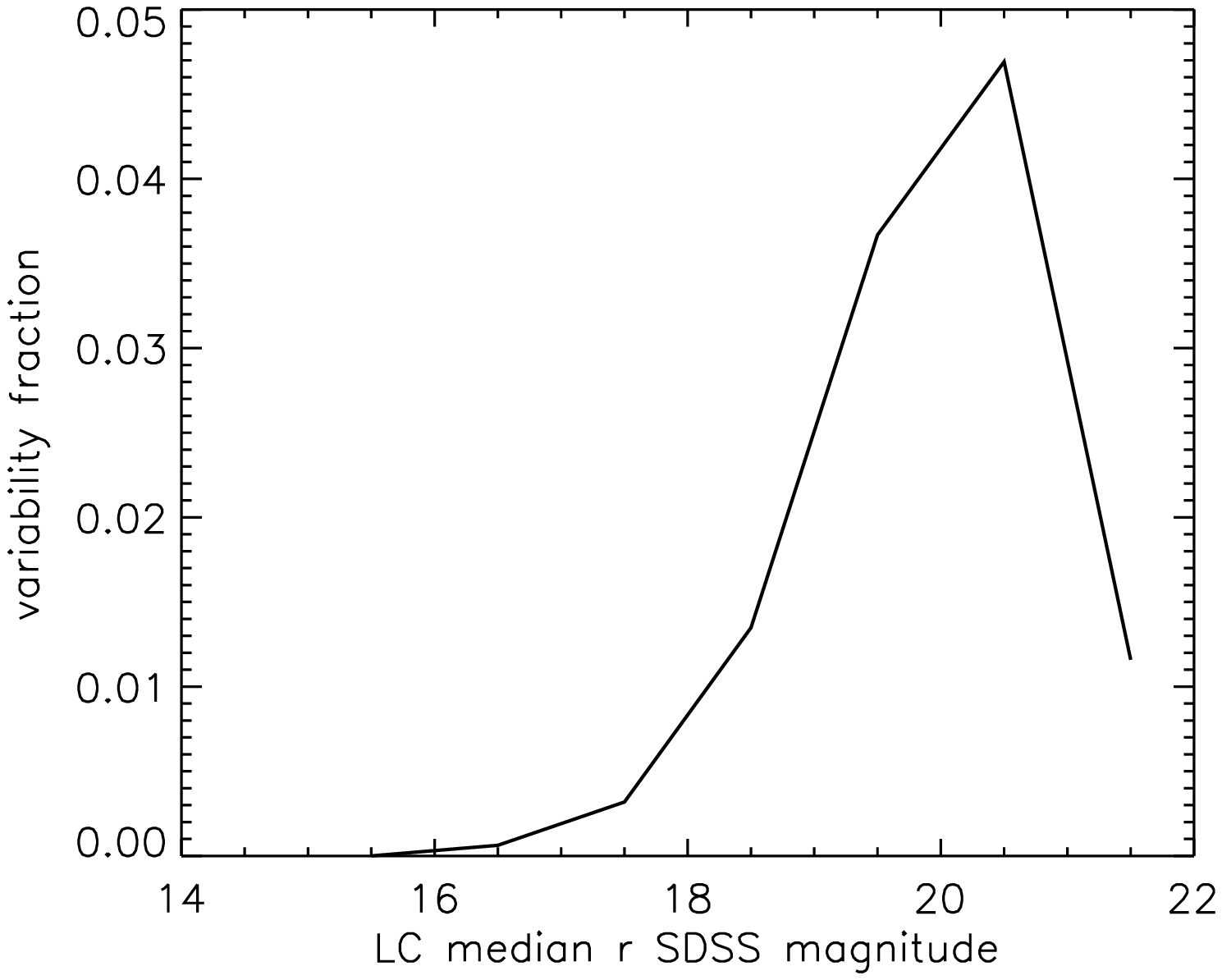}{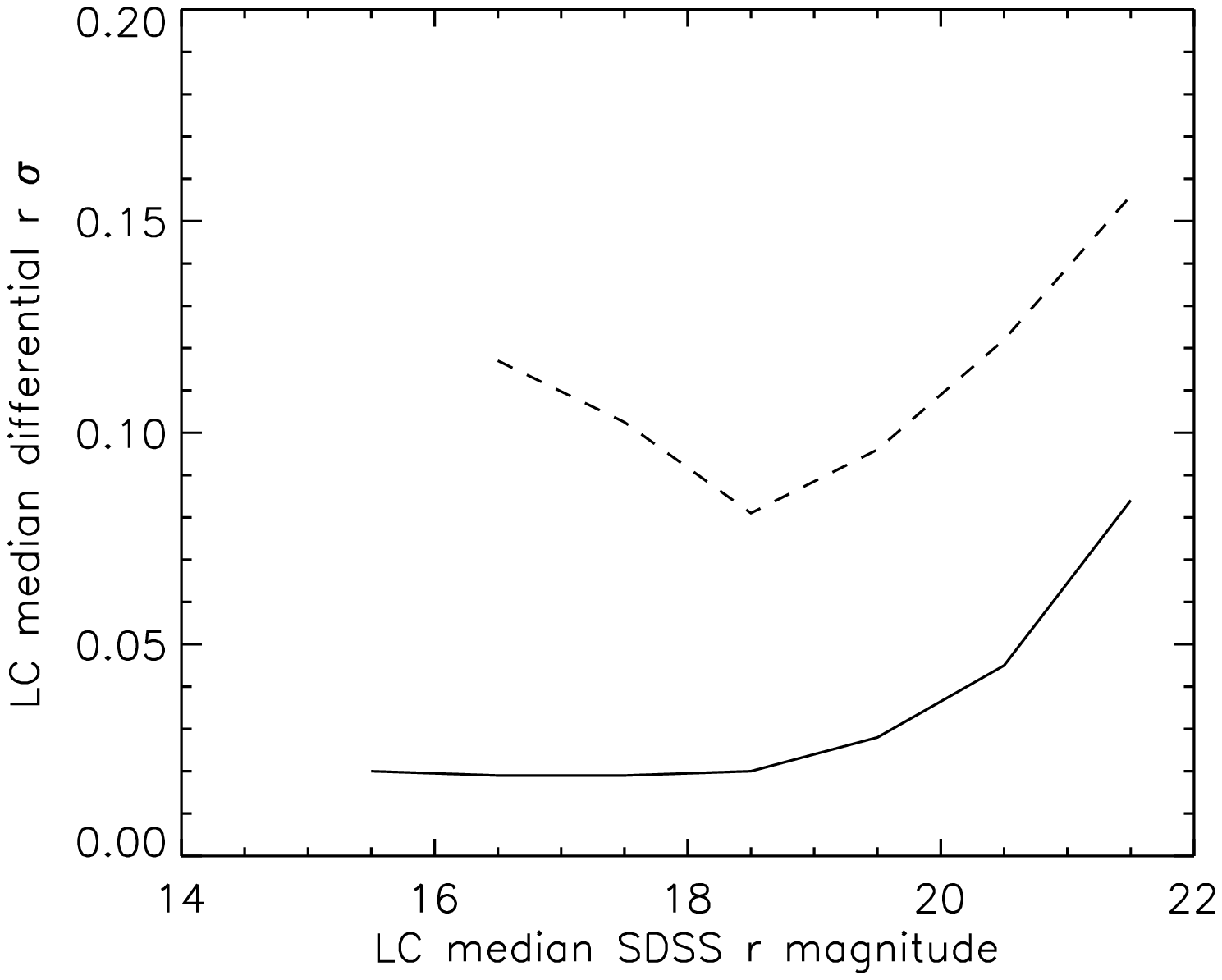}
    \caption{Left: The variability fraction as a function of $r$ median SDSS light-curve magnitude for the $r
    < 22.0$ sample of 221,842 objects. Right: The median differential magnitude light-curve $\sigma$ as a
    function of magnitude for objects marked nonvariable (solid line) and those tagged as probable variables
    (dashed line) for the $r < 22.0$ sample.}
    \label{fig_var_fraction}
  \end{figure}

  Our variable extraction routines produce 6,860 ($\sim3.1\%$ of the total objects) tentative variable
  candidates, and a final list of 6,520 ($\sim2.9\%$ of the total objects) probable variable candidates in our
  dataset. Figure \ref{fig_var_fraction} (left panel) presents the variable fraction as a function of SDSS $r$
  magnitude. This fraction rises to a maximum of $\sim$4.5\% at the level of $\sim 0.05$ mag for the bin $20.0
  < r \le 21.0$, which contains a significant number of point sources. The fall in the variability fraction
  observed in the bin $21.0 < r \le 22.0$ reflects the effect of increasing photometric noise on the
  determination of variability. In this magnitude bin, correlated variability across several bands is
  difficult to detect, thus the Stetson index trends towards zero, leading to a corresponding decrease in the
  variability fraction. Overall, we find that $\sim 2.9\%$ of the point sources in our dataset show
  variability at the level of $\sim 0.05$ mag, given a median $r$ magnitude of $\sim 19.9$. The right panel of
  Figure \ref{fig_var_fraction} shows the trends of the differential $r$ light-curve standard deviation with
  median SDSS $r$ light-curve magnitude for nonvariable (solid line) and probable variable (dashed line)
  objects respectively. The large difference between the two for all magnitude bins indicates that variables
  are robustly selected by our methods over this range of magnitudes.

\begin{figure}
    \plottwo{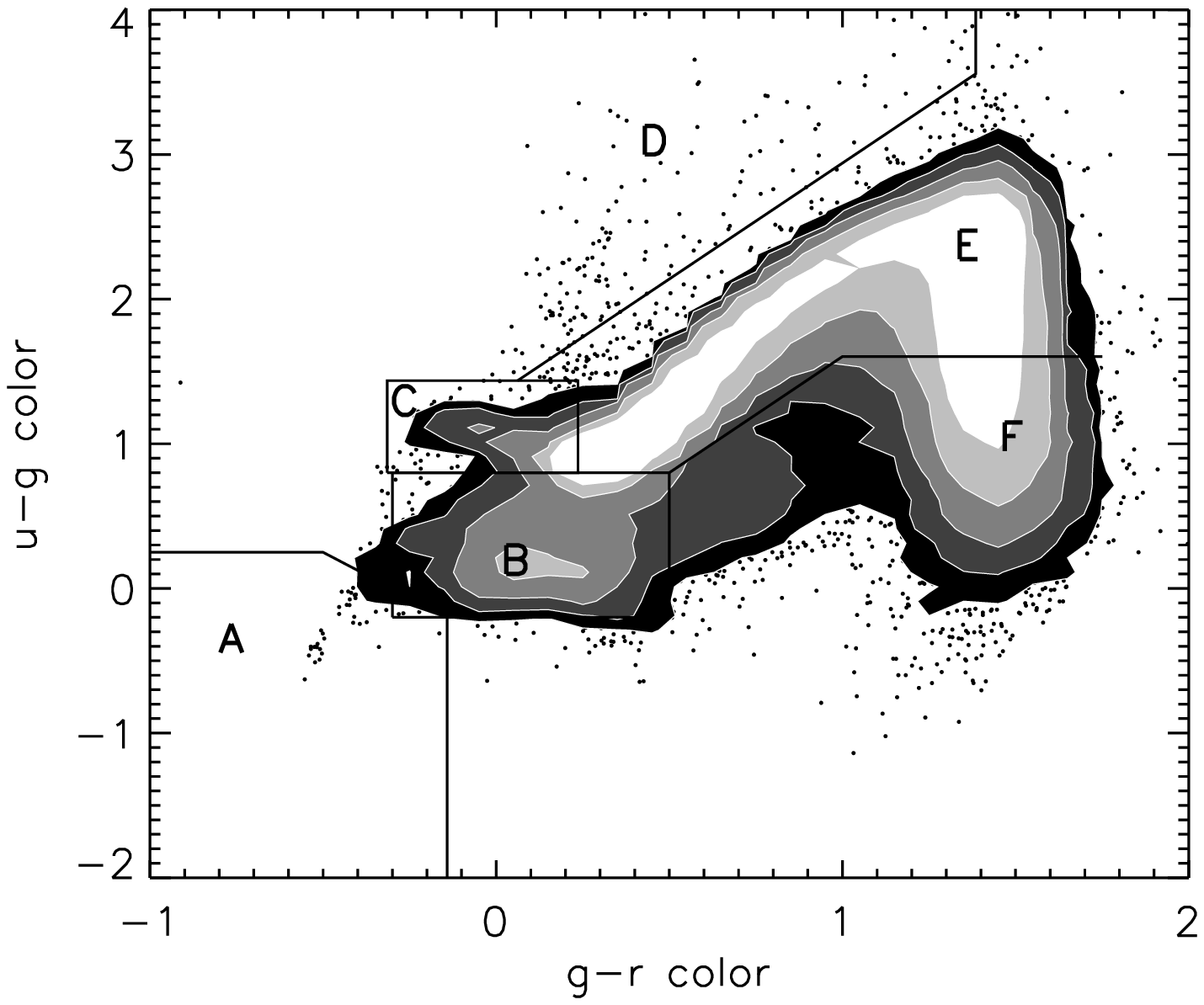}{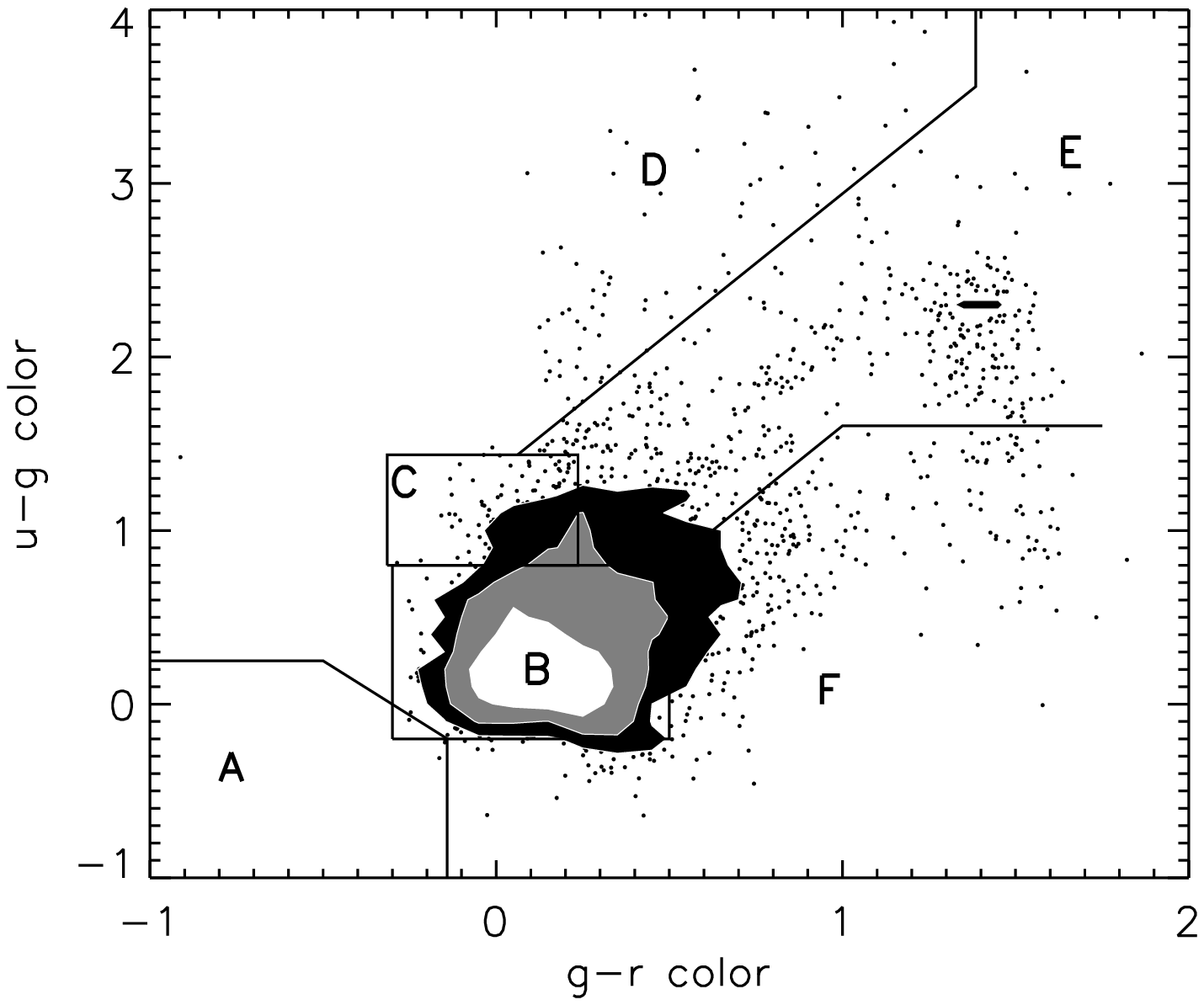}
    \caption{Left: SDSS \emph{g-r}/\emph{u-g} color-color diagram for 221,842 point sources in our final
    light-curve catalog for RA 0 to 4 h. The density contours are in half-dex increments ranging from 1.0 to
    3.0. Regions A--F indicate positions of various types of objects in color-color space: A. white dwarfs,
    B. low-$z$ quasars, C. A/BHB stars (including RR Lyrae), D. hi-$z$ quasars, E. main stellar locus,
    F. faint red objects (such as late M dwarfs and brown dwarfs). Right: The same diagram, this time only for
    6,520 probable variable point sources identified by our pipeline. Note that the QSOs in region B dominate
    the numbers of probable variables detected.}
    \label{fig_gr_ug}
  \end{figure}

  Figure \ref{fig_gr_ug} shows a $g-r$/$u-g$ color-color diagram for all point sources (left panel) and just
  the variable candidate point sources (right panel) in our final light-curve catalog. We can place all
  detected sources into six broad classification regions as depicted in this figure. Region A is where white
  dwarfs lie in $g-r$/$u-g$ color space, owing to their blue colors. Low-redshift quasars cluster in region B,
  also due to their blue colors. Region C contains A main sequence stars, blue horizontal branch stars, as
  well as blue stragglers. Region D includes mostly quasars with high redshifts ($z > 3.0$) as identified by
  cross-matching with the SDSS Quasar Catalog. Region E is the main stellar locus, with stars getting redder
  in color space as their mass decreases. This results in a progression of spectral types from approximately F
  to early M towards the top-right corner of the color-color diagram. The lowest-mass stars (the late M
  dwarfs), brown dwarfs, and other faint red objects with unreliable $u-g$ colors are found in region F. Their
  $u-g$ colors are unreliable due to the large photometric error in these bluer bands caused by the very red
  spectral energy distributions of these objects.

  The right panel of Figure \ref{fig_gr_ug} shows the $g-r$/$u-g$ color distribution of 6,520 probable
  variable candidates. The two dominant classes of variables appear to be the QSOs in region B, and a
  significant amount of faint red objects present along the stellar locus in region E. We note here that the
  majority of variable objects classified as `unknown' in Table \ref{type_table} are located within the color
  space defined by region B, and thus are themselves likely to be QSOs. We discuss these further in Section
  \ref{sec_sdss_qsos}.

\begin{figure}
    \plotone{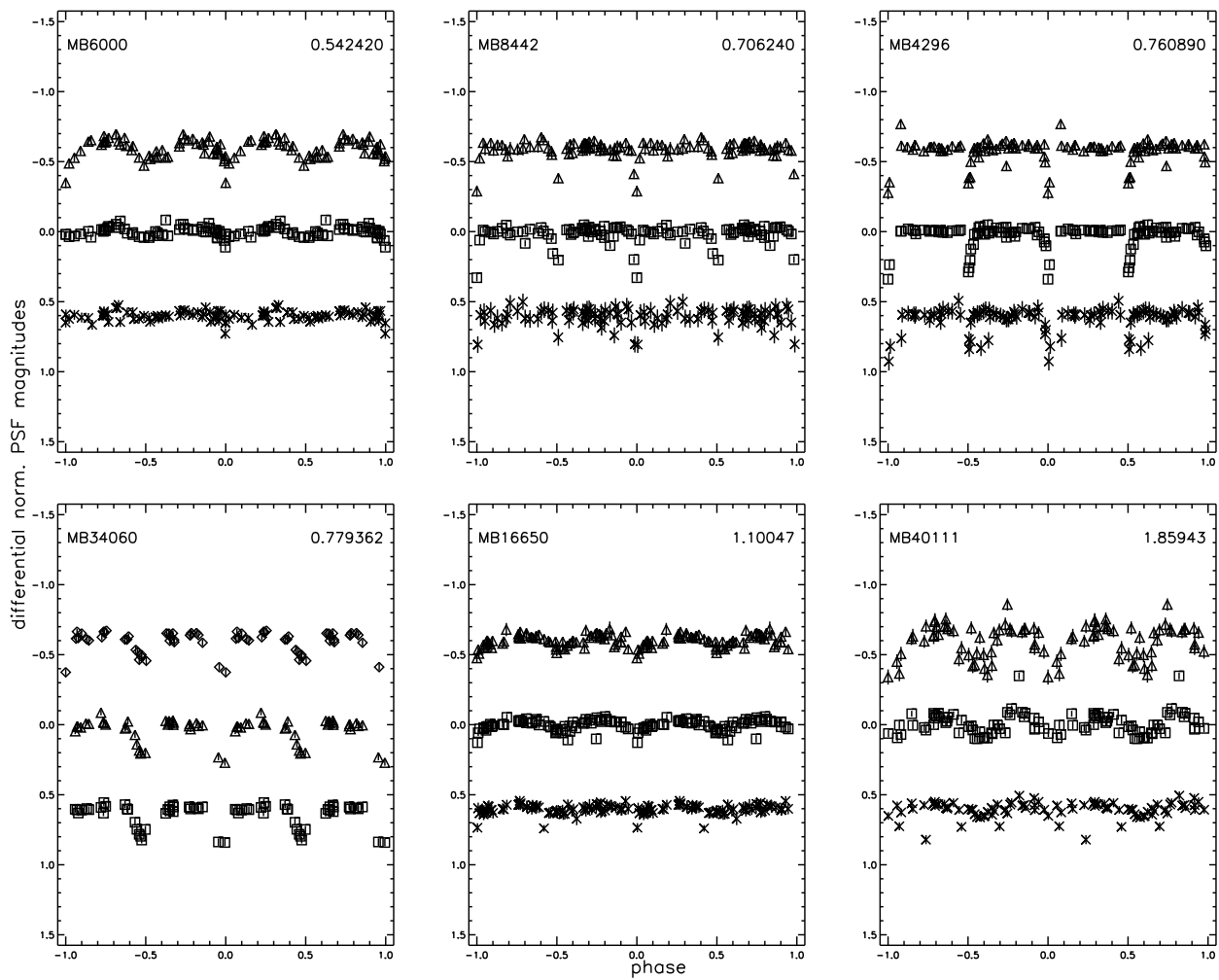}
    \caption{Phased differential magnitude light-curves for a sample of 6 objects with uncertain periods, in
    order of increasing period. The internal designation for an object and its best estimated period in days
    is shown in each panel. Diamond symbols are $g$ light-curve points, triangle symbols are $r$ light-curve
    points, box symbols are $i$ light-curve points and cross symbols are $z$ light-curve points. The
    light-curves are phased around the time of minimum light and repeated twice from phase -1 to +1 to clearly
    show variability.}
    \label{fig_bad_periods}
  \end{figure}

  We identify 143 periodic variables among the 6,520 probable variable candidates extracted from the
  catalog. Of these, 12 appear to show periodic variability upon inspection of their \emph{unphased}
  light-curves, but have extremely poor phase coverage or too much photometric noise to allow any period to be
  determined and assigned to the object. Thirty more objects, most of which are mid to late M-dwarfs, show
  eclipse-like periodic variability, but have ambiguous periods or phased light-curves that do not show
  convincing periodicity upon closer inspection\footnote{A list of these is available from
  \url{http://shrike.pha.jhu.edu/stripe82-variables}}. Figure \ref{fig_bad_periods} shows light-curves of a
  sample of 6 such problematic objects. We remove these 30 objects as well as the 12 that have no assigned
  periods from any further consideration.

\begin{figure}
    \plottwo{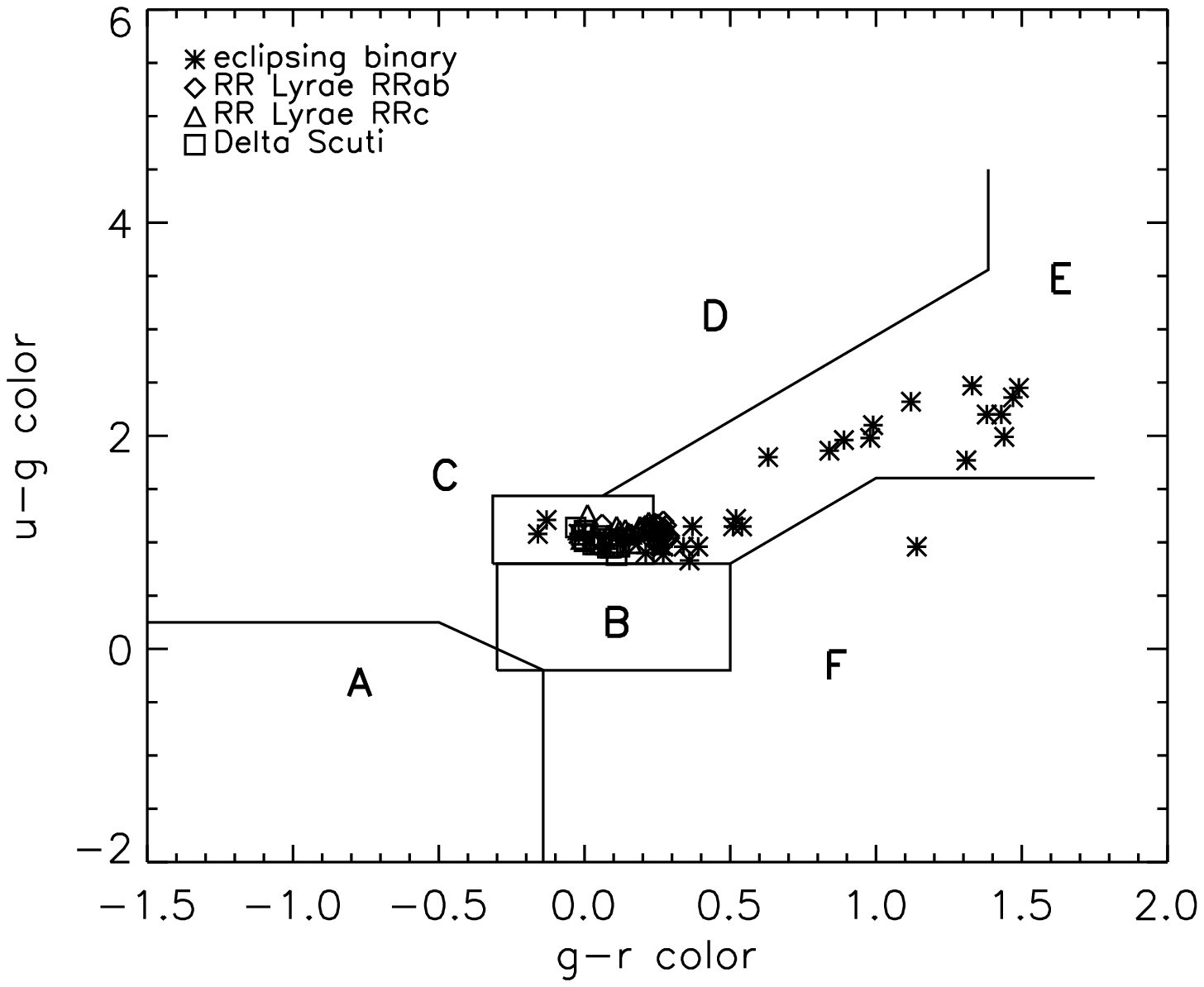}{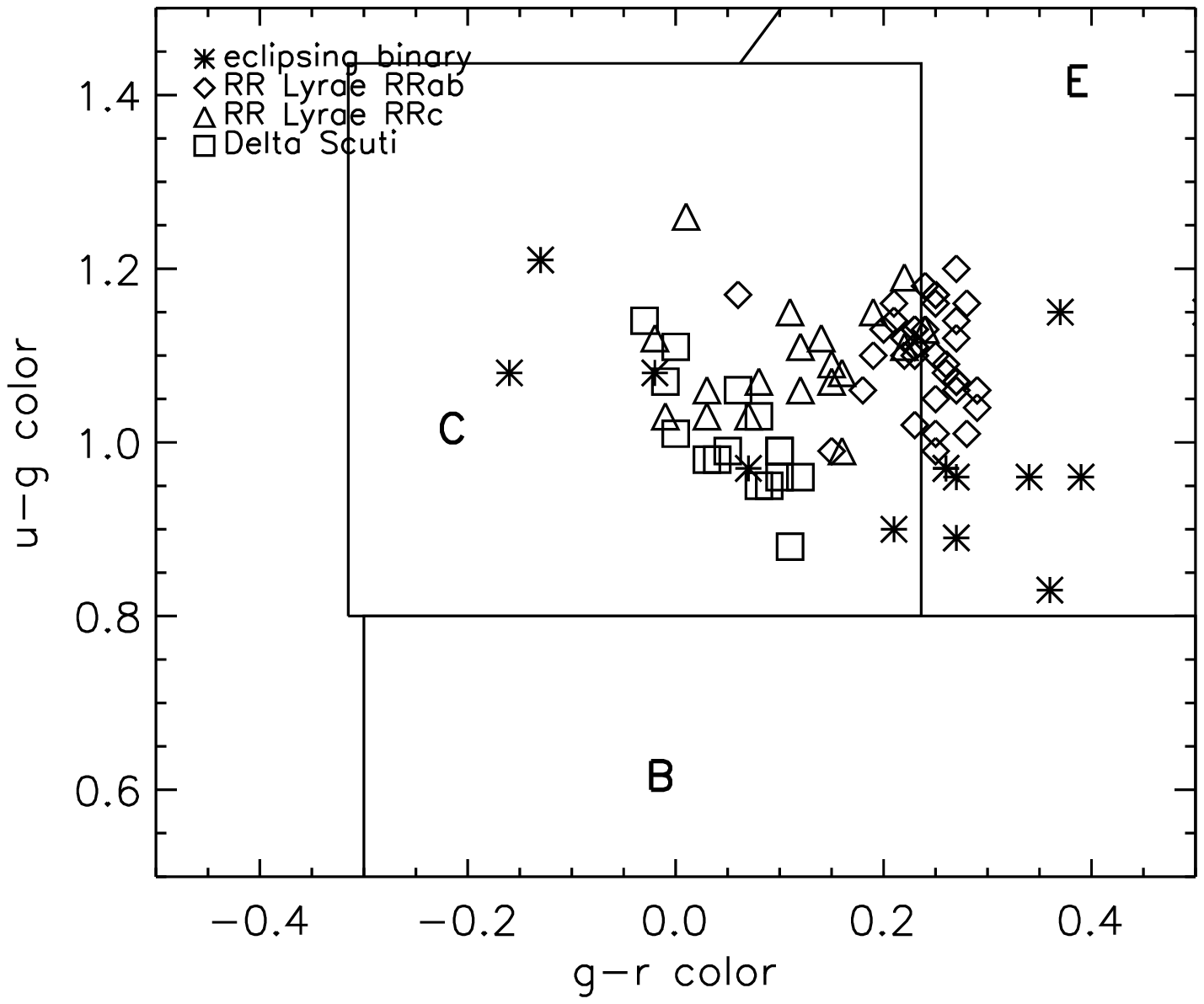}
    \caption{Left: Positions in $u-g$/$g-r$ color space of 101 periodic variables identified in this work. Regions
    A--F are the same as in Figure \ref{fig_gr_ug}. Right: Same as in the left panel, but zoomed into region C.}
    \label{fig_periodic_color}
  \end{figure}

  We are finally left with 101 periodic variables identified from our light-curve catalog. Table
  \ref{periodic_variable_table} lists these objects as classified by the shape of their light-curves. The
  positions of these periodic variables in $u-g$/$g-r$ color space are shown in Figure
  \ref{fig_periodic_color}. All of these objects are found on the stellar locus, and range in spectral type
  from A to late M. The RR Lyrae and Delta Scuti candidates are clustered in region C, where the blue
  horizontal branch stars may be found, as expected. In contrast, we find eclipsing variables all along the
  stellar locus. No periodic variables are found among the cross-matched white dwarfs or QSOs.

  \subsection{Variable Detection and Period Recovery Efficiency}\label{sec_efficiency}

  We test our general variable identification process by constructing a synthetic light-curve catalog
  representative of the objects present in our dataset. Running this light-curve catalog through our ensemble
  photometry, variable extraction, and period finding routines then provides upper limits for the efficiency
  and reliability of our detection methodology. We first draw median light-curve magnitudes, the associated
  standard deviations, the number, and dates (MJD) of observations for each object from probability
  distributions matching the observed distributions of these quantities. Light-curve measurements are then
  perturbed using Gaussian noise appropriate for the objects' assigned magnitudes obtained from the observed
  magnitude-$\sigma$ relation (for example, Figure \ref{fig_errmag_ensemble}). A catalog of 46,000 synthetic
  objects covering an area of 5 degrees in right ascension and 2.5 degrees in declination is constructed using
  these distributions and serves as a `background' for ensemble photometry of two classes of objects:
  sinusoidal variables and completely nonvariable objects. We insert 2,000 objects of each type into the
  synthetic light-curve catalog, making for a total of 50,000 objects that are then processed through our
  pipeline exactly in the same manner as real objects from the Stripe 82 dataset.

  The 2,000 artificial sinusoidal variables are inserted into the light-curve catalog at uniform random
  spatial coordinates, thus distributing them evenly over the 12.5 square degree area under
  consideration. Periods for these objects are chosen from a uniform distribution in $\log P$ (where $P$ is
  the period in days) ranging from 0.1 to 100.0 days. Amplitudes are assigned to these objects from a uniform
  random distribution ranging from 0.01 to 0.50 magnitudes. We simplify matters by assuming the same
  variability amplitude for each band. The epoch of minimum light for each periodic variable is also chosen
  from a uniform random distribution of observation dates present in the dataset. We then generate
  sinusoidally variable magnitude measurements distributed over the assigned observation dates using the
  periods, amplitudes, and epochs of minimum light thus chosen. Finally, each object's assigned magnitudes are
  perturbed by an amount chosen from a Gaussian distribution centered at 0.0 and with a standard deviation
  equal to the assigned artificial light-curve standard deviation, thus introducing `noise' to the
  light-curves of each object.

  Objects are required to have at least 10 observations in their timeseries to be eligible for ensemble
  photometry; 1,587 of the inserted variables survive this cut. Objects with color errors greater than 0.2 and
  magnitudes fainter than $r = 22.0$ mag are then discarded as in our usual processing. We then run ensemble
  photometry and variable object extraction routines on the remaining objects. This leaves us with 1,393
  objects to extract variables from. The results of these routines are summarized in Table
  \ref{fake_data_table}. The pipeline identifies 1,101 objects as \emph{tentative variables}; 1,091 of these
  are eventually tagged as \emph{probable variables}, resulting in an overall (ideal case) variable recovery
  rate of $\sim$ 78.3\%. The variable extraction procedures are most efficient at brighter magnitudes
  (reaching maximum efficiency near $r \sim 18.5$). They become progressively less so with fainter magnitudes,
  due to increasing photometric noise that makes it more difficult to identify variability.

\begin{figure}
    \plotone{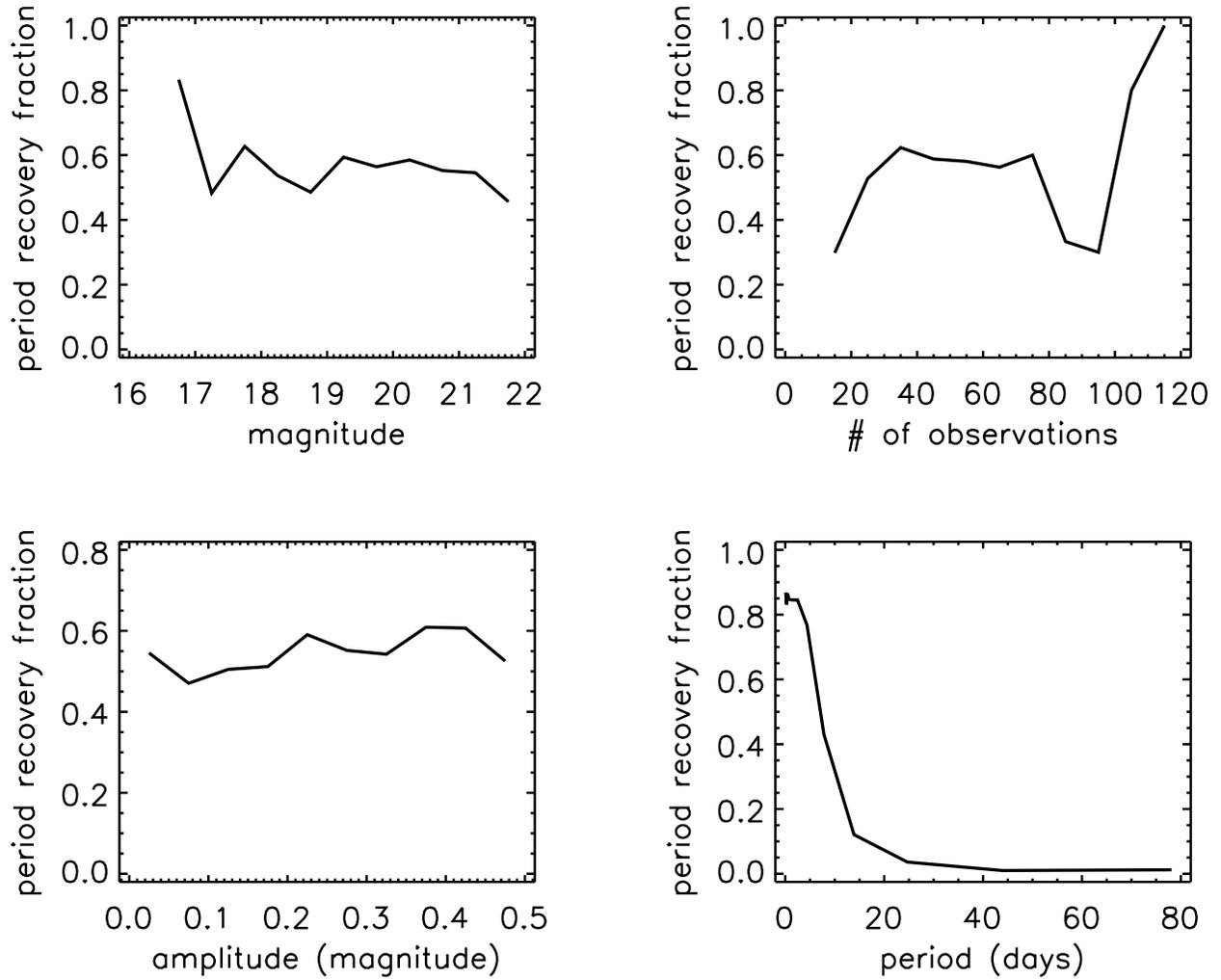}
    \caption{Top left: Period recovery fraction as a function of the median $r$ magnitude of synthetic
    probable variables. Top right: Period recovery fraction as a function of the number of
    observations. Bottom left: Period recovery fraction as a function of the amplitude. Bottom right: Period
    recovery fraction as a function of the input period.}
    \label{fig_sim_binplots}
  \end{figure}

  The final step in characterizing the efficiency of variable extraction is to test the recovery of the
  assigned periods for these sinusoidal variables. We carry this out by running all detected synthetic
  probable variables through our period finding routines as described in Section \ref{sec_periods}. We expect
  the recovery rate here to be quite low, given the limited phase coverage and small number of observations of
  each object. A periodic variable is considered to be recovered with the correct period if the difference
  between the recovered period and the original assigned period is less than 0.001 days. Larger differences
  introduce significant phasing errors when one attempts to construct phased light-curves for these
  objects. Overall, 600 out of 1,091 artificial sinusoidal probable variables have their periods recovered
  successfully, resulting in a recovery efficiency of $\sim$ 54.9 \%. Figure \ref{fig_sim_binplots} shows how
  this recovery fraction behaves as a function of median $r$ magnitude (top left), number of observations (top
  right), variability amplitude (bottom left), and input period (bottom right). The period of the variable
  object and the number of observations determine the phase coverage, and affect the period recovery fraction
  the most. Long period variables have poor phase coverage and a correspondingly small recovery fraction. The
  fraction drops with increasing magnitude, as expected, due to the increasing noise in the light-curves that
  makes it difficult to phase them correctly. The variable recovery fraction generally increases with the
  number of observations, except for the last few bins where the trend becomes ambiguous due to the small
  number of objects in these bins.

\begin{figure}
    \plotone{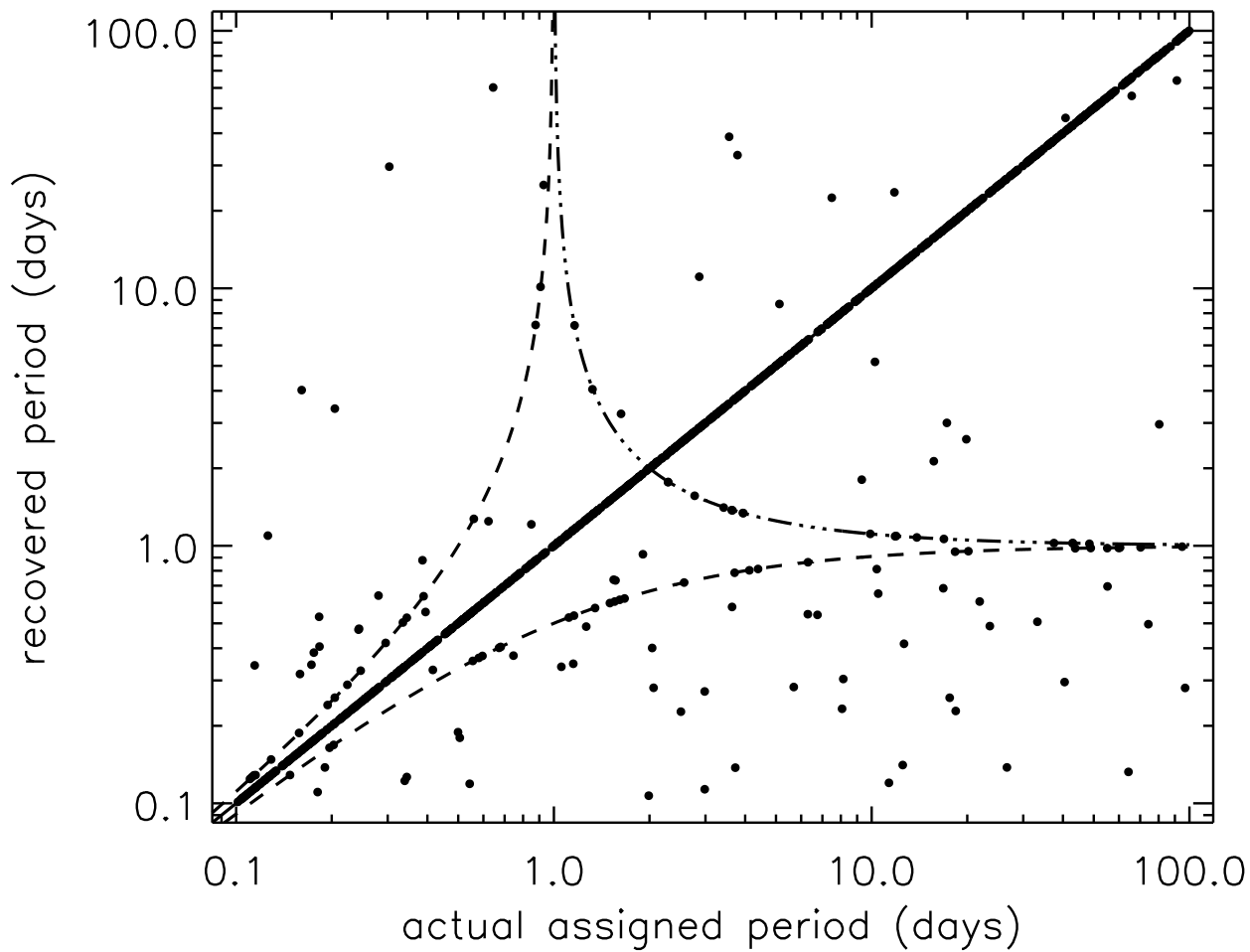}
    \caption{A comparison of assigned periods for artificial periodic variables and the period recovered by
    our pipeline. The solid line with unit slope indicates that the recovered and assigned periods are
    equal. The two dashed curves depict the relation $P_{recovered} = P_{assigned}/(1 \pm P_{assigned})$,
    while the dot-dashed curve depicts the relation $P_{recovered} = P_{assigned}/(P_{assigned} - 1)$. The
    solid circles are results of this experiment for each individual artificial periodic variable. Also see
    Figure 6 in \citet{2009MNRAS.398.1757W}.}
    \label{fig_sim_period_comp}
  \end{figure}

  Using results from our simulations, we can also address the question of how our period finding algorithms
  fail when they do. Figure \ref{fig_sim_period_comp} shows the relation between the assigned input periods
  and the actual recovered periods for all 1,091 artificial periodic variables identified by our pipeline. The
  curves depict the three most common types of aliases for our periods present in this dataset. We see that in
  most cases, the recovered period corresponds closely to the input period, but there are significant numbers
  of cases where the period finding algorithms latch on to these aliases in lieu of the actual input
  period. This is mostly a function of the phase coverage on each variable object; the better the balance
  between an object's period and the number of its observations, the more likely it is that we will recover
  the correct period.

  We see this effect most clearly in our sample of actual eclipsing binary candidates identified from Stripe
  82 data, where we discard nearly half of our initial sample due to ambiguous period recovery before settling
  on a final sample of 30 objects. The rapid changes in these objects' light-curves near eclipse and the very
  small amount of time each object spends in eclipse cause significant difficulties for our period finding
  algorithms. A photometric followup campaign to better characterize these objects would therefore be most
  successful if it used ephemerides generated using the periods reported here as well as several harmonics of
  these periods.

  Finally, we inject 2,000 artificial completely nonvariable objects into the synthetic light-curve
  catalog. These objects only have Gaussian noise added to their light-curves, but otherwise show no trends
  over time. These are used to test the false positive variable identification rate of our pipeline. 1,578
  such objects survive the initial ensemble pipeline cut on number of observations, while 1,399 of these
  survive our additional cuts on the color errors and imposed $r$ magnitude limit. Running the pipeline on
  these remaining objects yields 12 objects tagged as \emph{tentative variables}; only 2 of these are
  subsequently identified as \emph{probable variables}. Neither of these remaining objects appear to have to
  have any periodicity, as expected. On average, therefore, $\sim 0.15\%$ of all artificial nonvariable
  objects inserted into the catalog are expected to be misidentified as actual variables. The actual false
  positive rate will be higher than this number, but we believe our pipeline separates variables from
  nonvariables efficiently, based on the evidence from these simulations. We expect the contamination fraction
  of such falsely tagged nonvariable objects in our variable sample to be, at worst, near $10\%$.

  \subsection{Obtaining the Light-curve Catalogs}\label{sec_access}
  
  A complete set of light-curves for the 6,520 variable candidates, lists of the different types of periodic
  variables identified here, catalogs of the objects discussed in the following sections, and finally, summary
  files describing all 221,842 point sources in this dataset, are available at the following website:

  http://shrike.pha.jhu.edu/stripe82-variables/

  All object catalogs are in standard FITS binary table format. Light-curves are provided as individual FITS
  tables, CSV, and PDF files associated with each variable object, and include various related
  diagnostics. Details on the format and content of these data files are available at the website mentioned
  above. Finally, the software code and programs used to construct our light-curve catalog and carry out
  ensemble photometry are also provided.

  \subsection{Periodic Variables}\label{sec_periodic_variables}
  
  \subsubsection{Eclipsing and Ellipsoidal Binary Candidates}\label{sec_eclipsing_binaries}
  
\begin{figure}
    \plotone{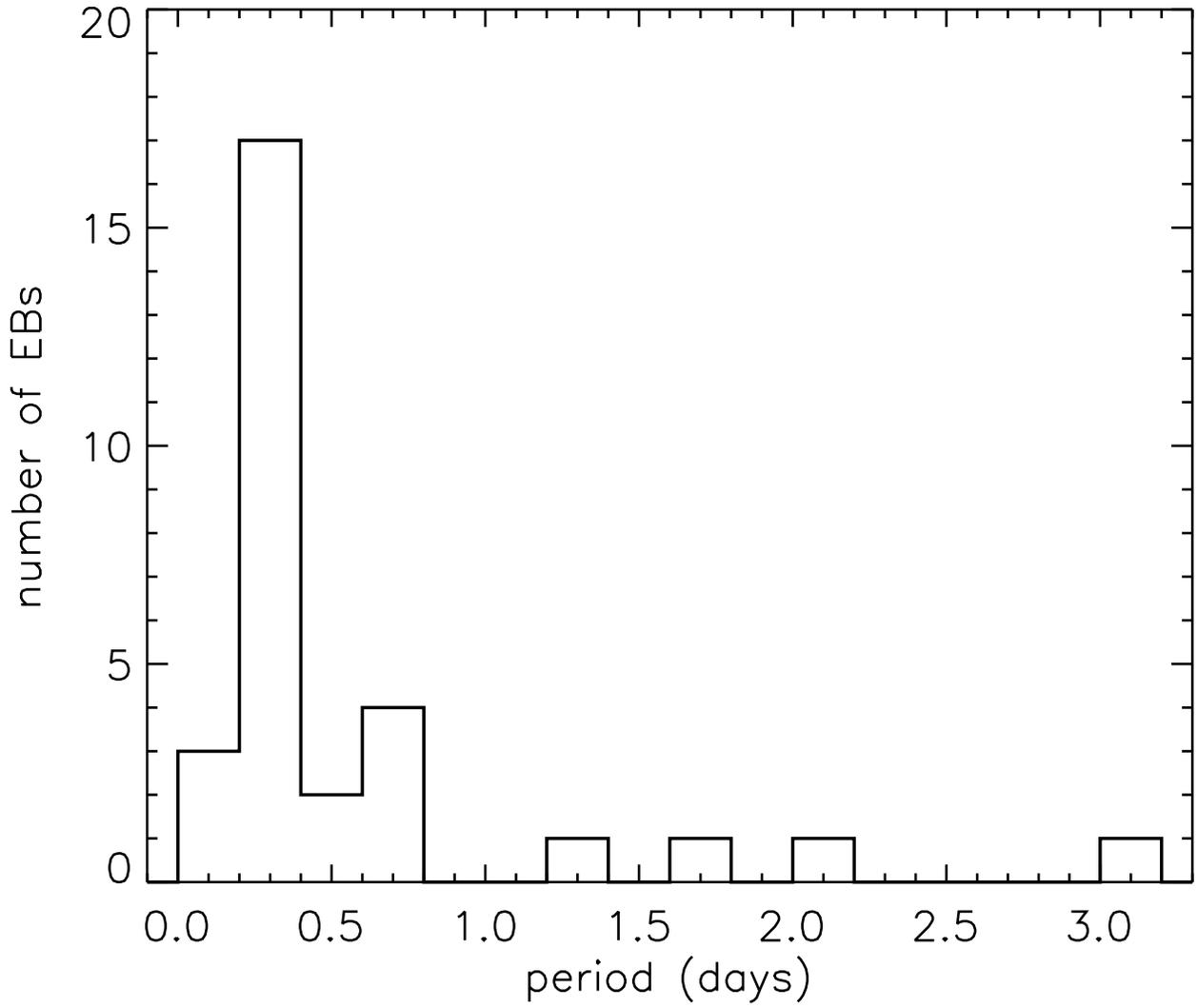}
    \caption{Period histogram for 30 eclipsing and ellipsoidal binary candidates from this dataset, using a
    bin width of 0.2 days. There is a distinct lack of objects with periods greater than 1.0 day.}
    \label{fig_periodhist_ebs}
  \end{figure}

\begin{figure}
    \plotone{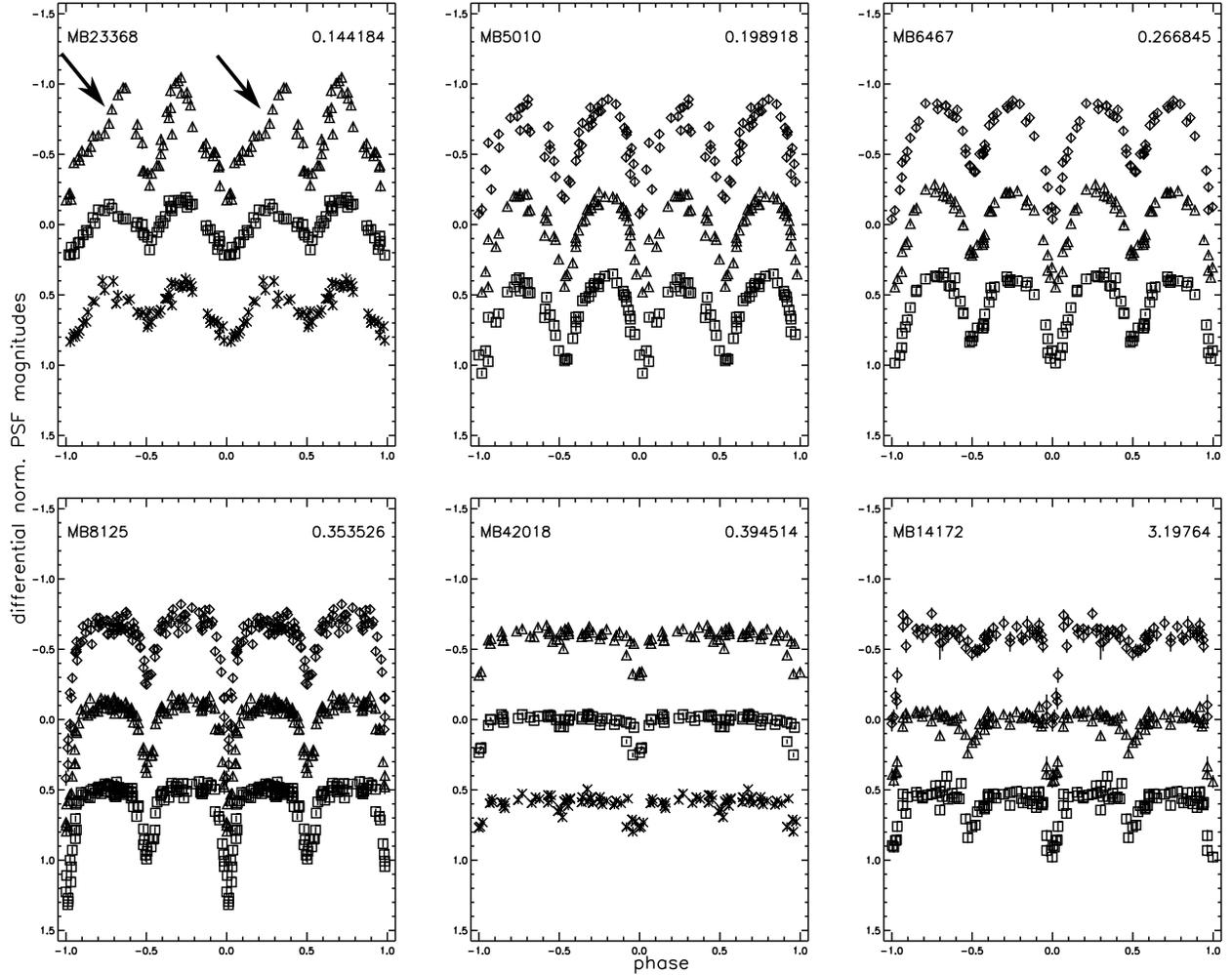}
    \caption{Phased differential magnitude light-curves for a sample of 6 eclipsing binary candidates
    identified by our pipeline, in order of increasing period. The internal designation for an object and its
    period in days is shown in each panel. Diamond symbols are $g$ light-curve points, triangle symbols are
    $r$ light-curve points, box symbols are $i$ light-curve points and cross symbols are $z$ light-curve
    points. The light-curves are phased around the time of minimum light and repeated twice from phase -1 to
    +1 to clearly show variability. The arrows point out the $r$-band out-of-eclipse variability observed for
    MB23368 mentioned in the text.}
    \label{fig_nonm_ebs}
  \end{figure}

  We find 30 eclipsing binary candidates in our dataset. The period distribution for all of these objects is
  shown in the form of a histogram in Figure \ref{fig_periodhist_ebs}. Example light-curves for these objects
  are presented in Figure \ref{fig_nonm_ebs}, and a listing of all eclipsing binary candidates found by our
  pipeline is presented in Table \ref{all_eb_table}. The comments column in that table indicates the type of
  object as classified by its color, as well as noteworthy features of its light-curve. Objects with periods
  greater than 1.0 day are relatively scarce. This is primarily due to the sampling cadence of the
  survey. Objects with longer periods spend little time in either primary or secondary eclipse, and sparse,
  uneven sampling of their light-curves, coupled with a small number of observations makes it difficult to
  obtain unambiguous periods. In contrast, objects with small periods are more suited to our sampling cadence;
  these are more likely to be in eclipse at any one time, and thus provide a better estimate of the
  period. The abundance of short period objects, therefore, is a result of observational bias, and cannot be
  ascribed to physical reasons.

  The binary candidates appear to be rather diverse in their nature. We find at least five W UMa type contact
  binary systems; two good examples of these are objects MB5010 (SDSS J035138.50-003924.5) and MB6467 (SDSS
  J031021.22+001453.9) in Figure \ref{fig_nonm_ebs}. Object MB23368 (see Figure \ref{fig_nonm_ebs}; SDSS
  J025953.33-004400.3) is a very short period late M-dwarf binary candidate, which shows an interesting out of
  eclipse variation in its light-curves, most prominently in the $g$ and $r$ band. This may be evidence of a
  star spot rotating in and out of view, but a more densely sampled light-curve for this object is required
  for confirmation.

  Other interesting binary candidates include objects MB42018 (SDSS J032515.05-010239.7) and MB14172 (SDSS
  J024255.78-001551.5). MB42018 is a mid M-dwarf object that appears to have a short period and rather shallow
  eclipses ($\sim 0.3$ mag for the primary, $\sim 0.1$ mag for the secondary), indicating a possible low mass
  companion. We note, however, that there is a third star present near the eclipsing binary candidate (within
  1.0$\arcsec$) and this may instead indicate that the shallow eclipses are caused by blended light from this
  system. At the other end of the size scale, MB14172 is a blue star with a long period and a deep primary
  eclipse and a much shallower secondary eclipse. It is tagged as a low-metallicity object and possibly a
  giant star by our pipeline, perhaps indicating a giant-dwarf binary system. Finally, we point out the
  well-sampled light-curve of the object MB8125 (SDSS J030753.52+005013.0, a short period K dwarf eclipsing
  binary candidate) as an example of the accuracy in period finding possible even with a relatively small
  number of observations (115 in this case) over the ten year length of the survey. Light-curves of all three
  objects are shown in Figure \ref{fig_nonm_ebs}.

  We note here that, in addition to the newly discovered binaries described above, we have also recovered two
  confirmed low mass eclipsing binaries present in the footprint of Stripe 82: the objects 2MASS
  J01542930+0053266 (\citealp{2008MNRAS.386..416B}; M0+M1), and SDSS J031823.88-010018.4 (also known as
  SDSS-MEB-1; \citealp{2008ApJ...684..635B}; M4). We were, however, unable to recover the periods of these
  objects as reported in the literature, because of limited phase coverage and a small number of observations,
  but our variable identification method was sufficient to pick out these objects as obvious variable
  candidates. These two objects are listed in Table \ref{all_eb_table}, along with our own eclipsing binary
  candidates.

  The poor phase coverage of most of our light-curve catalog, especially for our eclipsing binary candidates,
  precludes detailed analysis of these objects at this point. Dedicated photometric and spectroscopic followup
  will be required to draw meaningful conclusions about their physical properties as derived from their
  light-curves. We will discuss results of such followup for some of these binary candidates in a future
  paper.

  \subsubsection{RR Lyrae and Delta Scuti Variables}\label{sec_rr_lyrae}
  
\begin{figure}
    \plottwo{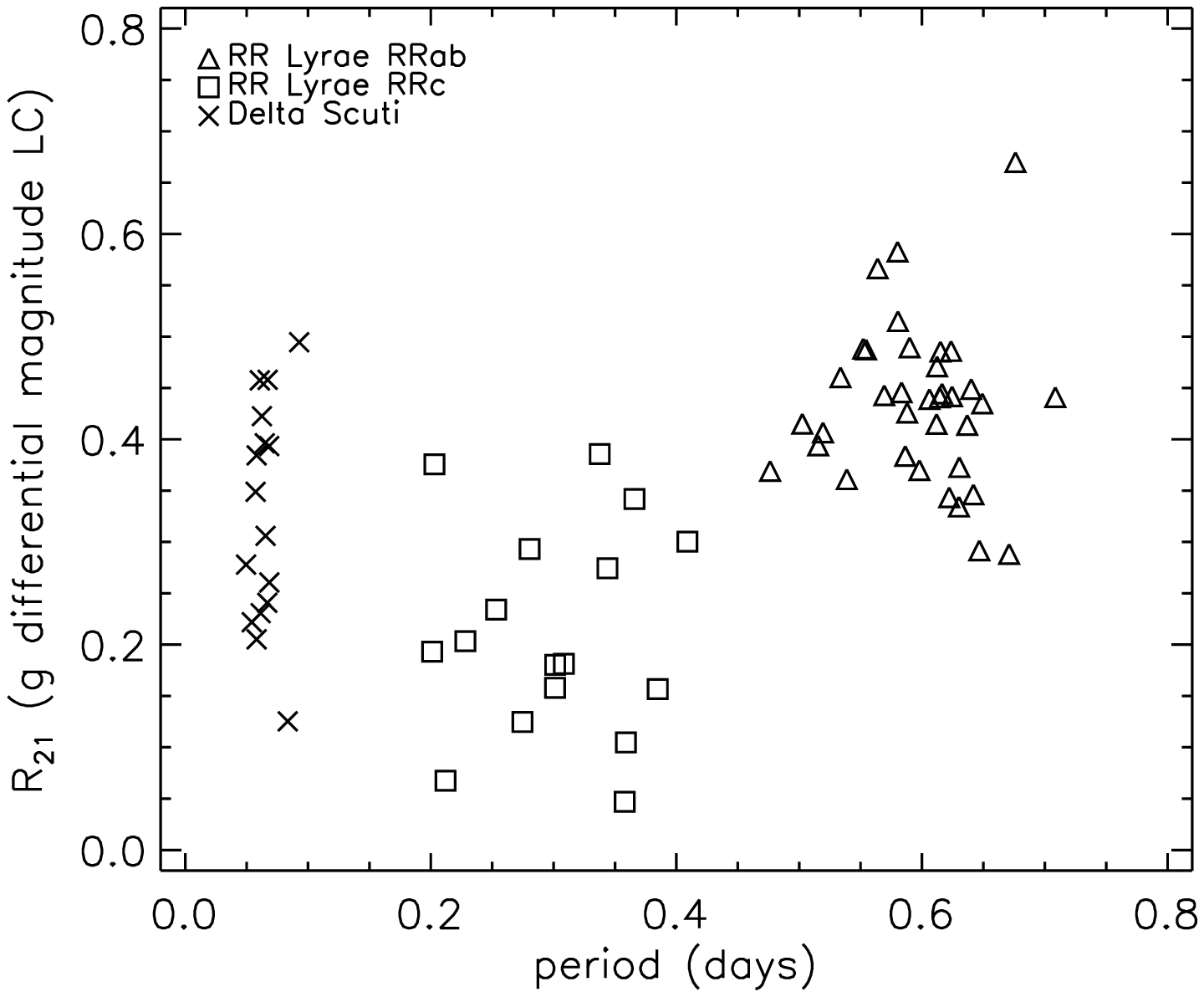}{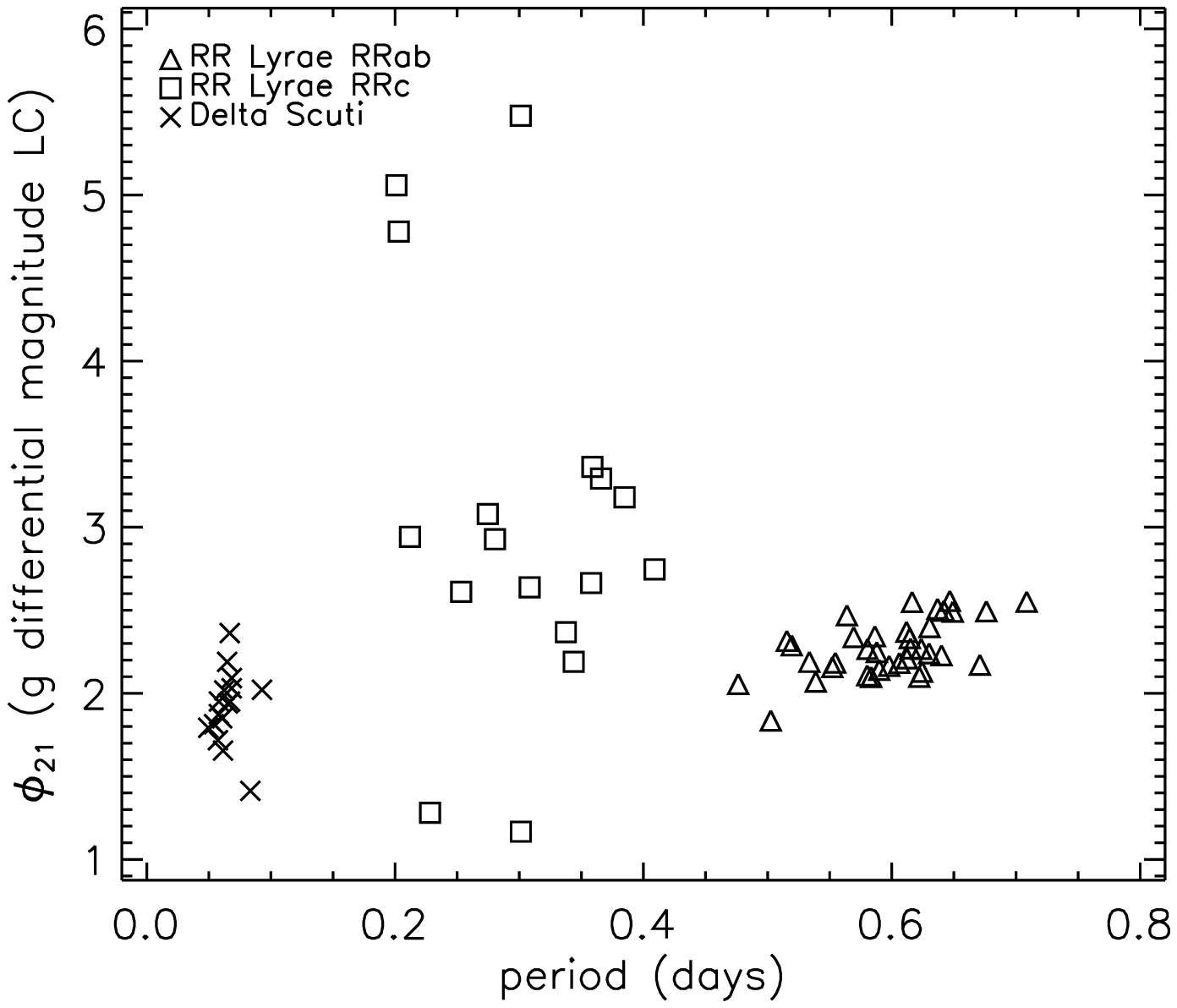}
    \caption{Left: Fourier parameter $R_{21}$ calculated for $g$-band differential magnitude light-curves
    plotted against the period for RRab (triangles), RRc (boxes), and Delta Scuti variable candidates
    (crosses). The three sinusoidal variable types are easily distinguishable from each other. Right: Fourier
    parameter $\phi_{21}$ calculated for $g$-band differential magnitude light-curves plotted against the
    period for the same objects in the left panel.}
    \label{fig_fourier_r21}
  \end{figure}

  We find 71 sinusoidal variables in our dataset. These are distinguished from the eclipsing binaries
  discussed previously by inspection of their phased differential magnitude light-curves. These variables can
  be further classified into three broad types based on light-curve shape, period, and amplitude; the RRab RR
  Lyrae, the RRc RR Lyrae, and high amplitude Delta Scuti variables. We fit Fourier components to the
  light-curves of all 71 sinusoidal variables to robustly classify them as belonging to one of the three
  sinusoidal variable types, thus
  \begin{equation}
    f(t) = A_{0} + \sum_{k=1}^{N} A_{k} \sin \left[ k \omega (t - t_{0}) + \phi_{k}\right]
  \end{equation}
  where $A_{0}$ is the mean magnitude, $A_k$ is the amplitude of the $k$th Fourier term, $\omega$ is the
  angular frequency and is given by $\omega = 2\pi/P$ for a period $P$, $t_0$ is the epoch, and $\phi_k$ is
  the phase for the $k$th Fourier term. We restrict the maximum order of the fit $N$ to 3, because of the low
  number of observations per object, which gives us poor phase coverage. We then calculate the Fourier
  parameters $R_{21}$ and $\phi_{21}$, given by the respective relations
  \begin{equation}
    R_{21} = \frac{A_2}{A_1}\ \mathrm{and}\ \phi_{21} = \phi_2 - 2\phi_1.
  \end{equation}
  The calculation of these two parameters allows us to quantitatively distinguish between RRab, RRc, and Delta
  Scuti variables \citep{2001A&A...371..986P}. We further require that an object be fit successfully using
  Fourier components to be classified as either one of these three sinusoidal variable types and be accepted
  as such into our periodic variable catalog. Figure \ref{fig_fourier_r21} shows plots of $R_{21}$ against
  period (left panel) and $\phi_{21}$ against period. The three classes of sinusoidal variable separate
  cleanly in both diagrams.

\begin{figure}
    \plotone{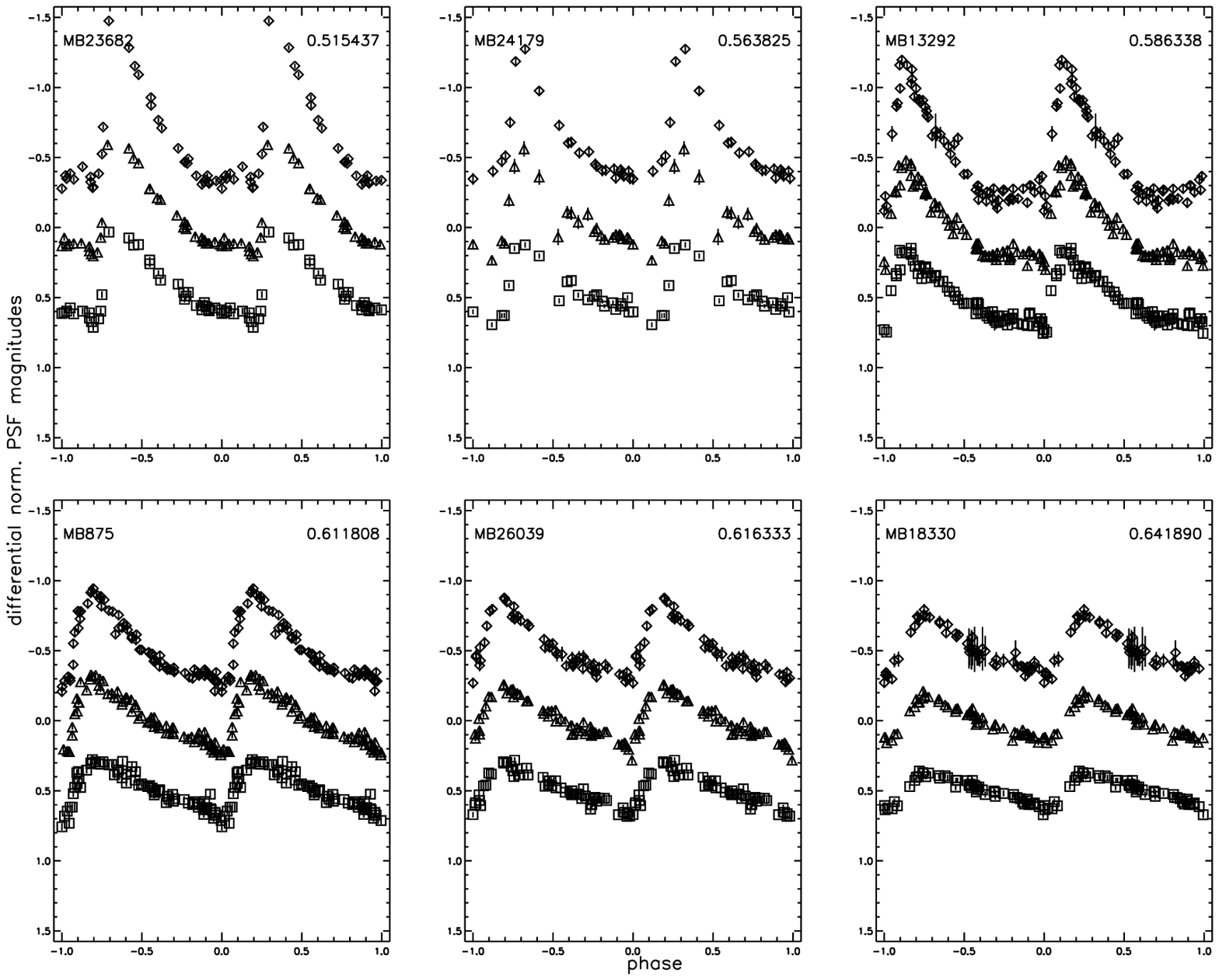}
    \caption{Phased $gri$ differential magnitude light-curves for a sample of 6 RRab RR Lyrae candidates
    identified by our pipeline, in order of increasing period. The internal designation for an object and its
    period in days is shown in each panel. Diamond symbols are $g$ light-curve points, triangle symbols are
    $r$ light-curve points, and box symbols are $i$ light-curve points. The light-curves are phased around the
    time of minimum light and repeated twice from phase -1 to +1 to clearly show variability.}
    \label{fig_rr_ab}
  \end{figure}

\begin{figure}
    \plotone{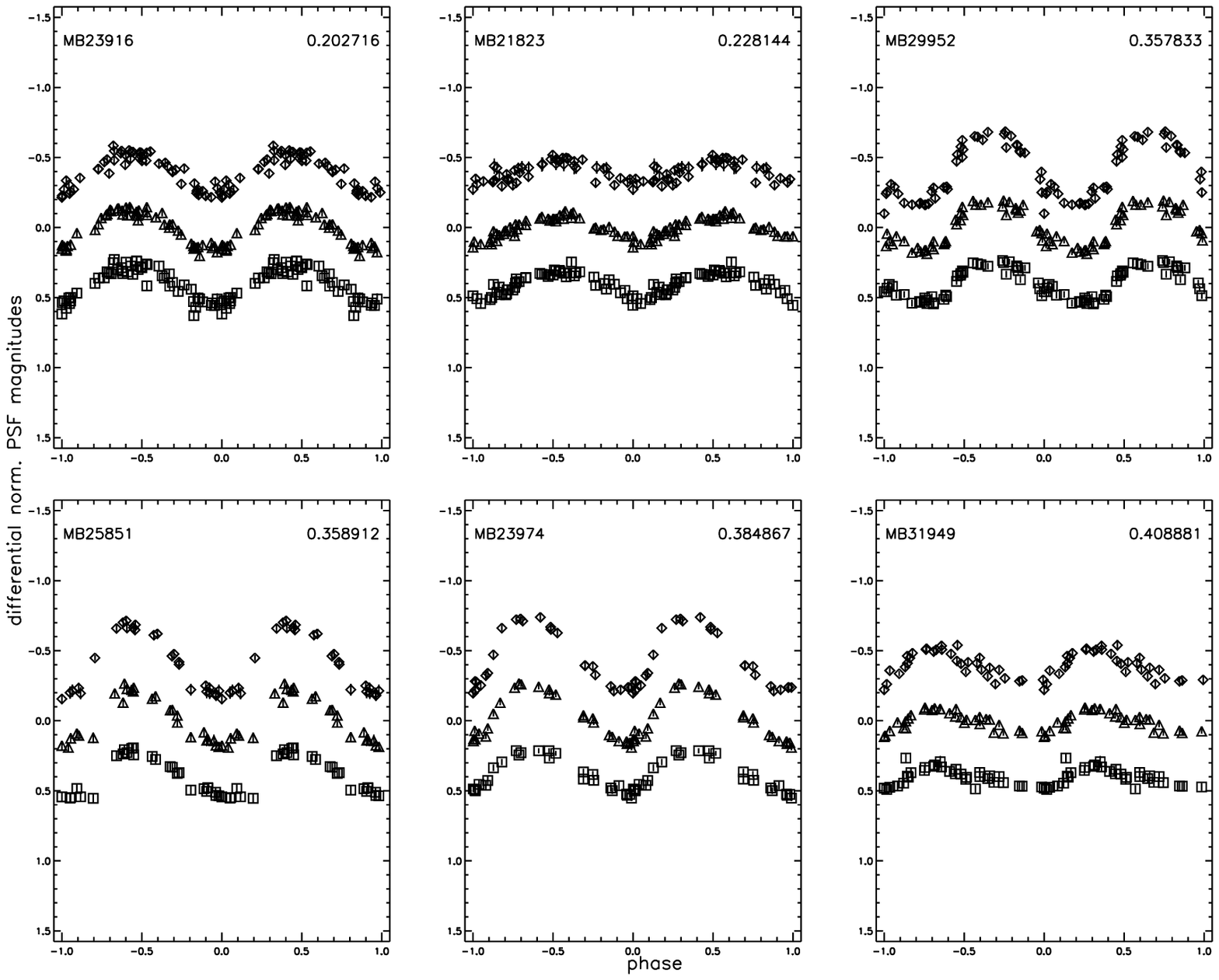}
    \caption{Phased $gri$ differential magnitude light-curves for a sample of 6 RRc RR Lyrae candidates
    identified by our pipeline, in order of increasing period. The internal designation for an object and its
    period in days is shown in each panel. Diamond symbols are $g$ light-curve points, triangle symbols are
    $r$ light-curve points, and box symbols are $i$ light-curve points. The light-curves are phased around the
    time of minimum light and repeated twice from phase -1 to +1 to clearly show variability.}
    \label{fig_rr_c}
  \end{figure}

  There are 55 RR Lyrae variables identified during this process. Of these, 36 are of the subtype RRab, and 19
  are of the subtype RRc. Table \ref{all_rr_table} gives a listing of these objects, their median SDSS
  light-curve magnitudes, and their periods. The periods for the RRab variables range between $\sim$ 0.476 and
  $\sim$ 0.709 days, with a median period of $\sim$ 0.608 days. These objects show the expected trend of
  decreasing amplitude of variation and more symmetric light-curves with increasing period. This trend is
  apparent in Figure \ref{fig_rr_ab}, which presents $gri$ differential magnitude light-curves of a sample of
  RRab variables identified in this dataset. The 19 identified RRc variables have periods ranging from $\sim$
  0.201 to $\sim$ 0.409 days, with a median period of $\sim$ 0.301 days. Figure \ref{fig_rr_c} presents
  example $gri$ differential magnitude light-curves for these objects.

  Although the selection of RR Lyrae by light-curve shape and period is more robust than selection by color
  alone, the completeness of our sample of such objects suffers due to poor phase coverage. The 55 objects
  selected for our sample are a meagre fraction of the 125 RR Lyrae probable variable candidates selected by
  color, and an even smaller fraction of the 1,664 RR Lyrae selected by color alone without regard to their
  variability. We quantify the completeness and efficiency of our selection process by carrying out
  simulations similar to those discussed in Section \ref{sec_efficiency}.

  We generate 2,000 RRab and 2,000 RRc synthetic objects using the magnitude, magnitude error, number of
  observations, and observation date distributions present in our dataset. Colors are assigned to these
  objects using the RR Lyrae candidate color cuts described in Section \ref{sec_class_point} and
  \citet{2007AJ....134..973I}. Periods are assigned to the RRab objects from a uniform distribution of periods
  between 0.5 and 0.8 days, and to the RRc objects from a uniform distribution of periods between 0.2 and 0.5
  days. Similarly, we assign variability amplitudes from uniform distributions of amplitudes between 0.4 and
  1.0 mag for RRab, and 0.1 to 0.5 mag for RRc objects respectively. Light-curves for these objects are
  generated and they are then inserted into separate catalogs of 48,000 nonvariable synthetic objects each,
  and then run through the ensemble photometry and variable extraction routines. We test the recovery of these
  artificial variables in two steps: (1) extracting them as \emph{tentative} and subsequently as
  \emph{probable} variables by using our variable identification methods, and (2) recovering their assigned
  periods by using our period finding algorithms.
  
\begin{figure}
    \plotone{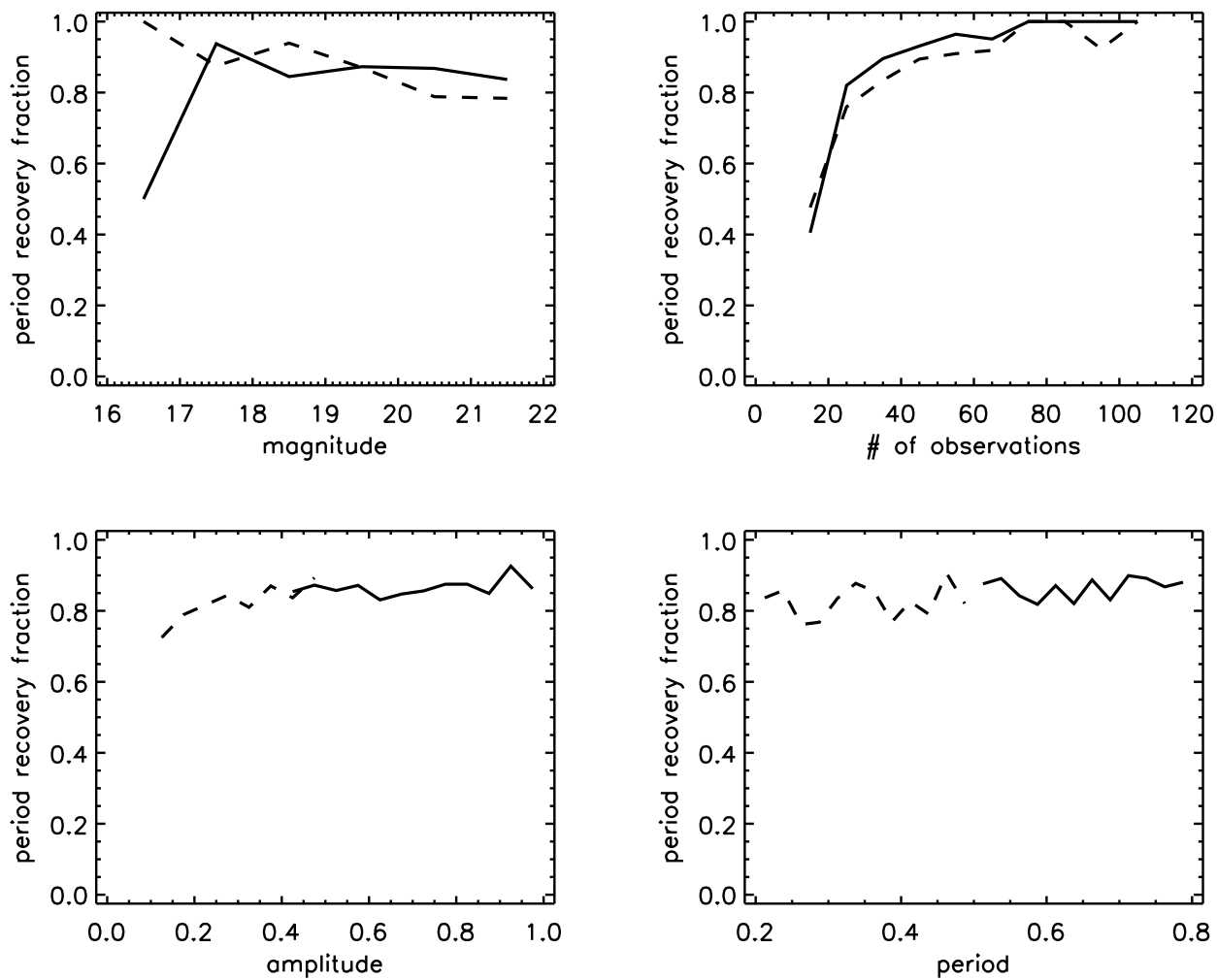}
    \caption{Top left: Period recovery fraction as a function of the median $g$ magnitude of synthetic RRab
      and RRc probable variables. Top right: Period recovery fraction as a function of the number of
      observations. Bottom left: Period recovery fraction as a function of the amplitude. Bottom right: Period
      recovery fraction as a function of the input period. The solid lines are for the RRab variables, and the
      dashed lines are for the RRc variables.}
    \label{fig_rr_binplots}
  \end{figure}

  Of the 2,000 synthetic RRab variables inserted into the catalog, 1,533 survive the ensemble process and
  subsequent conditions on color errors and magnitude limit. Of these objects, 1,517 ($\sim 98.9\%$) are
  successfully identified first as \emph{tentative variables} and then as \emph{probable variables}. Overall,
  1,312 of these variables are recovered with the correct periods, making for an average recovery fraction of
  $\sim 86.4\%$. Figure \ref{fig_rr_binplots} shows the trend of the period recovery fraction with $g$
  magnitude, number of observations, amplitude, and period for synthetic RRab variables (solid lines) as well
  as RRc variables (dashed lines). Period recovery quickly becomes more successful as the number of
  observations for a given object are increased, indicating the important role of good phase coverage. The
  recovery rate appears to be relatively insensitive to object magnitude, discounting the objects in the
  brightest magnitude bin, which are much fewer in number compared to those in the other magnitude bins. The
  period recovery fraction also appears to be insensitive to the variability amplitude; RRab variables have
  amplitudes that are large compared to our survey's sensitivity to variability. The weak trend with period is
  reflective of the small period range being explored.

  In contrast to the high variable recovery rate of synthetic RRab variables, the synthetic RRc variables
  suffer due to their small variability amplitudes. Of the 2,000 such objects inserted into the catalog, 1,543
  survive to the variability analysis stage after cuts on color error and magnitude limits. Only 1,132 of
  these objects are picked up as \emph{tentative variables}, and subsequently as \emph{probable variables},
  resulting in a relatively poor variable recovery rate of $73.4\%$. The variable recovery rate shows a
  significant decreasing trend with fainter magnitude after peaking in the magnitude bin $19.0 < g \le 20.0$,
  similar to the results of general variable recovery simulations presented in Section
  \ref{sec_efficiency}. The period recovery rate for synthetic RRc variables is similar to that for synthetic
  RRab variables: 934 of 1,132 of these ($\sim 82.5\%$) are recovered with the correct periods. Figure
  \ref{fig_rr_binplots} (dashed lines) shows how this rate is related to various input parameters. Once again,
  the most important factor is the number of observations, and by extension, the phase coverage for our
  variable objects. 

\begin{figure}
    \plotone{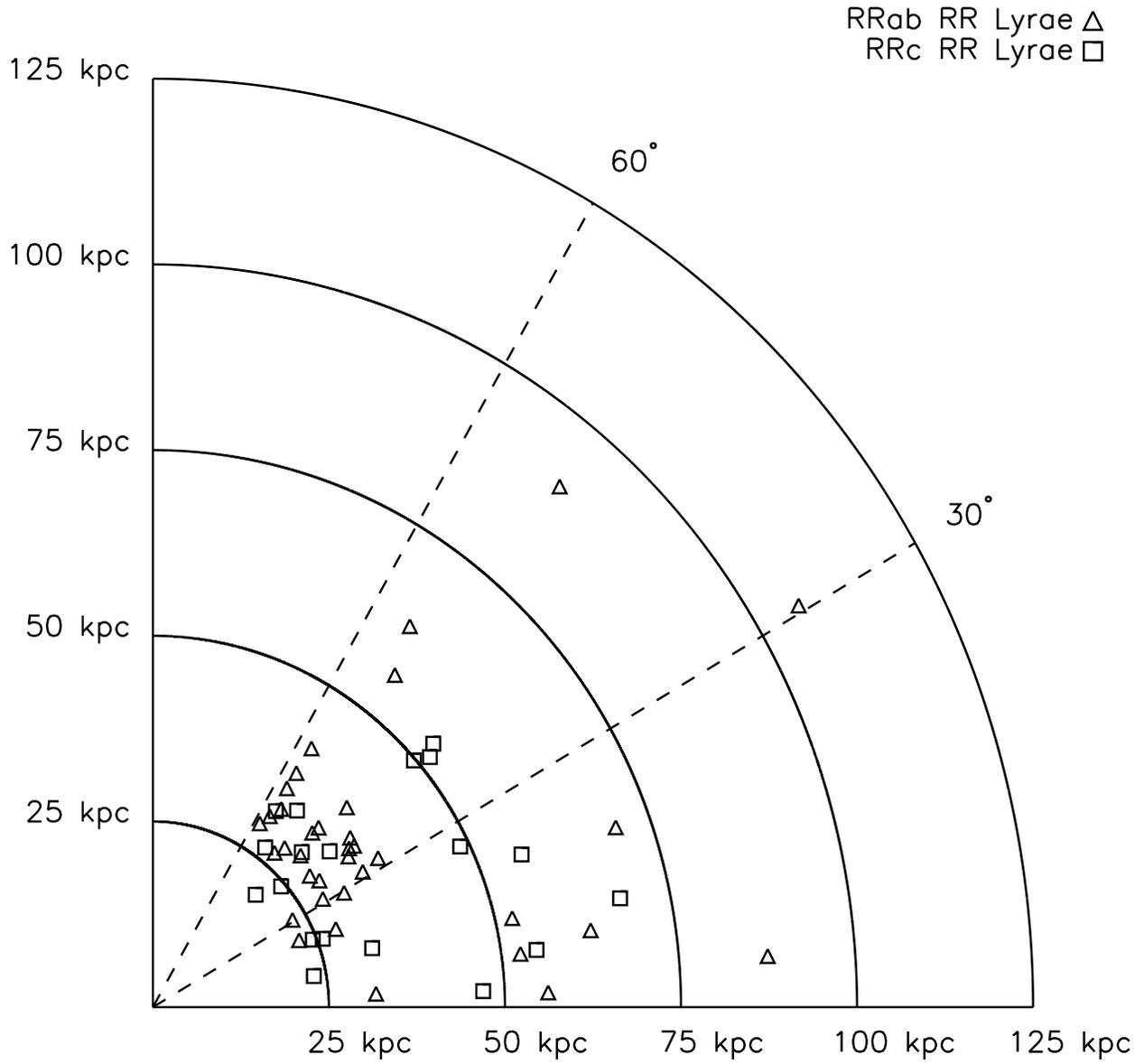}
    \caption{Distance vs. right ascension distribution of the 55 RR Lyrae in our dataset, which is restricted
    to a right ascension range of 0$\arcdeg$ to 60$\arcdeg$, and a declination range of -1.27$\arcdeg$ to
    +1.27$\arcdeg$.}
    \label{fig_rr_distance}
  \end{figure}

  Given the results of the variable and period recovery simulations above, we estimate a completeness of no
  better than $65\%$ for our RR Lyrae sample. Although this is low, the objects that are discovered are very
  likely to be real RRab and RRc variables due to successful classification by light-curve shape, period, and
  Fourier decomposition. We can, therefore, attempt to trace halo substructure using these objects for
  illustrative purposes. Assuming an absolute magnitude $M_V = 0.7$ for RR Lyrae stars, we calculate the
  distances to the 55 such objects in our catalog. Figure \ref{fig_rr_distance} shows a plot of the
  distribution of these RR Lyrae (36 RRab, and 19 RRc) as a function of the distance and the right
  ascension. The large clump near 25 kpc and ranging from 30$\arcdeg$ to 40$\arcdeg$ in right ascension is
  associated with part of the Sagittarius dwarf stream (S167-54-21.5 in \citealt{2002ApJ...569..245N}). Other
  clumps can be seen in the spatial distribution plot, and these may be associated with halo substructure as
  well. A detailed examination of RR Lyrae in Stripe 82 and how they relate to Milky Way substructure at large
  distances is, however, beyond the scope of this work, largely due to the small number of RR Lyrae in our
  sample. Excellent treatments of these topics may be instead be found in \citet{2008PhDT........10D} and
  \citet{2009MNRAS.398.1757W}.

\begin{figure}
    \plotone{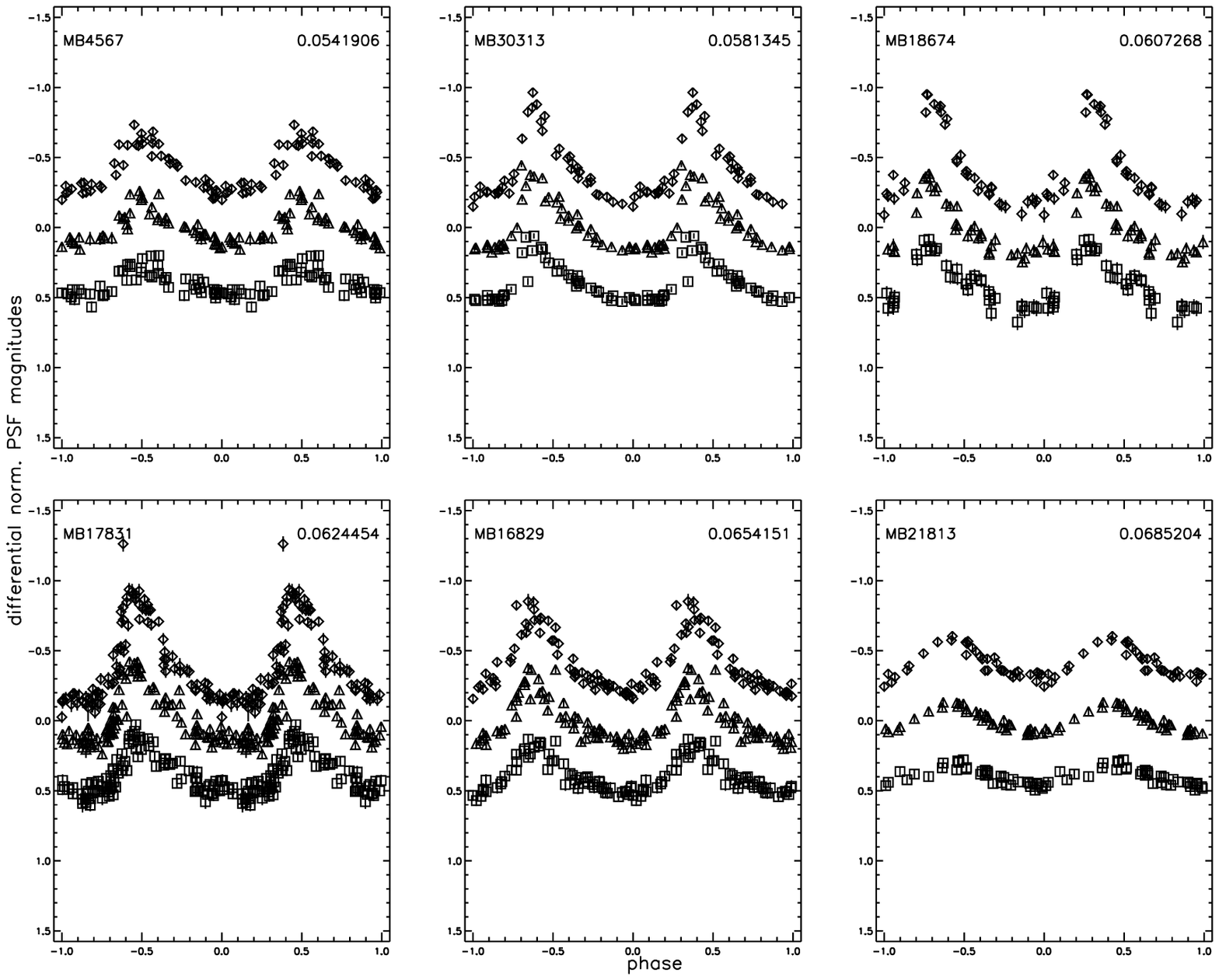}
    \caption{Phased $gri$ differential magnitude light-curves for a sample of 6 Delta Scuti variable
    candidates identified by our pipeline, in order of increasing period. The internal designation for an
    object and its period in days is shown in each panel. Diamond symbols are $g$ light-curve points, triangle
    symbols are $r$ light-curve points, and box symbols are $i$ light-curve points. The light-curves are
    phased around the time of minimum light and repeated twice from phase -1 to +1 to clearly show
    variability.}
    \label{fig_dsc}
  \end{figure}

  The final class of sinusoidal variables we discuss are the high amplitude Delta Scuti (HADS)
  variables. These have short periods ($< 0.3$ days) and amplitudes (up to 0.5 mag) greater than those of the
  usual Delta Scuti stars. We find 16 candidates for such objects in our dataset. Table \ref{all_dsc_table}
  gives a listing of these, along with median light-curve magnitudes and periods. Figure \ref{fig_dsc} shows
  example light-curves for 6 of these objects. The median period for these objects is $\sim$ 0.063 days, with
  a minimum period of $\sim$ 0.050 days, and a maximum period of $\sim$ 0.093 days. Longer period HADS
  variables can potentially be confused with short period RRc RR Lyrae in both period and color space (see
  Figure \ref{fig_periodic_color}), however, Fourier decomposition provides a robust mechanism of
  distinguishing between the two kinds of variables, as seen in Figure \ref{fig_fourier_r21}.

  \subsection{Variable Quasars}\label{sec_sdss_qsos}

%% maybe get rid of the Stetson plot and argue for these new QSO candidate variables based on color alone?

  The vast majority of quasars are expected to be variable on both short and long timescales (see
  \citealt{2002MNRAS.329...76H}, \citealt{2003AJ....126.1217D}, and references therein). The short term
  variability of these objects is likely to be associated with eruptive events and manifests as
  large-amplitude variations in light-curves (for example, BL Lac type objects). In contrast, the long term
  variability of quasars is dominated by small-amplitude variability, often resulting in increasing or
  decreasing trends in brightness on multi-year timescales. It is possible to use this intrinsic variability
  to distinguish between objects on the stellar locus and quasars (\citealt{2007AJ....134.2236S}, and
  references therein).

\begin{figure}
    \plotone{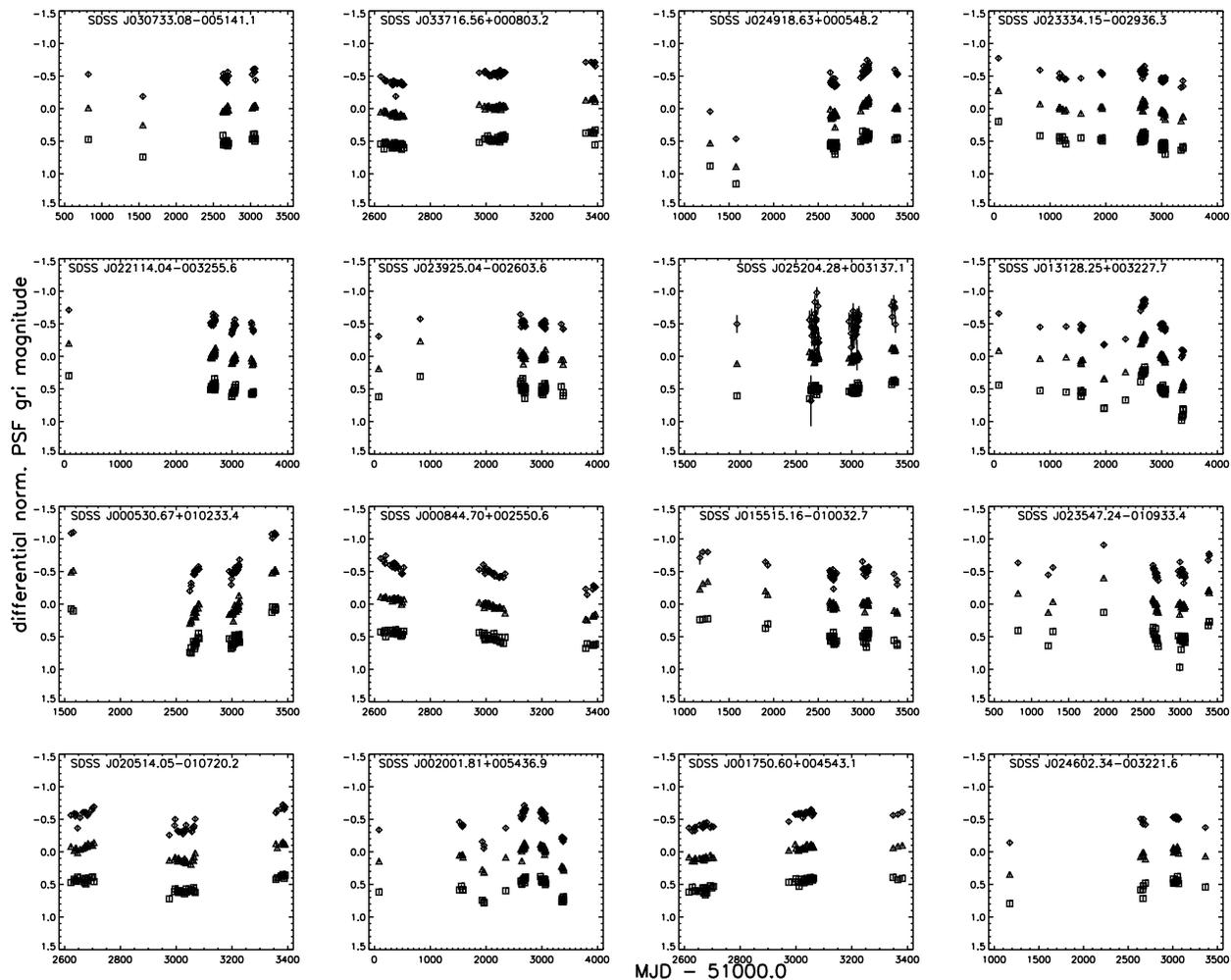}
    \caption{Differential magnitude $gri$ light-curves for a random sample of 16 objects from the 2,704
    quasars from the SDSS Quasar Catalog \citep{2007AJ....134..102S} tagged as variables by our
    pipeline. Diamond symbols are $g$ light-curve points, triangle symbols are $r$ light-curve points, and box
    symbols are $i$ light-curve points. The long term variability of these quasars is apparent over multi-year
    baselines.}
    \label{fig_qsos}
  \end{figure}

  Our light-curve catalog contains 4,196 quasars matched to the SDSS Quasar Catalog
  \citep{2007AJ....134..102S}. These objects mostly fall within regions B and D of the $g-r/u-g$ color-color
  diagram (see Figure \ref{fig_gr_ug}, left panel). The objects in the quasar catalog were selected using SDSS
  photometry and spectra and have measured redshifts available. We identify 2,704 objects among these matched
  quasars as probable variables tagged by our pipeline (see Figure \ref{fig_qsos} for example light-curves),
  resulting in a variable fraction of $\sim$ 0.64. Furthermore, these variable quasars make up a sizeable
  fraction, $\sim$ 0.41, of all identified candidate variable sources.

  New QSO candidates in our sample may be identified by taking advantage of their intrinsic variability and
  non-stellar colors. As noted in Section \ref{sec_var_gen_prop}, there are 8,463 objects tagged as `unknown'
  in our catalog due to the lack of any corresponding objects in any other catalogs, as well as no
  classification posssible by SDSS color alone. Of these objects, 1,102 are tagged as probable variables. A
  large fraction of these probable variable `unknown' objects have non-stellar colors and may be found in
  region A of Figure \ref{fig_gr_ug} (left panel). We identify new quasar candidates by requiring that they be
  tagged as probable variables and satisfy either one of two color cuts: low-$z$ quasars using $-0.3 < u-g <
  0.7$ and $-0.3 < g-r < 0.5$, and high-$z$ quasars using $u-g > 1.4$ and $u-g > 1.6(g-r) + 1.34$. These color
  and variability cuts result in the identification of 2,403 QSO candidates.

\begin{figure}
    \plottwo{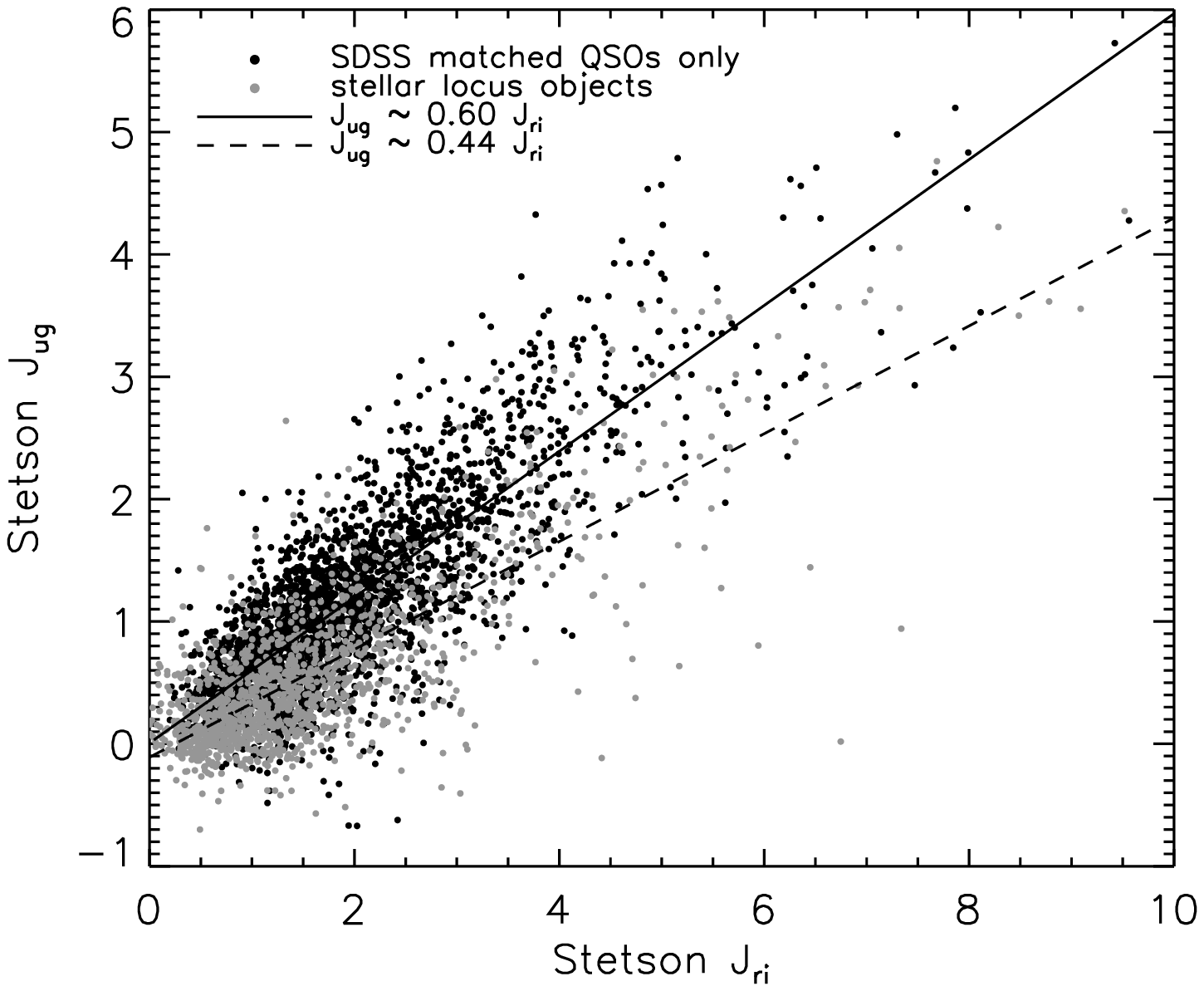}{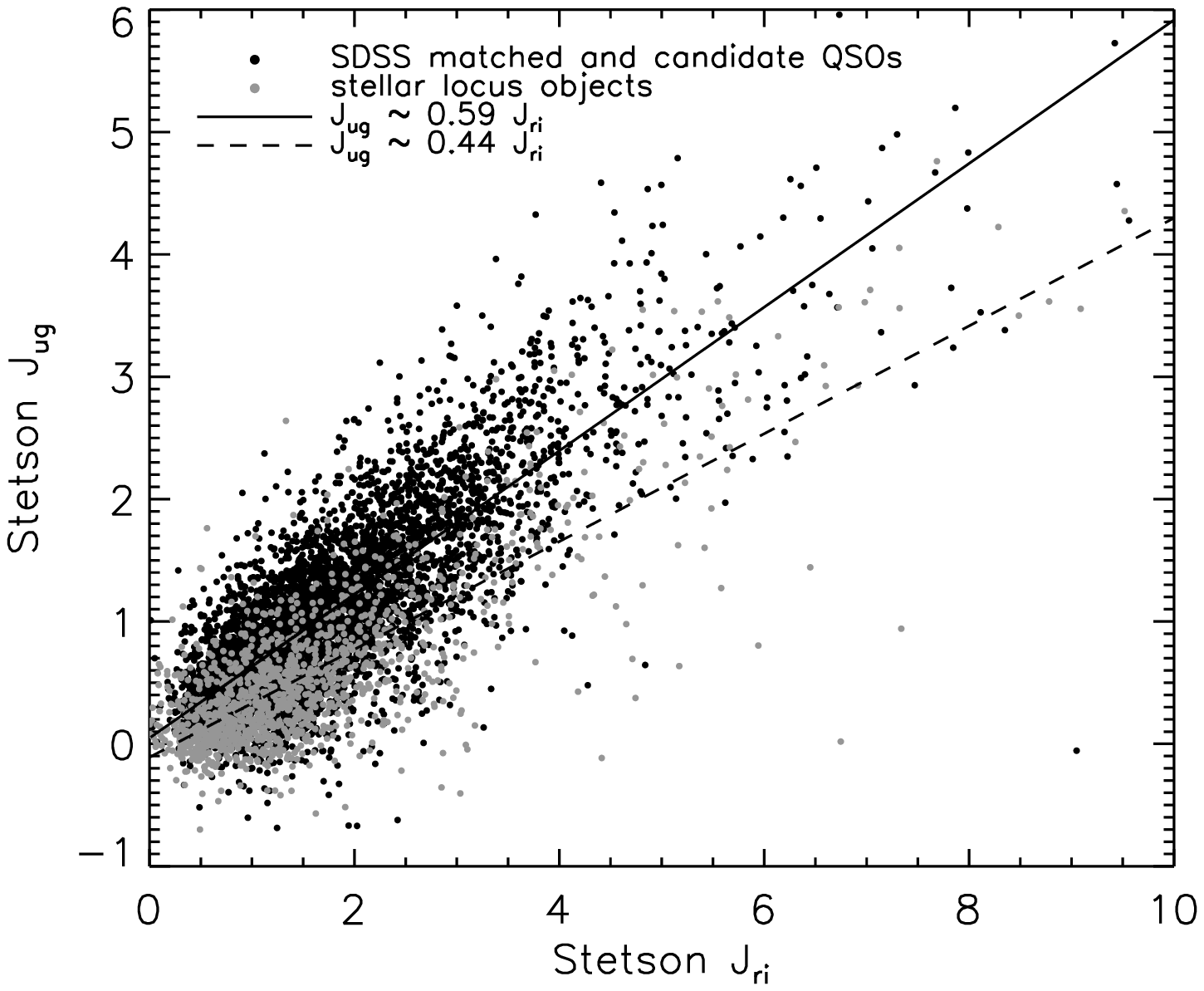}
    \caption{Left: Stetson variability index $J_{ri}$ against $J_{ug}$ for 2,704 variable QSOs (black circles)
    matched to the SDSS Quasar Catalog \citep{2007AJ....134..102S} and 1,413 variable stellar locus objects
    (grey circles). The 2,403 new variable QSO candidates identified in this work are not included in this
    plot. Right: Stetson variability index $J_{gr}$ against $J_{ug}$ for all 6,520 probable variables tagged
    by our pipeline. The black circles are 2,704 variable QSOs matched to the SDSS Quasar Catalog plus 2,403
    QSO candidates selected by color and variability. The grey circles are the remaining 1,413 variable
    stellar locus objects.}
    \label{fig_qso_stetson}
  \end{figure}

  Figure \ref{fig_qso_stetson} (left panel) shows the relation between Stetson indices $J_{ug}$ and $J_{ri}$
  for variable QSOs matched to the SDSS Quasar Catalog and stellar locus objects. Quasars appear to be more
  strongly variable at shorter wavelengths \citep{2004ApJ...601..692V}, and this tendency shows up in slopes
  of the $J_{ug}$ vs $J_{ri}$ trends plotted in the figure. Variable stellar locus objects show most of their
  variability at longer wavelengths, and thus have larger relative $J_{gr}$, and $J_{ri}$ values. The addition
  of 2,403 variable QSO candidates identified by our color and variability cuts to the plot (Figure
  \ref{fig_qso_stetson}, right panel) does not appreciably change the slopes of the two distributions,
  indicating that these objects largely follow the trend for matched QSOs from the SDSS Quasar Catalog. 

  Despite these candidate objects matching the colors of, as well having similar variability properties to the
  already known QSOs, confirmation of their quasar nature would require an extensive spectroscopic
  campaign. Lists of all objects matched to the SDSS Quasar Catalog, as well as those objects identified as
  candidate QSOs here are both available from the website mentioned in Section \ref{sec_var_gen_prop} above.

  \subsection{Nonvariable Objects}\label{sec_nonvariables}

  There are 214,982 objects in the final light-curve catalog (the $r < 22.0$ sample) that fail to meet the
  selection criteria for \emph{tentative variables}. Not all of these will be actual nonvariables. We can,
  however, use the tools developed here to define a sample of point sources that show no variability at the
  limits of our sensitivity. \emph{Probable nonvariables} are thus selected if they: (1) are not tagged as
  tentative variables by our pipeline, (2) do not match to any objects in the SDSS-I variable star catalog
  \citep{2007AJ....134.2236S}, (3) have a match in the SDSS standard star catalog \citep{2007AJ....134..973I},
  (4) do not match to any objects in the SDSS quasar catalog \citep{2007AJ....134..102S}, and finally, (5)
  have Stetson variability indices $J_{ug}$, $J_{gr}$, $J_{ri}$, and $J_{iz}$ all less than 0.05. These
  conditions select 19,704 objects. A further refinement can be made by demanding that these objects have at
  least 50 detections each, ensuring that the Stetson index selection for nonvariability is valid over at
  least three years of observations. With this final criterion in place, the probable nonvariable sample
  consists of 11,328 point sources. 

\begin{figure}
    \plotone{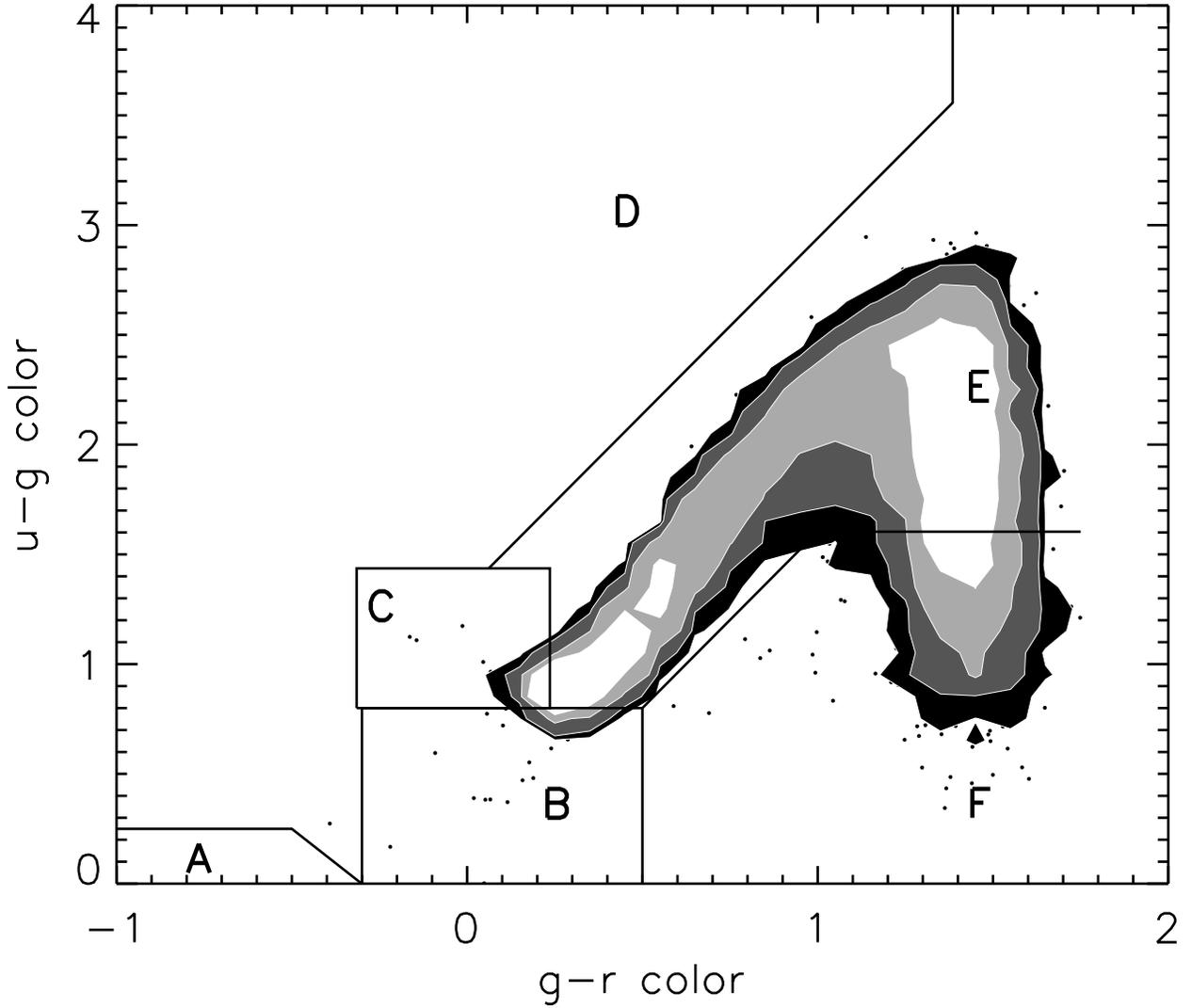}
    \caption{$g-r/u-g$ color-color diagram for 11,328 \emph{probable nonvariable} point sources identified by
    our pipeline. Regions A--F are the same as in Figure \ref{fig_gr_ug}. The density contours are in 0.5 dex
    increments from 0.5 to 2.0.}
    \label{fig_nonvar}
  \end{figure}

  Figure \ref{fig_nonvar} shows a $g-r/u-g$ color-color diagram of these objects. Nearly all of them are found
  on the stellar locus and are well-distributed in color from blue A stars to red M stars. These objects form
  a subset of the sources found in the SDSS standard star catalog and have been selected as nonvariables using
  a much longer timeseries baseline. Additional high cadence photometric monitoring would be required to rule
  out variability below the level of what we can detect with our methods ($ \sigma \le 0.05$ mag) before these
  are actually deemed suitable for use in a canonical standard star catalog. A list of these stars is
  available from this paper's accompanying website.

%%%%%%%%%%%%%%%%%%%%%%%%%%%%%%%%%
%% FUTURE WORK AND CONCLUSIONS %%
%%%%%%%%%%%%%%%%%%%%%%%%%%%%%%%%%

  \section{Conclusions}\label{sec_conclusion}

  We have constructed a light-curve catalog of 221,842 point sources in the RA 0 to 4 h half of Stripe 82, and
  identified 6,520 candidate variables. Of these, 2,704 turned out to be already identified quasars, while
  another 2,403 were classified as QSO candidates due to their colors and variability properties. We found 101
  periodic variables in this dataset, including 30 candidate eclipsing binary systems, 55 RR Lyrae and 16 high
  amplitude Delta Scuti candidates. We also identified a sample of 11,328 point sources that do not appear to
  be variable, based on observations over a long time baseline and rejection from our variable extraction
  algorithms.

  The use of inhomogeneous ensemble differential photometry and the Stetson variability index was crucial in
  identifying possible variables among the objects in the dataset, while removing many sources that appeared
  to be falsely variable. The poor phase coverage of our light-curves presented difficulties in extracting
  periodic variables from the candidate variable sources. Binless phase dispersion minimization methods, such
  as the Dworetsky and Stetson string length algorithms worked well on our sparse and unevenly sampled data,
  and had the added advantage of being unbiased with respect to the kind of variability being probed.

  We have made public our Stripe 82 variable object light-curve catalog, along with the software
  implementation of our ensemble photometry pipeline. In an upcoming paper, we will address variability in the
  remaining half of Stripe 82 (RA 20 to 0 h), and present our final catalogs for periodic and other types of
  variables present in this dataset. Unfortunately, the relative faintness of many interesting periodic
  variables, especially the binary candidates, presents difficulties in the confirmation of their nature and
  further study of their properties. We believe, however, that this dataset will be an important resource for
  variability studies of a large and diverse array of objects, serving as a prototype in advance of future
  large synoptic surveys such as Pan-STARRS and LSST.

%%%%%%%%%%%%%%%%%%%%%
%% ACKNOWLEDGMENTS %%
%%%%%%%%%%%%%%%%%%%%%

  \acknowledgments This research was supported by NASA grant NAG5-7697. We gratefully acknowledge the
  assistance of Brian Yanny (FNAL) and Ani Thakar (JHU) in accessing and downloading the SDSS-II Supernova
  Survey data. W. A. B. would like to thank Arti Garg (LLNL), Mark Huber (JHU), and Suvi Gezari (JHU) for
  useful discussions. We thank the anonymous referee for their careful reading and helpful comments and
  suggestions, which greatly improved the manuscript. This research has made extensive use of the IDL
  Astronomy Library and the University of Michigan/Chicago/NYU SDSS IDL Library. Finally, we acknowledge the
  use of data from the Sloan Digital Sky Survey (SDSS) and SDSS-II.

  Funding for the SDSS and SDSS-II has been provided by the Alfred P. Sloan Foundation, the Participating
  Institutions, the National Science Foundation, the U.S. Department of Energy, the National Aeronautics and
  Space Administration, the Japanese Monbukagakusho, the Max Planck Society, and the Higher Education Funding
  Council for England. The SDSS Web Site is http://www.sdss.org/

  The SDSS is managed by the Astrophysical Research Consortium for the Participating Institutions. The
  Participating Institutions are the American Museum of Natural History, Astrophysical Institute Potsdam,
  University of Basel, University of Cambridge, Case Western Reserve University, University of Chicago, Drexel
  University, Fermilab, the Institute for Advanced Study, the Japan Participation Group, Johns Hopkins
  University, the Joint Institute for Nuclear Astrophysics, the Kavli Institute for Particle Astrophysics and
  Cosmology, the Korean Scientist Group, the Chinese Academy of Sciences (LAMOST), Los Alamos National
  Laboratory, the Max-Planck-Institute for Astronomy (MPIA), the Max-Planck-Institute for Astrophysics (MPA),
  New Mexico State University, Ohio State University, University of Pittsburgh, University of Portsmouth,
  Princeton University, the United States Naval Observatory, and the University of Washington.

%%%%%%%%%%%%%%%%
%% REFERENCES %%
%%%%%%%%%%%%%%%%

%%%%%%%%%%%%
%% TABLES %%
%%%%%%%%%%%%

  \begin{deluxetable}{ll}
    \tablecaption{Data fields extracted from Stripe 82 object catalogs}
    \tablewidth{0pt}
    \tabletypesize{\footnotesize}
    \tablehead{\colhead{tsObj FITS/CAS Database Field} & \colhead{Description}}
    \startdata
    RUN & The SDSS run\\
    RERUN & Version of the SDSS FRAMES pipeline (40 or 41 in our dataset)\\
    CAMCOL & CCD camera column (ranges from 1 to 6) \\
    FIELD & Field number in SDSS run ($13\arcmin\times10\arcmin$ per field) \\
    ID & Non-unique ID number of a detection in a field \\
    RA & Right ascension J2000 ($\arcdeg$) \\
    DEC & Declination J2000 ($\arcdeg$) \\
    OBJC\_FLAGS/FLAGS & Quality flags associated with a detection \\
    TYPE\tablenotemark{a} & Detection type (3 = galaxy, 6 = star) \\
    PARENT/PARENTID\tablenotemark{a} & ID of parent object if current detection is a deblended child \\
    NCHILD\tablenotemark{a} & Number of deblended children if current detection is a blend \\
    STATUS\tablenotemark{a} & Status of the detection (see text for details) \\
    PSFCOUNTS/PSFMAG & PSF fitted magnitude in \emph{ugriz} bands \\
    PSFCOUNTSERR/PSFMAG\_ERR & Uncertainty in PSF fitted magnitude in \emph{ugriz} bands \\
    FIBERCOUNTS/FIBERMAG & Magnitude from flux in 3$\arcsec$ fiber radius in \emph{ugriz} bands \\
    FIBERCOUNTSERR/FIBERMAG\_ERR & Uncertainty in fiber magnitude in \emph{ugriz} bands\\
    MJD & MJD of detection of object in \emph{ugriz} bands\\
    REDDENING/EXTINCTION & Extinction in magnitudes for \emph{ugriz} bands
    \enddata
    \tablenotetext{a}{These fields are used to select suitable objects for extraction and are not saved in the
      output catalog}
    \label{param_table}
  \end{deluxetable}
  \clearpage
  
  \begin{deluxetable}{ll}
    \tablecaption{Color selection algorithms for classifying point sources}
    \tablewidth{0pt}
    \tabletypesize{\footnotesize}
    \tablehead{\colhead{Object Type} & \colhead{Color Selection}}
    \startdata
    RR Lyrae candidate\tablenotemark{a} & $0.98 < (u-g) < 1.30$ and $-0.05 < RR1 < 0.35$ and $0.06 < RR2 < 0.55$\\
                       & and $-0.15 < (r-i) < 0.22$ and $-0.21 < (i-z) < 0.25$\\
    Main-sequence + white-dwarf\tablenotemark{b}\tablenotemark{d} & $(u-g) < 2.25$ and 
    $-0.2 < (g-r) < 1.2$ and 
                                                   $ 0.5 < (r-i) < 2.0$ and\\    
                                & $-19.78(r-i)+11.13 < (g-r) < 0.95(r-i) + 0.5$\\
    sdO/sdB/white-dwarf\tablenotemark{b} & $-1.0 < (g-r) < -0.2$ and $-1.0< (u-g) < 0.7$ and \\
                & $u-g+2(g-r) < -0.1$\\
    low-metallicity\tablenotemark{b} & $-0.5 < (g-r) < 0.75$ and $ 0.6 < (u-g) < 3.0$ and $l > 0.135$\\
    AGB\tablenotemark{b}             & $2.5< (u-g) < 3.5$ and $0.9 < (g-r) < 1.3$ and $s < -0.06$\\
    A/blue horizontal branch\tablenotemark{b} & $-1.0 < (u-g) < -0.2$ and $-0.5 < (g-r) < 0.2$\\
    F/G\tablenotemark{b}                      & $0.20 < (g-r) < 0.48$\\
    F-turnoff/sdF\tablenotemark{b}            & $-0.7 < P1 < -0.25$ and $0.4 < (u-g) < 1.4$ and $-0.5 < (g-r) < 0.7$\\
    G\tablenotemark{b}                        & $0.48 < (g-r) < 0.55$\\
    K-dwarf\tablenotemark{b}                  & $0.55 < (g-r) < 0.75$\\
    K-giant\tablenotemark{b}  & $0.7 <(u-g) < 4.0$ and $0.35 <(g-r) < 0.7$ and \\
             & $0.15 <(r-i) < 0.6$ and $l > 0.07$\\
    sdM\tablenotemark{b} & $(g-r) > 1.6$ and $0.95 < (r-i) < 1.3$\\
    M-dwarf\tablenotemark{c} & $0.666 < 0.875(r-i) + 0.484[(i-z) + 0.00438] < 3.4559$\\
    brown-dwarf\tablenotemark{d} & $z < 19.5$ and $u > 21.0$ and $g > 22.0$ and $r > 21.0$ and $(i-z) > 1.7$
    \enddata
    \tablenotetext{a}{Using colors from \citet{2007AJ....134.2236S}.}
    \tablenotetext{b}{Using colors from \citet{2009AJ....137.4377Y}.}
    \tablenotetext{c}{Color locus computed from mean M-dwarf colors in \citet{2008AJ....135..785W}.}
    \tablenotetext{d}{Using magnitudes that are not dereddened.}
    \tablecomments{The various color indices used above are defined below:\\
      $P1$-color = $0.91(u-g)+0.415(g-r)-1.28$\\
      $l$-color = $-0.436u + 1.129g - 0.119r - 0.574i + 0.1984$\\
      $s$-color = $-0.249u + 0.795g - 0.555r + 0.124$\\
      $RR1$-color = $(u-g) + 0.67 (g-r) - 1.07$\\
      $RR2$-color = $0.45(u-g) - (g-r) - 0.12$}
    \label{color_table}
  \end{deluxetable}
  \clearpage

  \begin{deluxetable}{lrrr}
    \tablecaption{Distribution of point sources by type and variability ($r < 22.0$ sample)}
    \tablewidth{0pt}
    \tabletypesize{\small}
    \tablehead{\colhead{Object type} & \colhead{Probable variables} & \colhead{Total objects} & 
      \colhead{Variable fraction}}
    \startdata
    sdO/sdB/WD	&	4	&	235	&	0.017\\
    A/BHB	&	180	&	1841  &	0.097\\
    F-turnoff/sdF	&	507	&	26294	&	0.019\\
    F/G	&	1333	&	42901	&	0.031\\
    G dwarf	&	130	&	8346	&	0.016\\
    K giant	&	141	&	14790	&	0.009\\
    AGB	&	14	&	6001	&	0.002\\
    Low-metallicity	&	348	&	13522	&	0.026\\
    K dwarf	&	193	&	16141	&	0.012\\
    M subdwarf	&	4	&	332	&	0.012\\
    M dwarf	&	253	&	107483	&	0.002\\
    brown dwarf	&	0	&	0	&	0.000\\
    MS+WD	&	82	&	602	&	0.136\\
    unknown\tablenotemark{a}	&	1102	&	8463	&	0.130\\
    \hline
    RR Lyrae candidate\tablenotemark{b}	&	125	&	1664	&	0.075\\
    SDSS QSO\tablenotemark{c}	&	2704	&	4196	&	0.644\\
    SDSS white dwarf\tablenotemark{d}	&	0	&	419	&	0.000\\
    SDSS hot subdwarf\tablenotemark{d}	&	0	&	15	&	0.000\\
    SDSS-I variable\tablenotemark{e}	&	2766	&	3972	&	0.696\\
    SDSS standard\tablenotemark{f}	&	562	&	165591	&	0.003\\
    \hline
    All point sources  &  6520  & 221842 & 0.029
    \enddata
    \tablenotetext{a}{No color classification possible and no matches to other catalogs.}
    \tablenotetext{b}{Using color-selection from \citet{2007AJ....134.2236S}.}
    \tablenotetext{c}{Cross-matched to objects in \citet{2007AJ....134..102S}.}
    \tablenotetext{d}{Cross-matched to objects in \citet{2006ApJS..167...40E}.}
    \tablenotetext{e}{Cross-matched to objects in \citet{2007AJ....134.2236S}.}
    \tablenotetext{f}{Cross-matched to objects in \citet{2007AJ....134..973I}.}
    \tablecomments{Objects may have multiple types assigned, based on their colors.}
    \label{type_table}
  \end{deluxetable}
  \clearpage

  \begin{deluxetable}{lc}
    \tablecaption{Periodic variable candidates classified by light-curve shape}
    \tablewidth{0pt}
    \tablehead{\colhead{Periodic variable type} & \colhead{Number}}
    \startdata
    RR Lyrae type RRab  & 36\\
    RR Lyrae type RRc   & 19\\
    Delta Scuti         & 16 \\
    Eclipsing Binary    & 30\\
    \hline
    Total               & 101
    \enddata
    \label{periodic_variable_table}
  \end{deluxetable}
  \clearpage

  \begin{deluxetable}{lcccc}
    \tablecaption{Recovery rates for artificial sinusoidal variables as a function of $r$ magnitude bin}
    \tablewidth{0pt}
    \tabletypesize{\footnotesize}
    \tablehead{\colhead{Magnitude Bin} & \colhead{Inserted Objects} & \colhead{Tentative Variables} & 
      \colhead{Probable Variables} & \colhead{Recovery Fraction}}
    \startdata
    $16.0 < r \le 17.0$ & \phn\phn9 & \phn\phn6 & \phn\phn6 & 0.67\\
    $17.0 < r \le 18.0$ & 103 & \phn88 & \phn88 & 0.85\\
    $18.0 < r \le 19.0$ & 180 & 158 & 157 & 0.87\\
    $19.0 < r \le 20.0$ & 303 & 246 & 245 & 0.81 \\
    $20.0 < r \le 21.0$ & 454 & 362 & 360 & 0.79 \\
    $21.0 < r \le 22.0$ & 344 & 241 & 235 & 0.68\\
    \enddata
    \label{fake_data_table}
  \end{deluxetable}
  \clearpage

\begin{deluxetable}{lcccccccc}
\tablecaption{Eclipsing and Ellipsoidal Binary Candidates}
\tablewidth{0pt}
\tabletypesize{\footnotesize}
\rotate
\tablehead{\colhead{Object} & \colhead{Obs.} & \colhead{$u$} & \colhead{$g$} & \colhead{$r$} & \colhead{$i$}
  & \colhead{$z$} & \colhead{Period [days]} & \colhead{Comments}}
\startdata
SDSS J000845.39+002744.2 & 54 & 20.51 & 19.36 & 18.82 & 18.60 & 18.53 & 0.34419 & low-metallicity\\
SDSS J002719.16+002400.6 & 47 & 21.04 & 20.08 & 19.74 & 19.61 & 19.58 & 1.31839 & F/G \\
SDSS J002851.08+000751.0 & 49 & 22.70 & 20.50 & 19.07 & 18.04 & 17.48 & 0.59098 & M2 \\
SDSS J003042.11+003420.2 & 51 & 23.02 & 20.57 & 19.08 & 17.93 & 17.32 & 0.45705 & M3\\
SDSS J011155.73-002633.0 & 34 & 22.37 & 20.57 & 19.94 & 19.71 & 19.55 & 0.22758 & K dwarf, semi-detached?\\
SDSS J011156.52-005221.4 & 35 & 19.25 & 18.29 & 18.02 & 17.95 & 17.99 & 0.23002 & F/G, contact? \\
SDSS J011302.57+004822.9 & 52 & 21.20 & 19.34 & 18.50 & 18.14 & 17.96 & 0.31660 & F/G?\\
SDSS J011405.02+001138.5 & 44 & 20.78 & 19.88 & 19.67 & 19.62 & 19.64 & 0.28550 & F/G, contact?\\
SDSS J013536.05-011058.7 & 44 & 20.99 & 20.02 & 19.95 & 19.99 & 20.06 & 0.39226 & A/BHB, contact?\\
SDSS J015429.30+005326.7\tablenotemark{a} & 43 & 21.91 & 19.50 & 18.17 & 17.25 & 16.77 & 2.63902 & M1\\
SDSS J015940.01+010328.4 & 58 & 23.46 & 21.47 & 20.03 & 19.36 & 18.97 & 0.33779 & M0\\
SDSS J020540.08-002227.6 & 57 & 20.39 & 19.42 & 19.16 & 19.10 & 19.12 & 0.79789 & F/G, deep primary\\
SDSS J020816.51+003510.0 & 48 & 19.69 & 18.61 & 18.77 & 18.93 & 19.07 & 1.61420 & A/BHB\\
SDSS J021121.55-003808.3 & 40 & 21.57 & 19.10 & 17.77 & 17.07 & 16.63 & 0.31210 & M0\\
SDSS J021624.34-001817.7 & 45 & 21.95 & 20.80 & 20.43 & 20.34 & 20.28 & 0.63678 & F/G\\
SDSS J022733.94+002615.2 & 49 & 20.04 & 18.83 & 18.96 & 19.09 & 19.21 & 0.70626 & A/BHB\\
SDSS J022858.99-004120.4 & 37 & 20.61 & 19.53 & 19.55 & 19.63 & 19.73 & 0.64725 & A/BHB\\
SDSS J023621.96+011359.1 & 57 & 19.97 & 19.00 & 18.74 & 18.67 & 18.66 & 0.35666 & F/G, contact?\\
SDSS J024109.55+004813.6 & 52 & 21.27 & 19.29 & 18.31 & 17.90 & 17.64 & 0.27645 & deep primary\\
SDSS J024255.78-001551.5 & 62 & 21.42 & 20.20 & 19.68 & 19.48 & 19.39 & 3.19764 & low-metallicity, long period\\
SDSS J025953.33-004400.3 & 53 & 21.46 & 20.50 & 19.36 & 18.26 & 17.57 & 0.14418 & M2, contact?\\
SDSS J030753.52+005013.0 & 115 & 21.84 & 19.74 & 18.75 & 18.34 & 18.12 & 0.35353 & K dwarf, detached\\
SDSS J030834.42+005835.2 & 68 & 22.93 & 20.61 & 19.49 & 19.01 & 18.78 & 0.27080 & contact?\\
SDSS J031002.47-000916.2 & 54 & 19.79 & 18.83 & 18.44 & 18.31 & 18.29 & 2.19058 & F/G\\
SDSS J031021.22+001453.9 & 48 & 19.77 & 18.62 & 18.11 & 17.92 & 17.82 & 0.26684 & low-metallicity, contact\\
SDSS J031823.88-010018.4\tablenotemark{b} & 34 & 22.80 & 20.56 & 19.10 & 17.55 & 16.73 & 0.40704 & M4\\
SDSS J032515.05-010239.7 & 48 & 22.22 & 19.86 & 18.39 & 17.42 & 16.87 & 0.39451 & M2, shallow eclipses\\
SDSS J032949.18-001240.8 & 62 & 23.96 & 21.16 & 19.78 & 19.20 & 18.85 & 0.39196 & M0\\
SDSS J034256.26-000058.0 & 51 & 20.29 & 19.40 & 19.13 & 19.06 & 19.09 & 0.32034 & F/G, semi-detached?\\
SDSS J034757.70-001423.5 & 53 & 20.51 & 19.68 & 19.32 & 19.24 & 19.25 & 0.27509 & F/G\\
SDSS J035138.50-003924.5 & 54 & 21.03 & 19.07 & 18.18 & 17.88 & 17.70 & 0.19892 & K/M?, contact\\
SDSS J035300.50+004836.0 & 40 & 22.39 & 20.62 & 19.31 & 18.17 & 17.61 & 0.14855 & M2, ellipsoidal?\\
\enddata
\tablenotetext{a}{This is the confirmed eclipsing binary found by \citet{2008MNRAS.386..416B}.}
\tablenotetext{b}{This is the confirmed eclipsing binary found by \citet{2008ApJ...684..635B}.}
\tablecomments{Objects are presented in order of increasing right ascension.}
\label{all_eb_table}
\end{deluxetable}
\clearpage

\begin{deluxetable}{lcccccccl}
\tablecaption{RRab and RRc RR Lyrae Variables}
\tablewidth{0pt}
\tabletypesize{\footnotesize}
\tablehead{\colhead{Object} & \colhead{Obs.} & \colhead{$u$} & \colhead{$g$} & \colhead{$r$} & 
  \colhead{$i$} & \colhead{$z$} & \colhead{Period [days]} & \colhead{Type}}
\startdata
SDSS J000803.72-010557.9 & 49 & 20.61 & 19.45 & 19.17 & 19.11 & 19.11 & 0.64004 & RRab\\
SDSS J001031.09+010132.4 & 43 & 20.21 & 19.06 & 18.87 & 18.81 & 18.76 & 0.33755 & RRc\\
SDSS J001301.12-003502.6 & 27 & 19.41 & 18.21 & 17.94 & 17.81 & 17.82 & 0.61226 & RRab\\
SDSS J001800.72+010424.9 & 33 & 21.40 & 20.41 & 20.26 & 20.21 & 20.19 & 0.53879 & RRab\\
SDSS J003119.26+005055.8 & 48 & 20.41 & 19.31 & 19.12 & 19.09 & 19.05 & 0.62216 & RRab\\
SDSS J003209.57-000445.1 & 26 & 20.52 & 19.40 & 19.42 & 19.48 & 19.62 & 0.20442 & RRc\\
SDSS J003748.95-000312.0 & 49 & 20.74 & 19.70 & 19.41 & 19.32 & 19.31 & 0.62441 & RRab\\
SDSS J004125.63+010859.3 & 14 & 18.65 & 17.53 & 17.39 & 17.41 & 17.48 & 0.25317 & RRc\\
SDSS J004951.22-000705.8 & 26 & 20.85 & 19.86 & 19.70 & 19.67 & 19.81 & 0.20296 & RRc\\
SDSS J005253.90-002141.8 & 27 & 20.37 & 19.30 & 19.03 & 18.95 & 18.92 & 0.58011 & RRab\\
SDSS J005712.29-000557.7 & 27 & 19.33 & 18.24 & 18.09 & 18.08 & 18.10 & 0.38487 & RRc\\
SDSS J012052.63-010411.8 & 46 & 21.10 & 19.93 & 19.68 & 19.62 & 19.56 & 0.58336 & RRab\\
SDSS J012333.48+010420.7 & 15 & 18.89 & 17.76 & 17.52 & 17.46 & 17.43 & 0.28045 & RRc\\
SDSS J012543.58+011032.7 & 43 & 20.51 & 19.45 & 19.42 & 19.45 & 19.57 & 0.20111 & RRc\\
SDSS J012714.38+004403.6 & 14 & 18.79 & 17.64 & 17.53 & 17.50 & 17.54 & 0.27462 & RRc\\
SDSS J012813.56+011357.4 & 28 & 19.12 & 17.94 & 17.70 & 17.63 & 17.59 & 0.58792 & RRab\\
SDSS J013352.87-004502.7 & 18 & 18.64 & 17.47 & 17.22 & 17.10 & 17.10 & 0.60606 & RRab\\
SDSS J014528.54+005133.7 & 47 & 20.17 & 19.14 & 19.11 & 19.17 & 19.24 & 0.22814 & RRc\\
SDSS J015813.75+010143.5 & 37 & 19.31 & 18.17 & 17.90 & 17.82 & 17.79 & 0.70862 & RRab\\
SDSS J020206.96-003534.2 & 34 & 21.97 & 20.84 & 20.61 & 20.53 & 20.39 & 0.59791 & RRab\\
SDSS J020245.96-000001.7 & 24 & 18.61 & 17.51 & 17.26 & 17.18 & 17.18 & 0.56383 & RRab\\
SDSS J020430.98+002005.4 & 19 & 19.12 & 17.95 & 17.89 & 17.83 & 17.82 & 0.63040 & RRab\\
SDSS J020548.43+000144.4 & 40 & 19.54 & 18.42 & 18.20 & 18.14 & 18.15 & 0.55212 & RRab\\
SDSS J020825.13-003444.4 & 35 & 19.68 & 18.58 & 18.35 & 18.30 & 18.26 & 0.51544 & RRab\\
SDSS J022309.05+005734.2 & 43 & 19.15 & 18.02 & 17.82 & 17.71 & 17.70 & 0.63008 & RRab\\
SDSS J022429.81+000725.8 & 34 & 19.44 & 18.38 & 18.20 & 18.16 & 18.16 & 0.47630 & RRab\\
SDSS J022857.91+005359.8 & 44 & 19.52 & 18.47 & 18.22 & 18.13 & 18.13 & 0.61506 & RRab\\
SDSS J023001.99-011146.9 & 39 & 19.55 & 18.43 & 18.16 & 18.11 & 18.11 & 0.64910 & RRab\\
SDSS J023346.13-001901.7 & 22 & 19.07 & 17.97 & 17.75 & 17.69 & 17.68 & 0.50251 & RRab\\
SDSS J023635.42+003005.0 & 44 & 19.65 & 18.49 & 18.24 & 18.17 & 18.16 & 0.53365 & RRab\\
SDSS J023932.33+002650.0 & 45 & 19.53 & 18.27 & 18.26 & 18.31 & 18.35 & 0.30838 & RRc\\
SDSS J024216.46+003438.6 & 52 & 20.30 & 19.27 & 19.28 & 19.35 & 19.41 & 0.20272 & RRc\\
SDSS J024647.88+010944.7 & 47 & 20.36 & 19.33 & 19.26 & 19.32 & 19.34 & 0.30117 & RRc\\
SDSS J024701.75-010814.8 & 20 & 18.72 & 17.64 & 17.48 & 17.49 & 17.51 & 0.21192 & RRc\\
SDSS J024727.34-000026.1 & 35 & 20.26 & 19.19 & 19.11 & 19.21 & 19.26 & 0.30126 & RRc\\
SDSS J025658.30-010918.3 & 43 & 19.09 & 18.03 & 17.76 & 17.70 & 17.69 & 0.51943 & RRab\\
SDSS J025717.50+004706.9 & 38 & 19.79 & 18.63 & 18.42 & 18.36 & 18.36 & 0.55448 & RRab\\
SDSS J025828.94-003635.1 & 27 & 19.13 & 18.06 & 17.91 & 17.94 & 17.95 & 0.35891 & RRc\\
SDSS J030304.27+003029.6 & 53 & 19.47 & 18.34 & 18.11 & 18.03 & 18.07 & 0.64189 & RRab\\
SDSS J030413.34+005828.1 & 54 & 19.34 & 18.27 & 18.00 & 17.93 & 17.94 & 0.61633 & RRab\\
SDSS J030413.80-011314.3 & 19 & 18.37 & 17.31 & 17.19 & 17.15 & 17.17 & 0.36579 & RRc\\
SDSS J031537.87-005341.9 & 30 & 19.08 & 17.97 & 17.74 & 17.65 & 17.66 & 0.62367 & RRab\\
SDSS J032109.31+000705.7 & 31 & 18.99 & 17.86 & 17.62 & 17.53 & 17.51 & 0.56928 & RRab\\
SDSS J032202.45-000151.6 & 12 & 21.63 & 20.49 & 20.28 & 20.13 & 20.09 & 0.67613 & RRab\\
SDSS J032910.97+003614.3 & 55 & 19.51 & 18.32 & 18.10 & 18.09 & 18.11 & 0.34381 & RRc\\
SDSS J032953.69-011413.0 & 64 & 20.45 & 19.46 & 19.21 & 19.11 & 19.11 & 0.58634 & RRab\\
SDSS J033343.92+004457.6 & 32 & 18.95 & 17.84 & 17.62 & 17.56 & 17.59 & 0.40888 & RRc\\
SDSS J033817.21-011112.9 & 62 & 20.70 & 19.69 & 19.41 & 19.29 & 19.32 & 0.67096 & RRab\\
SDSS J034239.98-000009.9 & 74 & 19.33 & 18.25 & 17.99 & 17.91 & 17.93 & 0.61181 & RRab\\
SDSS J034602.52-003342.2 & 41 & 19.31 & 18.20 & 18.08 & 18.06 & 18.13 & 0.35783 & RRc\\
SDSS J034830.07+004042.4 & 39 & 19.68 & 18.57 & 18.34 & 18.27 & 18.24 & 0.61444 & RRab\\
SDSS J034830.94-002320.9 & 56 & 19.88 & 18.79 & 18.53 & 18.49 & 18.50 & 0.63686 & RRab\\
SDSS J034836.34+005334.7 & 53 & 19.48 & 18.42 & 18.13 & 18.07 & 18.08 & 0.58038 & RRab\\
SDSS J034837.58+000804.8 & 57 & 19.15 & 18.13 & 17.90 & 17.84 & 17.87 & 0.64671 & RRab\\
SDSS J035400.50+002110.2 & 32 & 19.02 & 18.01 & 17.76 & 17.71 & 17.73 & 0.58990 & RRab\\
\enddata
\tablecomments{Objects are presented in order of increasing right ascension.}
\label{all_rr_table}
\end{deluxetable}
\clearpage

\begin{deluxetable}{lccccccc}
\tablecaption{High Amplitude Delta Scuti Variables}
\tablewidth{0pt}
\tabletypesize{\footnotesize}
\tablehead{\colhead{Object} & \colhead{Obs.} & \colhead{$u$} & \colhead{$g$} & \colhead{$r$}
  & \colhead{$i$} & \colhead{$z$} &\colhead{Period [days]}}
\startdata
SDSS J001142.02+005946.8 & 42 & 20.13 & 19.17 & 19.05 & 19.05 & 19.09 & 0.06837 \\
SDSS J001544.29+003735.1 & 84 & 20.03 & 18.92 & 18.92 & 18.97 & 19.03 & 0.04953 \\
SDSS J005312.91-000522.5 & 37 & 21.14 & 20.19 & 20.10 & 20.13 & 20.21 & 0.06073 \\
SDSS J013945.70-003325.6 & 50 & 20.80 & 19.73 & 19.74 & 19.77 & 19.82 & 0.06480 \\
SDSS J014200.19-001756.6 & 49 & 20.70 & 19.64 & 19.58 & 19.65 & 19.72 & 0.06682 \\
SDSS J014225.68+001549.6 & 51 & 20.27 & 19.28 & 19.18 & 19.18 & 19.23 & 0.09279 \\
SDSS J014442.67-003741.7 & 94 & 20.55 & 19.57 & 19.54 & 19.55 & 19.54 & 0.06245 \\
SDSS J014531.34+011211.9 & 56 & 20.09 & 19.21 & 19.10 & 19.08 & 19.12 & 0.08348 \\
SDSS J020841.38-002038.3 & 50 & 19.60 & 18.57 & 18.49 & 18.54 & 18.62 & 0.06145 \\
SDSS J021541.11-000750.1 & 47 & 20.50 & 19.55 & 19.47 & 19.53 & 19.57 & 0.05419 \\
SDSS J024132.57+002324.1 & 37 & 20.27 & 19.28 & 19.18 & 19.21 & 19.25 & 0.06852 \\
SDSS J024751.89-002434.5 & 47 & 21.13 & 20.17 & 20.07 & 20.10 & 20.07 & 0.06713 \\
SDSS J030256.28-004150.7 & 41 & 18.84 & 17.86 & 17.82 & 17.87 & 17.93 & 0.05813 \\
SDSS J030758.97+000751.7 & 58 & 20.42 & 19.41 & 19.41 & 19.44 & 19.51 & 0.06542 \\
SDSS J031956.18-000238.9 & 49 & 20.00 & 19.01 & 18.96 & 19.00 & 19.08 & 0.05815 \\
SDSS J034507.01+010040.1 & 30 & 21.30 & 20.16 & 20.19 & 20.21 & 20.13 & 0.05733 \\
\enddata
\tablecomments{Objects are presented in order of increasing right ascension}
\label{all_dsc_table}
\end{deluxetable}
\clearpage

%%%%%%%%%%%%%
%% FIGURES %%
%%%%%%%%%%%%%

  \clearpage

  \clearpage

  \clearpage
  
  \clearpage

  \clearpage

  \clearpage

  \clearpage

  \clearpage

  \clearpage

  \clearpage

  \clearpage

  \clearpage

  \clearpage

  \clearpage

  \clearpage

  \clearpage

  \clearpage

  \clearpage

  \clearpage

  \clearpage

  \clearpage

  \clearpage

  \clearpage

  \clearpage

  \clearpage

  \clearpage

  \clearpage

\end{document}